\documentclass[useAMS,usenatbib]{mn2e}
\usepackage{epsfig,natbib}
\usepackage{eufrak}
\usepackage{amsmath}
\newcommand{\prd}{Phys.~Rev.~D}
\newcommand{\aap}{A.\&Ap.}
\newcommand{\mnras}{MNRAS}
\newcommand{\apj}{ApJ}
\newcommand{\apjl}{ApJ}
\newcommand{\physrep}{Physics Report}
\newcommand{\nat}{Nature}
\newcommand{\om}{\Omega_{\rm m}}
\newcommand{\ob}{\Omega_{\rm b}}

\newcommand{\beq}{\begin{equation}}
\newcommand{\eeq}{\end{equation}}
\newcommand{\ev}[1]{{\mathbf e}_{#1}(z)}
\newcommand{\Mlim}{M_{\rm lim}}
\newcommand{\Msun}{M_\odot}

\newcounter{remcount}

\title[Future Probes of Dark Energy]
{Complementarity of Future Dark Energy Probes}
\author[J.~Tang et al.]
{Jiayu Tang\thanks{Email: jtang@star.ucl.ac.uk}, Filipe B. Abdalla
  and Jochen Weller\\Department of
Physics \& Astronomy, University College London, Gower Street, London WC1E 6BT, UK.}
\date{Accepted ???, Received ???; in original form \today}
\pagerange{\pageref{firstpage}--\pageref{lastpage}}
\pubyear{2007}
\begin{document}
\setcounter{remcount}{0}
\maketitle
\label{firstpage}
\begin{abstract}
In recent years a plethora of future surveys have been suggested to
constrain the nature of dark energy. In this paper we adapt a binning
approach to the equation of state factor ``w'' and discuss how future
weak lensing, galaxy cluster counts, Supernovae and baryon acoustic
oscillation surveys constrain the equation of state at different
redshifts. We analyse a few representative future surveys, namely DES, PS1, WFMOS, PS4, EUCLID, SNAP and SKA,
and perform a principal component analysis for the
``w'' bins. We also employ a prior from Planck cosmic microwave
background measurements on the remaining cosmological parameters. We
study at which redshifts a particular survey constrains the
equation of state best and how many principal components are
significantly determined. We then point out which surveys would be
sufficiently complementary. We find that weak lensing surveys, like
EUCLID, would constrain the equation of state best and would be able to
constrain of the order of three significant modes. Baryon acoustic
oscillation surveys on the other hand provide a unique opportunity to
probe the equation of state at relatively high redshifts. 
\end{abstract}
\section{Introduction}
One of the most challenging puzzles in modern physics is the cause for
the observed accelerated expansion of the Universe. The
renewed interest in accelerating cosmologies was born out of Type Ia
Supernovae observations a decade ago
\citep{Perlmutter:97,Riess:98,Perlmutter98}. In the meantime the
cosmic concordance model \citep{Ostriker:95} has been confirmed by complementary
observations of the Cosmic Microwave Background (CMB) \citep{Spergel03}, the
large scale distribution of galaxies \citep{Efstathiou02,Tegmark:04}
and clusters of galaxies \citep{Allen:04,Rapetti:04}. In addition observations of
large scale CMB anisotropies \citep{Scranton:03, Padmanabhan:04} and
Baryon Acoustic Oscillations (BAO) in the matter distribution
\citep{Cole:05,Eisenstein:05} also indicate the presence of late time
acceleration in the Universe.

The simplest way to model an accelerated expansion in the Universe is
by including a cosmological constant in Einstein's equations of
gravity. However, the introduction of a cosmological constant requires
an extreme fine tuning of the initial conditions of our Universe to about
120 orders of magnitude. This motivates the search for alternative
explanations for the speeding up of the expansion rates, although there
have been recent successes in obtaining universes with the observed
cosmological constant in a range of the multi verses of the meta-stable vacua of
the string theory landscape \citep{Kachru:03}. This, in connection with
the anthropic principle \citep{Weinberg:87,Efstathiou:95,Bousso:06,Peacock:07,Cline:07},
could motivate the observed value of the
cosmological constant. Nevertheless, it is still valid to seek
other alternatives to a cosmological constant. One approach is to introduce
a dynamical scalar field, usually dubbed Quintessence, which is either slowly rolling down a
potential or is trapped in a false vacuum state
\citep{Wetterich:88,Peebles:88,Ratra:88,1995PhRvL..75.2077F,Ferreira:98,Zlatev:98,Albrecht:00}. These
models can lead to cosmic
acceleration with less fine tuning. Another possibility was that the
Universe is permeated by a network of cosmic defects, which also acts
like a source in the energy-momentum tensor and can lead to
accelerated expansion \citep{Battye:99,Battye:05b}.
The dynamical evolution of the Universe in Quintessence or defect based models
is governed by the equation of state of the dominant components. The
equation of state is given by the ratio of the pressure to the density
of these fluids. For a cosmological constant this is
-1. However dynamical dark energy models can deviate from this value
and the ratio can also vary with time. Typically the ratio is named
``$w=p_{\rm de}/\rho_{\rm de}$''. The precise measurement of the value of ``$w$'' is hence one of
the foremost tasks for observational cosmology. Any significant measurement of a
deviation of ``$w$'' from -1, would be a major result. Currently
conservative constraints on ``$w$'' are in the region of $w=-1\pm 0.15$
\citep{Tegmark:04,Astier:06}.

Up to now, it has been difficult to make any connection between suggested Quintessence fields
and fundamental theories. Another possibility to obtain accelerated
expansion is the modification of Einstein's gravity on large
scales. There are two approaches along these lines. One is to modify a
proposal by \cite{Starobinsky:80}, which includes higher order
curvature terms in the gravitational
action to obtain accelerated expansion in the early Universe. This has
been adapted to explain the observed accelerated expansion
\citep{Carroll:04,Carroll:05}. Another approach is to obtain effective
four dimensional Friedman equations, from a five or higher dimensional
theory, where our standard model is confined to a brane, with only
gravity allowed to leak out into the extra dimension(s)
\citep{Dvali:00,Deffayet:02,Dvali:03}. Although these models might
have structural problems such as ghosts \citep{Koyama:05,DeFelice:06} it is a worthwhile
exercise to explore their cosmological phenomenology. In particular
they are different to Quintessence models in the way structures grow
in the Universe \citep{Song:05,Koyama:06}. Finally it is possible, at
least in theory, that the observed acceleration is an effect, which
results from an averaging procedure to describe the large scale
dynamics of an intrinsically inhomogeneous Universe
\citep{Buchert:05}.

All the models described above have in common that they can, but don't
have to, deviate from the evolution of a so called $\Lambda$-cold dark
matter ($\Lambda$CDM)
Universe consisting of a cosmological constant and dark
matter. Effectively the background evolution for all these models can
be mimicked by a suitable chosen additional component in the
energy-momentum tensor. Hence the quest for
``$w$'' goes beyond its usual application to Quintessence models,
although they are the only models where there is actually a physical
meaning associated with ``$w$''. It is therefore an interesting question
to ask if ``$w$'' is different from -1 the value for a cosmological
constant. The first attempt to address this question is to look for
different {\em  constant} values of ``$w$''. However, this approach is
very restrictive given that dark energy models allow for evolving
equation of state factors ``w''
\citep{Weller:00,Weller01,Maor:2001ku}. Hence it is important to allow
for a time varying equation of state parameter ``$w$''. A simple Taylor
expansion in redshift \citep{Weller01} is too restrictive if we want
to include data sets which are sensitive to high redshifts, like the CMB.
A more successful approach is to parameterise a time varying equation
of state with a Taylor expansion in the scale factor $a$, like
$w=w_0+w_a(1-a)$ \citep{Chevallier:01,Linder:03}. This was also the
parameterisation of choice adapted by the Dark Energy Task Force
(DETF) \citep{DETF:06}. The DETF chose this parameterisation to be
able to compare different proposed surveys on an equal footing. In
addition this parameterisation allows to some extend the distinction
of two subclasses of Quintessence models, namely freezing and thawing
out models \citep{Linder:06}.

In this paper we want to pursue a more general approach. In order to
figure out the redshift sensitivity to the dark energy density, or
better its time derivative, we will bin the equation of state in
redshift \citep{Huterer:03,Crittenden:05,Huterer:07a,Albrecht:07,Sullivan:07}. Although, there
are some ambiguities how to exactly perform the binning
\citep{dePutter:07}, the binning approach is in general the most
model independent approach to fit for the background evolution of the
Universe. Note that this approach is inherently related to the
weighting function method \citep{Saini:03,Simpson:06} in the limit of
an infinite number of bins.

We are now faced with the difficult task of selecting surveys to
pursue our analysis. We emphasize that our selection is
subjective and our aim is that we have at least one
representative survey for each of the probes we are discussing.
In this introduction we will mention briefly the surveys and what
their technical specification is expected to be, while we describe how they exploit
particular cosmological probes in the sections where we discuss
constraint from the different probes.

We start with Stage II surveys according to the classification by
the DETF \citep{DETF:06}. Already online in its simplest
configuration is the PANoramic Survey Telescope and Rapid Response
System (PAN-Starrs)\footnote{see at: {\tt
    pan-starrs.ifa.hawaii.edu}}. PAN-Starrs in it's full configuration
will consist of four 1.8 metre telescopes, equipped with a wide
field camera with a field of view of 7 arcminutes and a total of
1.4 Gigapixels. Currently only one of the four planned telescopes
is deployed. We refer to this setup as PAN-Starrs1 (PS1) in
contrast to the full setup PAN-Starrs4 (PS4). The PAN-Starrs
survey is planned  to take exposures of $\sim 60$ seconds in each of
the grizy filters in the PS1 configuration and should go much deeper in the PS4
configurations with grizy filters. One of the surveys
with PS1 will be a nearly all sky, $20.000\,{\rm deg}^2$,
survey up to a median redshift of $z\approx 0.5$. The nearly all
sky survey is hoped to be exploited for large scale structure and
BAOs \citep{Peebles:70,Hu98;3,Blake:03} experiments as well as
weak gravitational lensing observations. The medium deep survey will be
suitable for cosmological constraints with Type Ia SNe out to
redshifts of $z\sim 0.8$. PS1 in the DETF category is a
Stage II survey, while PS4 is Stage III.

Another survey already deployed, which is relevant to dark energy
science, is the South Pole Telescope (SPT) \citep{Ruhl:04}. The
SPT is a 10 metre telescope with a bolometer array in its focal
plane, designed to detect thousands of clusters of galaxies via
their Sunyaev-Zel'dovich \citep{Sunyaev:72} decrement in the
sub-millimeter domain over $4.000\,{\rm deg}^2$. Current plans are
to operate at 150, 219 and 274 GHz. This survey in addition with a
redshift survey will allow to constrain dark energy with the
clustering and redshift distribution of galaxy clusters.

We describe now near-future surveys, which are currently not
deployed. The Dark Energy Survey (DES) \citep{DES}
is an optical-near infrared survey under construction. The optical part will
go down to 24$^{\rm th}$
magnitude in the grizy bands on the 4 metre Blanco telescope at
Cerro Tololo Inter-American Observatory (CTIO). The IR counterpart will
come from part of the Vista Hemisphere Survey (VHS) survey performed by ESA's 4m Visible and
Infrared Survey Telescope for Astronomy (VISTA)
in the JHK bands to $\sim$20{\rm th} magnitudes. The DES camera has a
2.2 deg field of view, which allows to obtain imaging and
photometric informations of millions of galaxies. It is planned to
observe $5.000\,{\rm deg}^2$, which overlap with the area of the
SPT survey. This will allow to constrain dark energy via weak
lensing tomography, galaxy clustering and galaxy cluster redshift
distribution. In addition there is a wide-area survey where 10\% of
the allocated time will be used to follow Type Ia SNe
light-curves.

The Wide Field MultiObject Spectrograph (WFMOS)
\citep{Glazebrook:05} will consist of a new wide-field (1.5 deg) 4000
fibers spectrograph with a passband of 0.39-1.0 microns which would
be deployed in a 8{\rm m} class telescope. Current
suggestions are that there will be 10 low dispersion spectrographs
with R=1800 in the blue and R=3500 in the red. This will enable
the measurements of BAOs in $0.5<z<1.3$ and $2.3<z<3.3$ using the
redshifts of millions of galaxies over $\sim 2,000\,{\rm deg}^2$
at low z and $\sim 500\,{\rm deg}^2$ at high z.

We will now introduce two proposed satellite missions, which might
come online in the long term. Both NASA and ESA in their ``Beyond
Einstein'' and ``Cosmic Vision'' studies have proposed to pursue
missions which are able to probe dark energy\footnote{see at: 
{\tt universe.nasa.gov/home.html} and {\tt www.esa.int/esaSC/SEMA7J2IU7E\_index\_0.html}}. One of the
contenders for NASA is the SuperNovae Acceleration Probe (SNAP)
\citep{SNAP:05a,SNAP:05b}. SNAP is a 2 metre telescope in space
with a 0.7 square-degree wide field imager and a $R\sim 100$
spectrograph. Both are sensitive in the 0.4-1.7 $\mu$m wave-band.
SNAP is designed to probe dark energy with a SuperNovae and weak
lensing survey, where the weak lensing takes advantage of the 9
filters with a depth of 26.6AB magnitude. We will discuss the
particulars of the two surveys in the relevant sections later in
the paper. The EUCLID survey is a combination of the former
SPACE \citep{space} and DUNE proposals.
The Dark UNiverse Explorer (DUNE) \citep{Refregier:06}
is a proposed wide field space imager on a 1.2 metre
telescope with a 1 deg$^2$ visible/near-IR field of view. It is
designed to measure cosmic lensing over 20.000 deg$^2$ of the sky
and will exploit this as its main cosmological probe.
DUNE is designed to use one broad visible band (R+I+Z) for accurate
shape measurement for weak lensing and Y,J,H
in the near infrared to complement optical photometry which should be available by 2017
for accurate photometric redshifts in the range $0<z<2$. 

The Square Kilometer Array (SKA) is a radio interferometer planned to operate
in a large range of frequencies (60{\rm MHz} - 35{\rm GHz}) with a sensitivity
of 20,000${\rm m^2/K}$ in the range 0.5-1.4{\rm GHz} with a field of view
of $\sim$ 200{\rm deg$^2$} at 500{\rm MHz} \citep{2004astro.ph..9274C}.
This interferometer would allow to locate galaxies in the Universe given their
21{\rm cm} line emission and would allow us to perform large surveys for galaxy
evolution and Cosmology \citep[][Abdalla et al. in preparation]
{2004NewAR..48.1013R,2004NewAR..48.1063B,2005MNRAS.360...27A,
2007MNRAS.381.1313A}. The project is planned to be an on-going
project which should build up from $1\%$ demonstrators in the following
years to a $10-15\%$ core in $\sim 5-10$ year time to its full completion in
$\sim 2020$.

This concludes our preview of the surveys being discussed in this
paper. However we will exploit one more survey, not for its
ability to constrain dark energy, rather for its ability to
constrain other cosmological parameters to high precision. The
forthcoming Planck satellite mission will observe the sky in 9 radio wave bands in order to measure the anisotropies to in the
CMB \citep{Planck}. Given that the primary anisotropies are mainly a probe of the
angular diameter distance to the surface of last scattering,
Planck alone is not a strong probe of dark energy. However it puts
strong constraints on other cosmological parameters\citep{Planck}.
We will hence use forecasts for the Planck surveyor to put prior
constraints on the remaining cosmological parameters.

We start the paper by discussing the method of principal
components analysis (PCA) in the context of the binning of the
equation of state in Section \ref{sec:pc}. In Section
\ref{sec:cmb} we discuss constraints arising from the Planck
surveyor and introduce the covariance matrix we will use in the
later sections of the paper. In Section \ref{sec:sne} we study the
principal components of Type Ia Supernovae as cosmological probes.
In this section we will also analyse the impact of marginalization
of the remaining cosmological parameters. In Section \ref{sec:wl}
we will study the PCA in the context of cosmic lensing. Section
\ref{sec:cc} will analyse cluster counts as cosmological probes. In the Section \ref{sec:bao} we will examine the
ability of BAOs to constrain dark energy. Before concluding in
Section \ref{sec:conc} we will discuss how to perform a joint
principal component analysis between complementary surveys.

\section{Principal Components of the Equation of State}\label{sec:pc}
We will now introduce the method we use to constrain the equation of
state of dark energy. As outlined in the introduction we will pursue
a binning approach to the equation of state. Binning in this context
was first introduced by \citet{Huterer:03}, but has since been studied
by many authors
\citep{Crittenden:05,Huterer:07a,Albrecht:07,Sullivan:07,dePutter:07}.
There are different possibilities to bin the equation of state, but
the one we follow here is given by

\begin{equation}
w(z)=\left\{\begin{array}{ll}
  w_{\rm i}, & z_{\rm i}-\frac{\Delta
z_{\rm i}}{2}\le z \le z_{\rm i}+\frac{\Delta z_{\rm i}}{2} \\
  w_{h}, & z>z_{\rm max}
\end{array}\right.\;,
\label{eq:w_z}
\end{equation}
where $w_i$ is value of the equation of state of dark energy in a
given redshift bin $z_{\rm i}\pm\frac{1}{2}\Delta
z_{\rm i}$. Note that beyond a maximum redshift $z_{\rm max}$ we
assume a constant equation of state factor $w_h$. Although the binning
of $w$ in redshift leads to a quasi model independent fitting
procedure, the increased number of parameters in general lead to a
better fit but with the drawback of greatly increased
error bars. Typically we will choose the redshift bins in the region of
$\Delta z_i = 0.05$, hence obtaining tens of new parameters for a
given survey. Occam's razor tells us that this large number of
parameters do not lead to significant improvement of the fit in
general. However just increasing the bin width or cutting of all
surveys at a given redshift does not do justice to all surveys we are
going to discuss. This is because in general the error bars between
different w bins are highly correlated. What we want is to extract
information described in a decorrelated way. This can be achieved by
diagonalising the correlation matrix of the w bins and then expressing
the fit in terms of the eigenmodes. This is essentially a Principal
Component Analysis (PCA). By ordering the eigenmodes with respect to
the size of their corresponding eigenvalues, we build a hierarchy
from the best constrained modes to the least constrained ones.

For any given experiment we look at the parameter vector, which
includes the standard cosmological parameters, like the matter
contents $\om$, the baryon contents $\ob$, the Hubble constant $H_0$, the spectral index $n_s$ of
primordial perturbations , the amplitude of the primordial power
spectrum $\sigma_8$. Note that we restrict our analysis to flat cosmologies.
In addition, as we
shall see later, each experiment has also a few nuisance parameters.
We introduce the parameter vector
\beq
\Theta = (\theta_i) = \left\{\om, \ob, H_0, ..., w_1,
w_2, w_3, ..., w_N, w_h\right\}\; ,
\eeq
where we assume $N$ bins in redshift to fit $w$. In order to obtain
the correlation matrix for a given survey we need to know the
likelihood of the data vector $\mathbf x$ given the parameters
$\Theta$: $p({\mathbf x},\Theta)$. From this we can estimate the
correlation matrix with the help of the Fisher information matrix
\beq
F_{ij}^{\rm obs} \equiv
\left<\frac{\partial^2\cal{L}}{\partial\theta_i\partial\theta_j}\right>\; ,
\eeq
with ${\cal{L}} = -2\,\ln\, p({\mathbf x},\Theta)$. We will show how
to calculate the Fisher matrix for the particular surveys in the
corresponding sections.

In order to study the ability of a given survey to constrain dark
energy in different redshift bins we first have to marginalize over
the other cosmological parameters and the nuisance terms. This
usually involves the addition of prior information on the other
cosmological parameters. For the work presented here, this means in
general to add a prior from the forthcoming Planck CMB surveyor and
this probe will be discussed in the next section. This means we just
have to multiply the likelihood $p$ with the prior to obtain the
posterior, or in terms of the Fisher information matrix
\beq
\tilde{\mathbf F} \equiv {\mathbf F}^{\rm prior}+{\mathbf F}^{\rm obs}\;.
\label{eq:marg_f}
\eeq
The remaining procedure is to integrate the posterior over the
nuisance parameters and the cosmological parameters not involving the
equation of state. In terms of the Fisher matrix approximation this can
be achieved by inverting $\tilde{\mathbf F}$, projecting this inverse
to the space involving the equation of state parameters $w_i$ and
inverting again, i.e.
\beq
{\mathbf F} \equiv \left[ \left(\tilde{\mathbf F}^{-1}\right)_{\{i,
    j\}}\right]^{-1}\; ,
\eeq
where the index $\{i, j\}$ runs over the indices of the $\theta_k$
corresponding to the w bin parameters.

The matrix ${\mathbf F}$ is
the estimator of the correlation matrix we want to decorrelate. We
will hence calculate the eigenvalues $\lambda_i$ and the eigenvectors
$\ev{i}$, where the vector has the dimension of the number of bins
$N$. In the basis of the eigenvectors $\ev{i}$ the Fisher matrix is
diagonal and hence decorrelated. The eigenvectors represent the {\em principal components} of the
corresponding survey. We can now write the underlying $w(z)$ in terms
of the eigenvectors
\beq
w(z) = \sum_{i=1}^N\alpha_i\ev{i}\, ,
\label{eq:w}
\eeq
The
error bars on the $\alpha_i$ coefficients are then given in terms of
the eigenvalues $\Delta\alpha_i = \lambda^{-1/2}_i$. Hence the error
on $w$ is given by
\beq
\delta w(z)^2 =
\sum_{i=1}^N\Delta\alpha_i^2\ev{i}^2=\sum_{i=1}^N\frac{\ev{i}^2}{\lambda_i}\;,
\label{eqn:dw}
\eeq
where $\ev{i}^2\equiv (e_{i1}^2,e_{i2}^2,...,e_{iN}^2)^T$ is a vector.

Since we have now a hierarchy of modes, with increasing error bars, we
can in fact use a criteria similar to Occam's razor to decide how many
modes are constrained significantly by the
experiment. \citet{Huterer:03} used a risk factor to decide
this. This includes the increasing variance with the number of modes,
but also a decreasing bias factor. In a modern Bayesian approach this
problem can be addressed with the Bayesian information criterion or
evidence \citep{Saini:04,Liddle:04}. We will come back to this in the discussion section
of the paper. In the meantime we will concentrate on studying the
structure of the eigenmodes and their associated eigenvalues. While
\citet{Albrecht:07} concentrated on the area of the posterior
probability for a given number of modes to assign a {\em figure of
  merit} to an experiment, we will in addition study the redshift
dependence of the eigenvectors. This will allow to shed
light on the redshift sensitivity of a given survey. In order to study
this dependence and accuracy we will plot in the forthcoming sections
the quantity
\beq
\phi_i(z)\equiv N\left|\sqrt{\lambda_i}\ev{i}\right|\;,
\label{eqn:lev}
\eeq
where the amplitude of this expression is representative to the
accuracy of the mode and $\ev{i}$ is encoding the redshift
sensitivity. Note that the prefactor $N$ appears in
Eqn. (\ref{eqn:lev}) in order to make this quantity quasi-independent to the
number of bins.

We will now proceed to forecast the parameter constraints from the
Planck surveyor, which we subsequently will use as prior information
for the other surveys.

\section{Cosmic Microwave Background}
\label{sec:cmb}
In this section we describe our principal
component analysis for the case of the CMB as a cosmological
probe. The Fisher matrix for CMB power spectrum is given by
\citep{Zaldarriaga97,Zaldarriaga:97b}
\begin{equation}
  F_{ij}^{CMB}=\sum_{l}\sum_{X,Y}\frac{\partial
    C_{X,l}}{\partial\theta_{i}}\mathrm{COV^{-1}_{XY}}\frac{\partial
    C_{Y,l}}{\partial\theta_{j}},
  \label{eq:cmb_fisher}
\end{equation}
where the variables X and Y represent different cross and auto correlations in the CMB power spectrum, i.e.
$C_{\rm X,l}$ represent the TT, EE or TE power spectra where $T$ represents
temperature anisotropy and $E$ represents E-mode polarization anisotropy.
The covariance between auto and cross power spectra is given by ${\rm COV_{XY}}$\citep{Zaldarriaga97,Zaldarriaga:97b}.

The covariance ${\rm COV_{\rm XY}}$ will depend on the experimental noise of a given CMB experiment.
In this paper, we choose the Planck experiment as a baseline to define the noise attributes that will determine
${\rm COV_{\rm XY}}$. We refer the reader to the Planck blue book for these values \citep{Planck}.
The Planck mission is designed to measure the CMB anisotropies in
nine frequencies
over the entire sky. We assume here that the different frequencies will provide enough information so that foregrounds
can be removed over as much as 80\% of the sky in order to perform cosmological analysis over this area.
We assume only {\em one} science frequency is used after foreground
cleaning, e.g. the 143 GHz channel.

Usually one can use all the information including low multipoles to determine cosmological parameters from the CMB.
However, in this paper we are concentrating on the role of the
dark energy equations of state. In order to treat low CMB multipoles
correctly one requires to calculate the integrated Sachs-Wolfe (ISW)
effect. For a consistent treatment of the ISW it is necessary to
include dark energy perturbations \citep{Weller:03}. There
is a singularity in the perturbation equations (not necessarily
essential), when the equation of state crosses the line $w=-1$
\citep{Hu:05,Caldwell:05}, which of course for arbitrary bins can
happen. In addition there arises a problem for the
binning approach in context of the ISW. Since the perturbations in the
dark energy component depend on the time derivative of the equation of
state, any step binning in w involves singular derivatives
\citep{Weller:03}. This, of course, could be resolved by imposing
continuous bins \citep{Crittenden:05}.
In light of these difficulties we choose not to include CMB multipoles
below $\ell_{\rm min}=20$. The
drawback of this approach is that the constraints from the CMB on the
optical depth are poor, but we prefer to have a conservative estimate
rather than treat this perturbations incorrectly.

If we ignore foreground noise and the late time ISW effect, Eqn.~(\ref{eq:cmb_fisher})
forecasts the errors on parameter set $\theta_{\rm i}$ given by the Planck experiment. For the purposes of this paper we will
treat Eqn.~(\ref{eq:cmb_fisher}) as a prior for other dark energy
probes. Note that in addition to the cosmological parameters
introduced in section \ref{sec:pc}, the CMB analysis requires the
optical depth $\tau$ due to reionization of the Universe as a free
parameter.

\begin{table}
  \begin{center}
    \begin{tabular}{|c|c|c|}
      \hline
      Parameter     & Fiducial Value   \\
      \hline
      $\Omega_{\rm m}$   &   $ 0.3$  \\
      $\Omega_{\rm b}$    &    $0.04$  \\
      $\sigma_{\rm 8}$   &    $ 0.78  $  \\
      $H_{\rm 0}$   & $72$   \\
      $n_{\rm s}$   &    $ 1.0 $   \\
      $\tau$   &   $  0.09 $  \\
      $w(z)$ & $-0.9$  \\
      \hline
    \end{tabular}
    \caption{The fiducial cosmological model used for the the
      forecasting analysis in this paper.}
\label{tb:cosmo_fiducial}
  \end{center}

\end{table}
The fiducial values of the cosmological parameters we use for the
analysis are listed in Table~(\ref{tb:cosmo_fiducial}). In the
remainder of the paper we refer to the cosmological
parameters excluding the $w$-bin parameters as {\em standard} cosmological parameters.
In this paper we use a modified version of CAMB \citep{Lewis:1999bs} to calculate CMB
power spectrum; we modified the code by allowing for binning in $w$ as
defined in Eqn. (\ref{eq:w_z}). In the case of the CMB alone we choose
$z_{\rm max}=3$. This has been chosen at a relatively
low redshift compared to the CMB redshift because for a general
equation of state $w(z)<0$ it is mainly the low redshift behaviour
that has the highest impact on CMB observables \citep{Saini:03}.
The choice of a higher $z_{\rm max}>3$ will have very
little impact on the principal components that we study.
For the convenience of joint analysis on different probes, we set the
redshift width of each bin as a constant $\Delta z=0.05$. We emphasize that our
results do not change if we decrease the size of the bins. With this set
up it is simple to perform a joint analysis between two experiments.
Even though two experiments may have two different maximum redshifts
and hence a different total number of bins N, in the overlapping
redshift range the $w_{\rm i}$ are the same and it is straight forward to add their Fisher matrices and combine them to
produce principal components for joint experiments. In order
to do this, we simply add corresponding columns in the Fisher
Matrices. For parameters, which are only relevant for one experiment
but not for a second experiment, we insert zeros in the Fisher matrix
given $\partial O_{\rm j}/\partial \theta_{\rm i}=0$ for the
relevant observable $O_{\rm j}$ and $\theta_{\rm i}$ for the second experiment.

\begin{figure}
\includegraphics[width=8.0cm,angle=0]{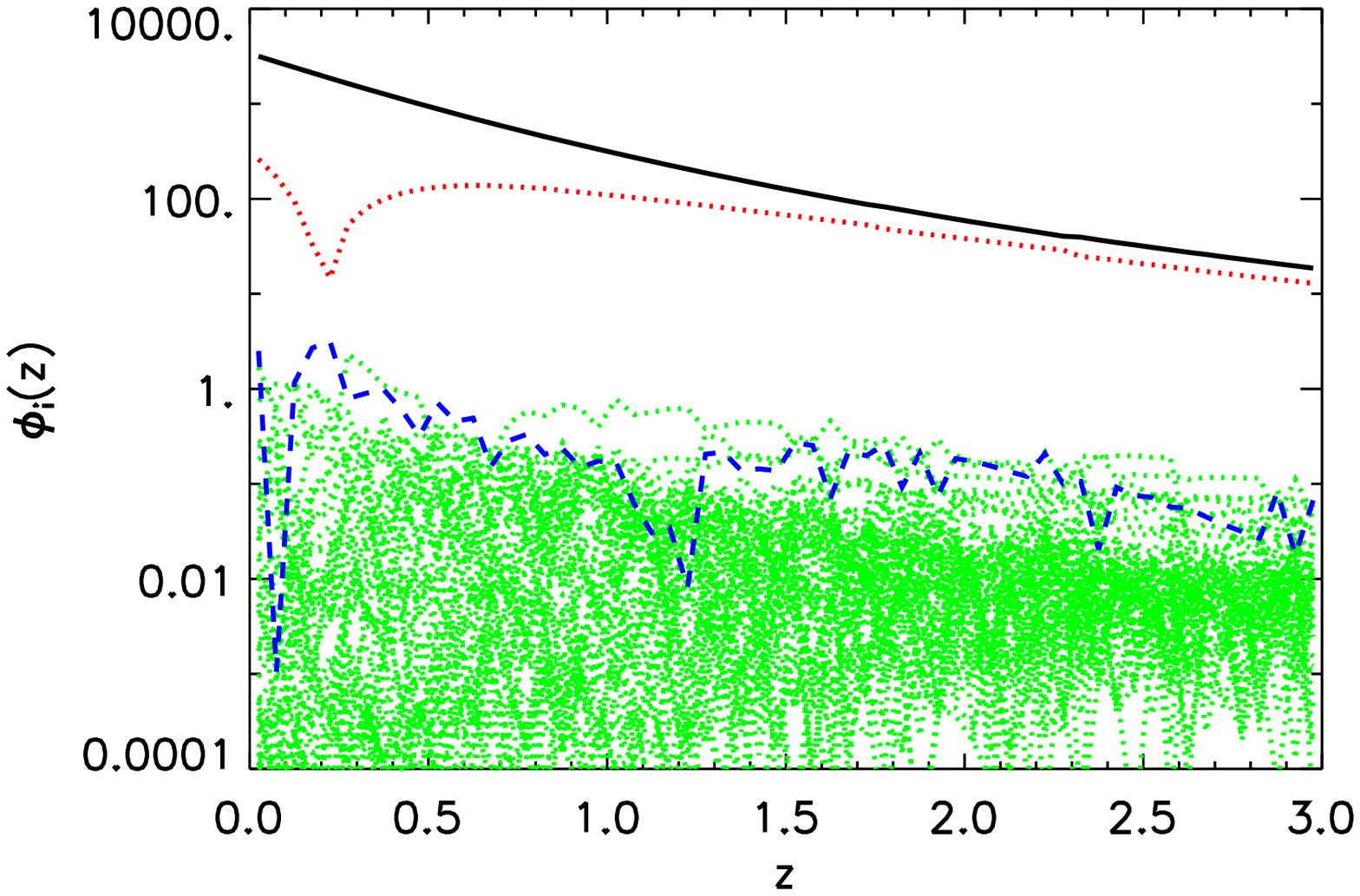}
\includegraphics[width=8.0cm,angle=0]{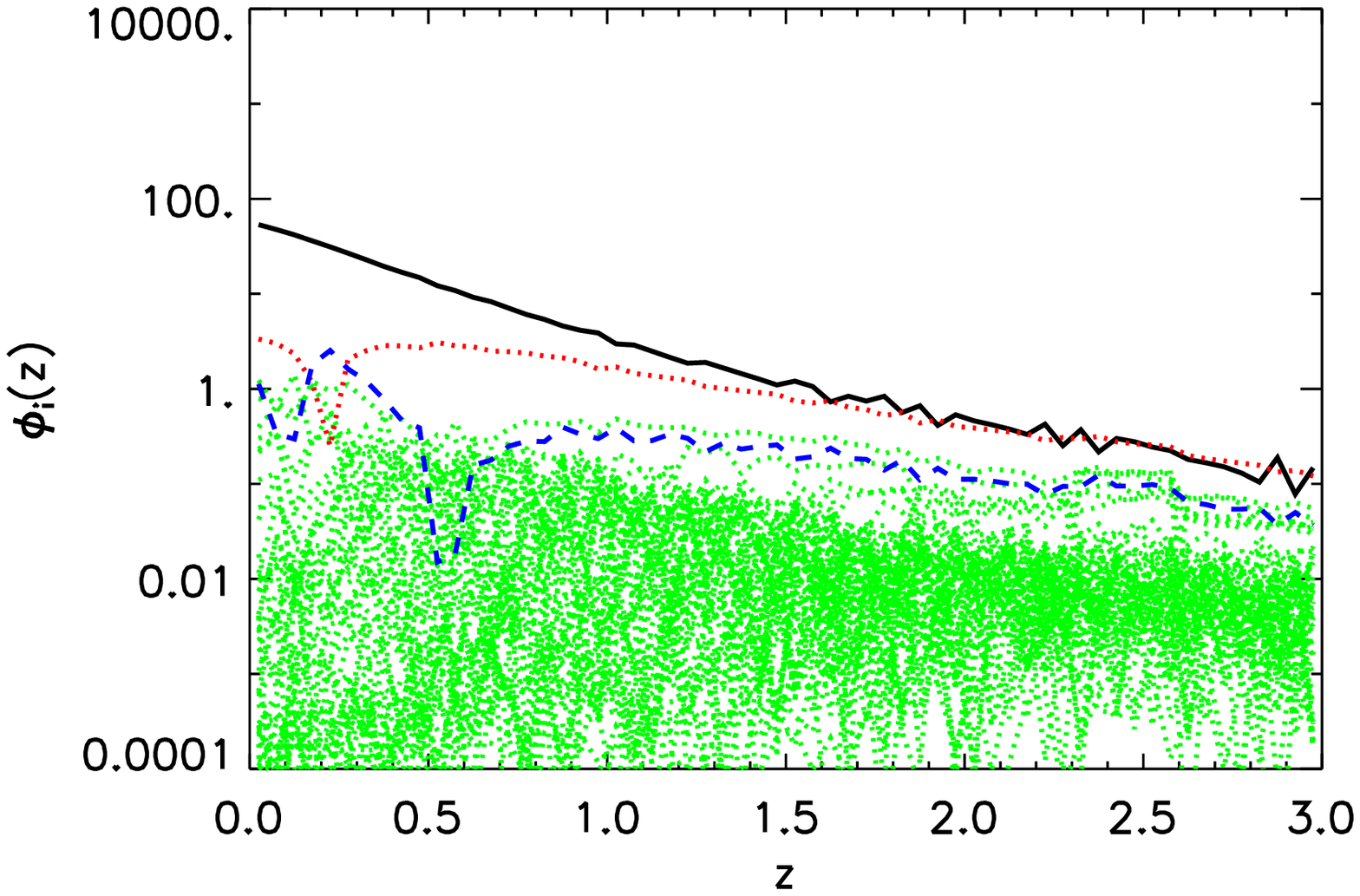}
\caption{We plot $\phi_i(z)=N\sqrt{\lambda_{\rm i}}|e_{\rm i}|$ for
  the Planck experiment.
The solid line indicates the first best estimated principal
component (PC). The dotted line shows the second best estimated PC and the dashed line
indicates the third one. The faint dotted lines show the remaining
PCs. The upper panel represents the PCs with fixed cosmological parameters, while the lower
panel shows the PCs after cosmological parameters have been
marginalized. \label{fig:planck_evec}}
\end{figure}

In Fig. \ref{fig:planck_evec} we plot $\phi_i(z)$ as given in Eqn. (\ref{eqn:lev}) for the Planck
experiment. The upper panel shows $\phi_i(z)$ when cosmological parameters are fixed.
In this panel, the first and second components are three orders of magnitude higher than the rest of $\phi_i$.
No sign transition occurs on the first mode, while on the second mode
$\phi_2$ changes sign around $z=0.25$. This is indicated by the kink
in the logarithmic plot. In the lower
panel we plot $\phi_i(z)$ after the standard cosmological parameters have been marginalized. Compared with
the upper panel, all $\phi_i$ are one order of magnitude smaller. This is due to the degeneracy between $w(z)$ and
the cosmological parameters, in particular $\Omega_{\rm m}$.

\begin{figure*}
\begin{center}
\begin{minipage}[c]{1.00\textwidth}
\centering
\includegraphics[width=5.7cm,angle=0]{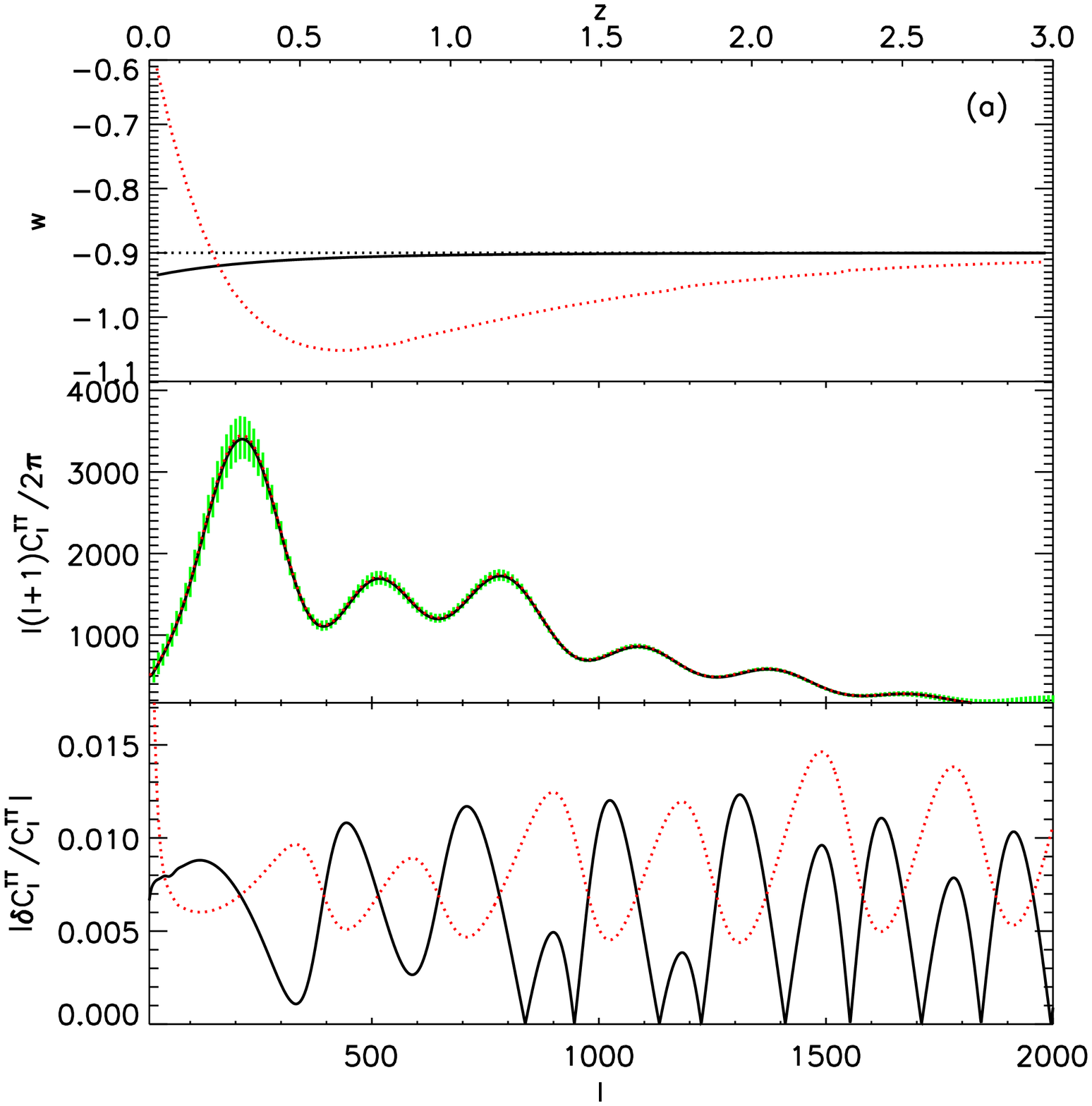}
\includegraphics[width=5.7cm,angle=0]{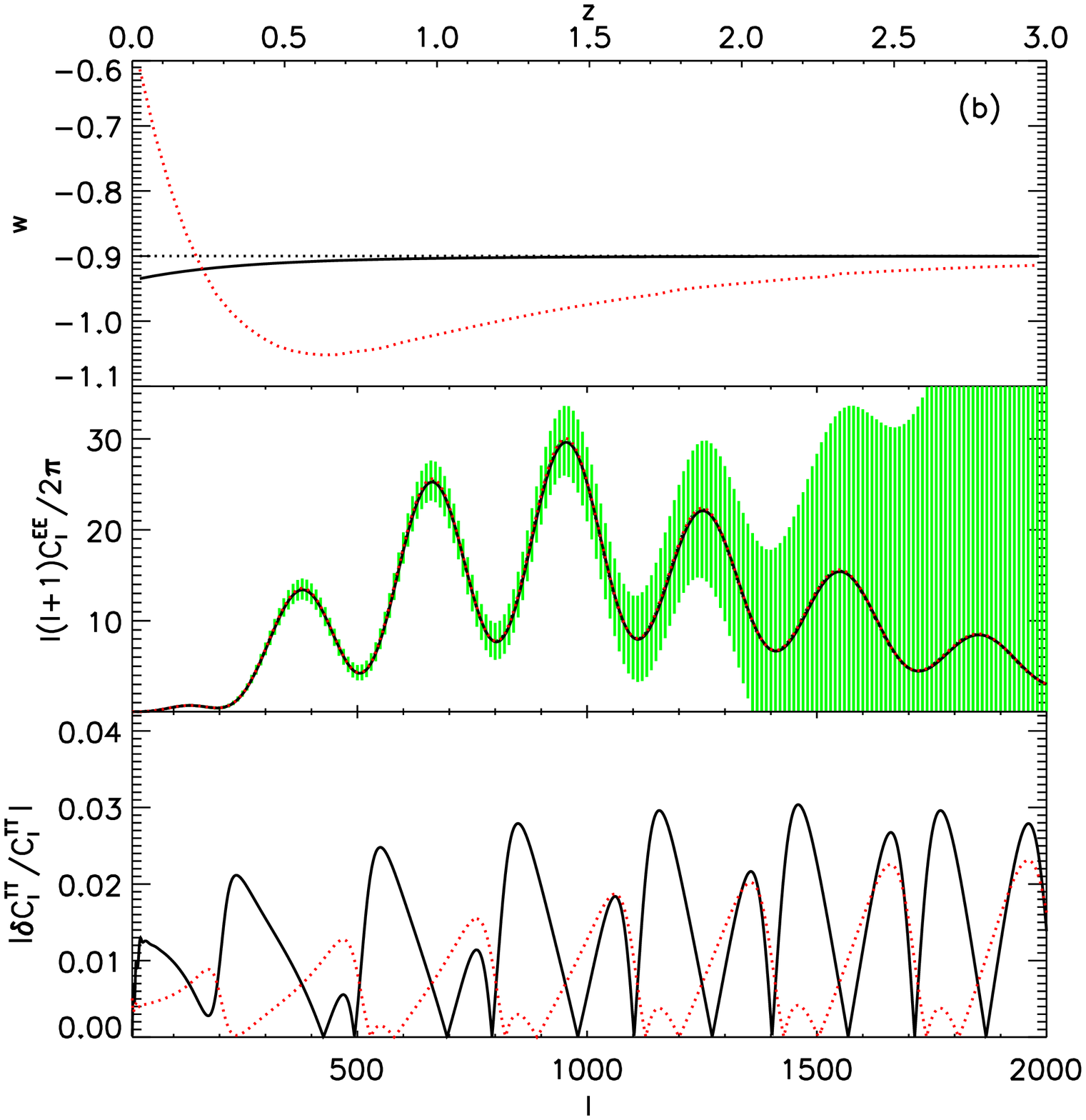}
\includegraphics[width=5.7cm,angle=0]{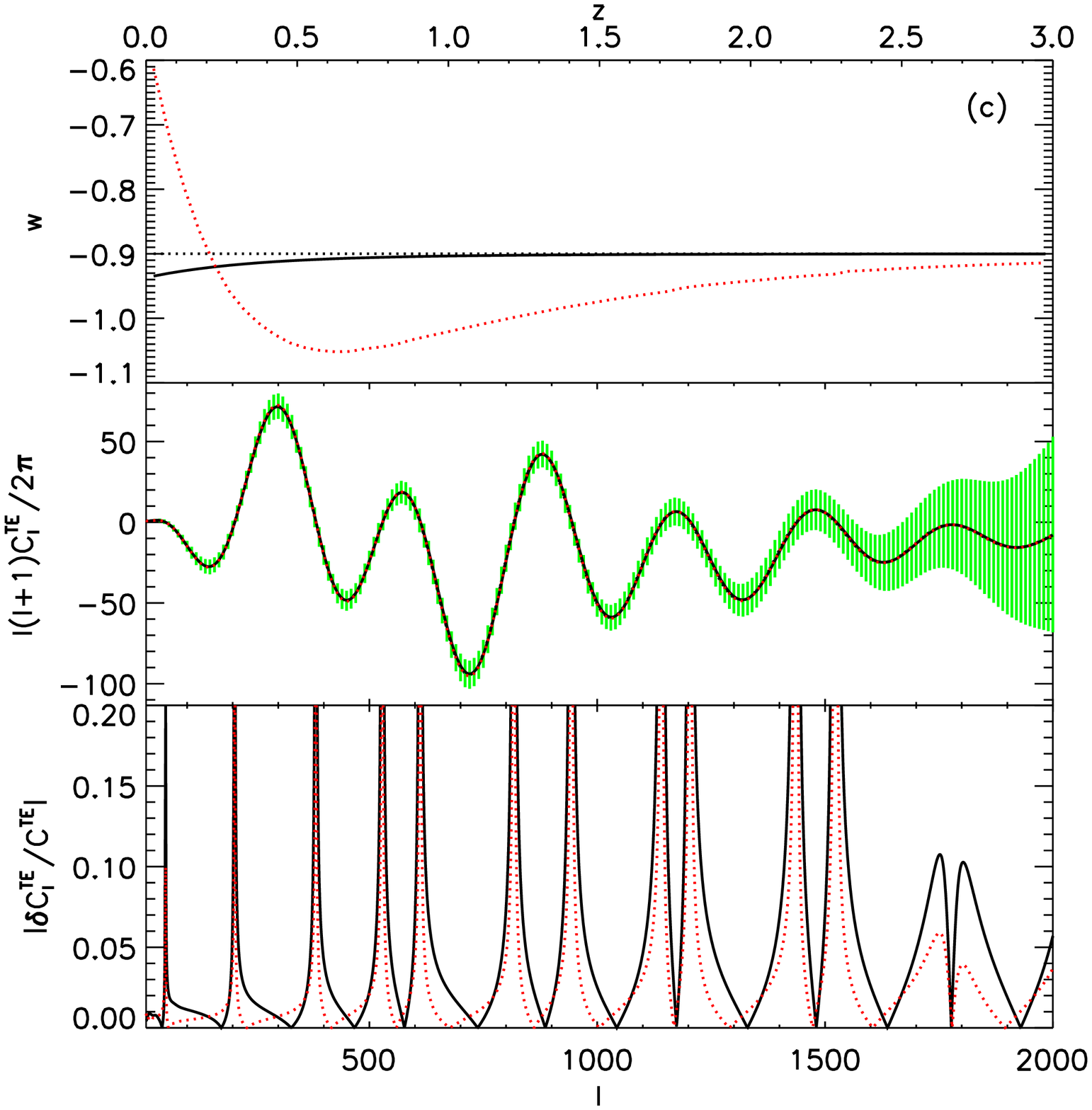}
\end{minipage}
\caption{The changes of $C_{\rm X,l}$ as we perturb $w$ around the
  fiducial model along the eigenmode directions by 1-$\sigma$ with standard cosmological parameters.
In (a), (b) and (c) we show the changes of $C_{\rm TT}$, $C_{\rm EE}$
and $C_{\rm TE}$, respectively.
The solid and dotted line show the first and second eigenmode,
respectively. The faint dotted line is the fiducial model. In the top plot of each panel, we show $w(z)$ given the
change along the different principal components. In the middle plots the
light area shows the observational error on each spectrum per
multipole $l$. In the bottom we show
$\delta C_{\rm X,l}/C_{\rm X,l}$.
\label{fig:planck_pc}}
\end{center}
\end{figure*}
To visualize the impact of the principal component on the observable,
we plot in Fig.~ \ref{fig:planck_pc} the change of $C_{\rm X,l}$ when
$w(z)$ changes along the direction of the principal components, i.e,
we construct $w(z)$ as
\begin{equation}
w(z)=w_{\rm fid}(z)+\Delta\alpha_{\rm i} \ev{i}\; ,
\end{equation}
where $\Delta\alpha_i$ is the error on $\alpha_{\rm i}$. For a data
set with 60 parameters, $\Delta\alpha_i=8.2\sqrt{\lambda_{\rm i}^{-1}}$
give the 1-$\sigma$ boundary along the $i$-th eigenmode direction
from the center. Note that the prefactor in this relation is there to
represent a 1-$\sigma$ errorbar in a 60-dimensional parameter space.

Figures \ref{fig:planck_pc} (a),(b) and (c) presents the changes of $C_{\rm TT}$, $C_{\rm EE}$
and $C_{\rm TE}$, respectively. The solid and dotted line represent changes in the first and second eigenmodes, respectively.
In the top plot of each panel, we show how $w$ deviates from the
fiducial value (faint dotted). For $i=1$ (solid), $w$ only deviates slightly from $w_{\rm fid}$ at $z<0.5$; while
for $i=2$ (dotted), $w$ descends from $-0.6$ at $z=0.$ to $-1.1$
around $z=0.6$ and then goes slowly back to $-0.9$ at higher
redshift. The second plot from the top in each panel
shows how $C_{\rm X,l}$ behaves. The light area shows the observational error. For every spectrum, the deviation from the fiducial
$C_{\rm l}$ is very small compared with the observational error, in fact too small to be seen in the graph.
For clarity, we presents in the bottom plot in each
panel the relative difference $\Delta C_{\rm X,l}/C_{\rm X,l}$
to show how $C_{\rm X,l}$ changes.
The peaks of $\Delta C_{\rm X,l}/C_{\rm X,l}$ have the same height,
which indicates that $C_{\rm X,l}$ shifts along $l$ when we add the
principal components. This is because $w(z)$ changes the expansion
history of the Universe and therefore the angular positions of
acoustic peaks. $C_{\rm TT}$, $C_{\rm EE}$ and $C_{\rm TE}$ all have
the same qualitative behaviour with changing eigenmodes. Notice in the bottom panel of 
(c) that the peaks become infinity because $C_{\rm TE}$ crosses zero at the same position.

The `constraint' on $w(z)$ from CMB power spectrum mainly come from the estimation on the angular positions
of acoustic peaks. Since the angular positions also depend on $\Omega_{\rm m}$, there is a large degeneracy between
$\Omega_{\rm m}$ and $w(z)$\citep{Bean01}. Therefore, $\lambda_{\rm
  i}^{1/2}$ in the marginalized case is two orders of magnitude
smaller than the one with fixed cosmological parameters.
However, the first mode is still three orders of magnitude higher than the rest of $\phi_{\rm i}$.
The second mode drops down two orders of magnitude, but it still changes sign around $z=0.25$.
The lower panel is consistent with the lowest panel of Fig.(1) in
\citet{Crittenden:05}.

From the discussion above, one can tell that the information for $w$ from CMB power spectrum is very limited. However, since
the Planck surveyor will be able to pin down the other cosmological parameters to a percentage level,
we will choose it as a prior on cosmology
when we evaluate $w$ from other dark energy probes.

\section{Type Ia Supernovae}
\label{sec:sne}
Type Ia SNe are so called standardisable candles and can be used to probe
cosmological models \citep{Perlmutter98,Riess:98}. Their luminosity
distance-redshift relation provides a straightforward way to
measure the expansion rate of the Universe.
The magnitude-redshift relation for Type Ia SNe is given by
\begin{equation}
m\equiv M+5\log_{10} d_{\rm L}(z)+25,
\label{eq:m_z}
\end{equation}
where $m$ is the apparent magnitude and $M$ is the
intrinsic magnitude of the Supernovae. In order to forecast a given
SNe survey, we have to choose
a fiducial value for M, and we assume $M_{\rm fid}=-19.3$
\citep{Perlmutter98}. The luminosity distance $d_{\rm L}$ in a flat Universe
is given by
\begin{equation}
d_{\rm L}(z)=c(1+z)\int_{0}^{z}\frac{dz'}{H(z')}
\label{eq:dl}
\end{equation}
where $c$ is the speed of light and $H(z)$ is the Hubble parameter. For a flat universe, $H(z)$
is defined as
\begin{equation}
H(z)=H_{\rm 0}\left[\Omega_{\rm m} (1+z)^{3}+\left(1-\Omega_{\rm
m}\right)\,\exp\left(3\int_{0}^{z}\frac{1+w(z')}{1+z'}dz'\right)\right]^{1/2}\;.
\end{equation}

The Fisher matrix for the Supernovae survey is given
by \citep{1998astro.ph..4168T}
\begin{equation}
F_{ij}=\int_0^{z_{\rm max}}\frac{1}{\sigma_{\rm
m}^2}\frac{\partial m_{\alpha}}{\partial \theta_i}\frac{\partial
m_{\alpha}}{\partial \theta_j}n(z)dz,
\label{eq:fisher_sn}
\end{equation}
where $n(z)$ is the density distribution of
the Supernovae satisfying
\begin{equation}
\int_{\rm 0}^{z_{\rm max}}n(z){\rm d}z=N,
\end{equation}
where $N$ is the total number of SNe Ia in the survey, $z_{\rm max}$
denotes the survey depth and $\sigma_{\rm m}$ is the error on the
magnitude $m$. We
assume that
$\sigma_{\rm m}=0.15$ for all Type Ia SNe surveys, which is a simplistic,
but common assumption neglecting, for example, measurement errors and cadence differences.

We start our discussion with the SNAP survey \citep{SNAP:05a,SNAP:05b}.
SNAP is a proposed space mission designed to measure the light curves
and spectra of Type IA SNe and the spectra of their host galaxies to
estimate their redshifts. It is estimated that up to $10,000$
SNe Ia could be found out between $z=0.3$ and $1.7$, $2,000$ out of which
will have well-measured light curves \citep{SNAP:05a,SNAP:05b}. 
Note that the value of $z_{\rm max}$ in Eqn.~(\ref{eq:w_z}) will be different depending on
different survey parameters and for the SNAP SNe survey is $z_{\rm
  max}=1.7$.

For a flat Universe the only standard cosmological parameters relevant for
Supernovae as cosmological probes are the total matter density
$\Omega_{\rm m}$ and $H_{\rm 0}$.  The intrinsic magnitude $M$ can be
combined with $H_0$ and be treated as a `nuisance' parameter and
marginalized over. However, we treat both of them separated given that
$H_{\rm 0}$ can be constrained to a certain extent by the Planck
surveyor.

\begin{figure}
\includegraphics[width=8.0cm,angle=0]{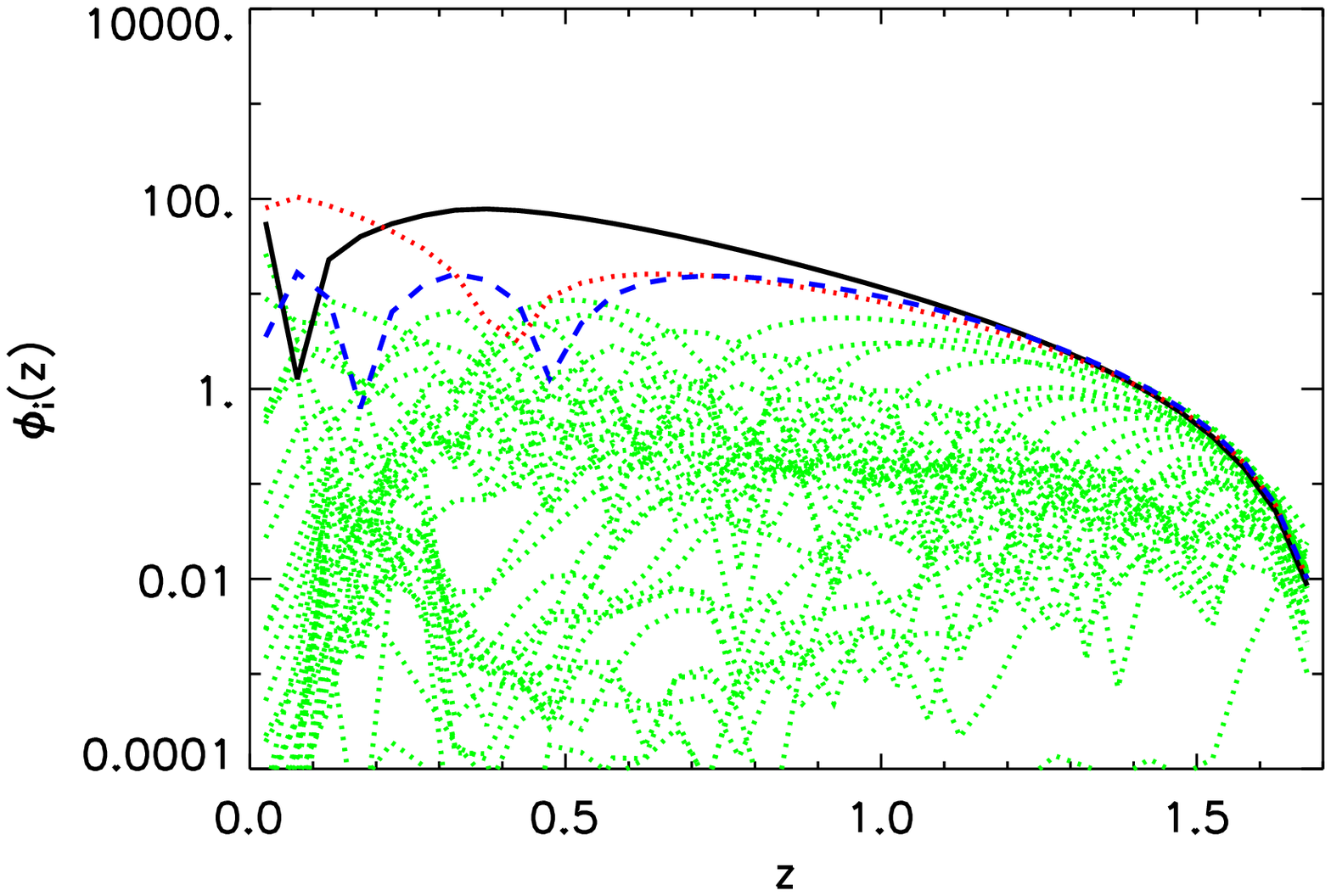}
\includegraphics[width=8.0cm,angle=0]{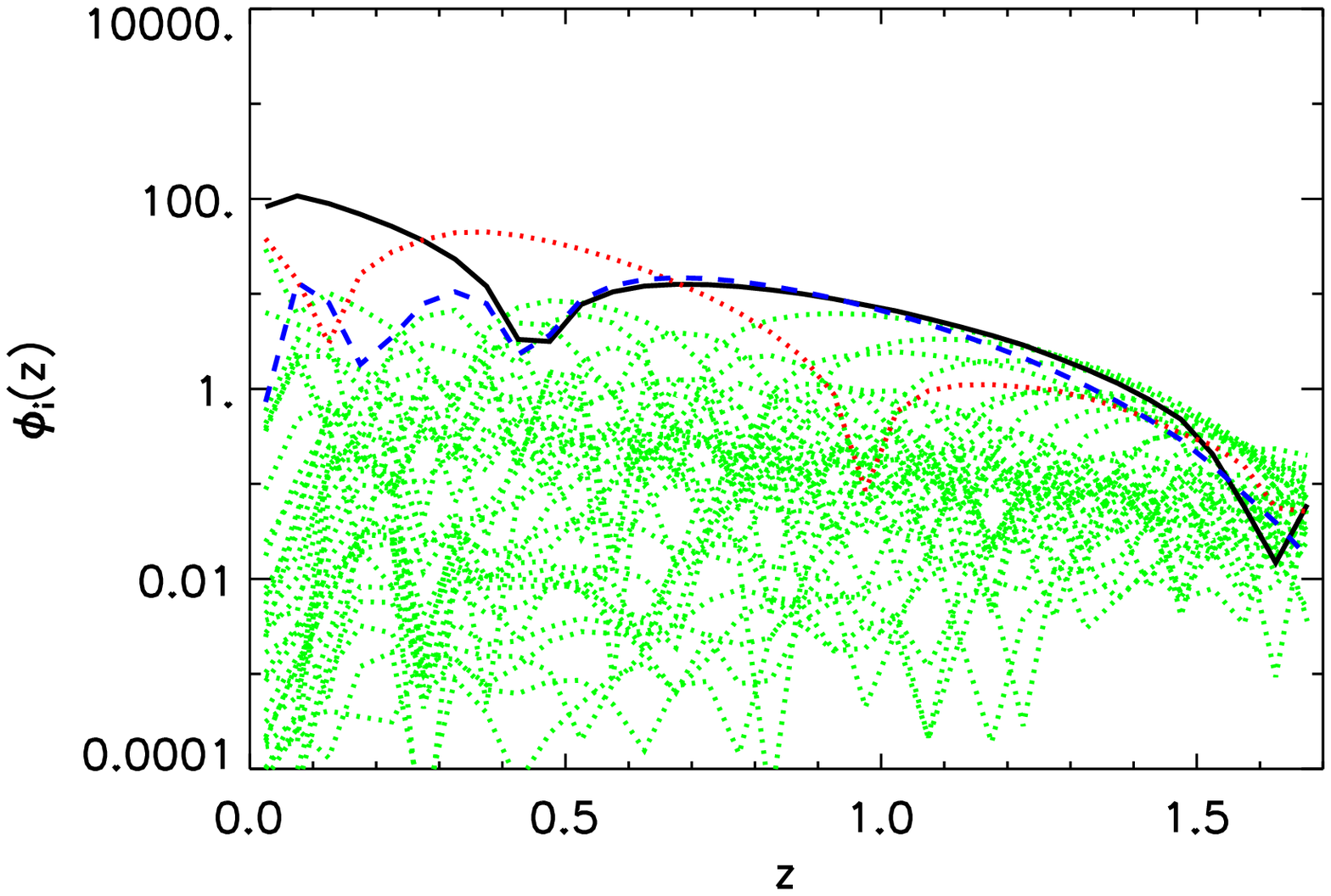}
\caption{The $\phi_i$ for the SNAP SNe Ia survey.
The solid line shows $\phi_1(z)$.  The dotted line
and dash line show $\phi_2(z)$ and $\phi_3(z)$, respectively. The
  faint (green) dotted lines
show the remaining modes. The upper panel is for fixed cosmological parameters, while the lower
panel is after cosmological parameters have been marginalized
  including the prior from Planck.  \label{fig:sne_uni_snap}}
\end{figure}

We conservatively assume that we have $50$ calibration SNe Ia between $z=0-0.3$.
In addition, we assume that the rest of SNe Ia are uniformly distributed in redshift. 

The evolution of the eigenmodes in $z$, therefore, will mainly depend on
the derivatives of the apparent magnitude $m(z)$ with respect to $w_{\rm i}$. Fig.~(\ref{fig:sne_uni_snap})
shows all $\phi_{\rm i}$ for this survey. There are $34$ eigenmodes in Fig.~(\ref{fig:sne_uni_snap}).
As in the CMB case, we highlight the first three eigenmodes.
The (black) solid line, the (red) dotted line and the (blue) dashed line represent $\phi_1$,
$\phi_2$ and $\phi_3$, respectively. The faint (green) dotted lines
show all higher order eigenmodes. The upper panel shows $\phi_i$ maintaining $\Omega_{\rm m}$ and $H_{\rm 0}$ fixed, while
the lower panel represents the eigenvectors after marginalizing over
$\Omega_{\rm m}$ and $H_{\rm 0}$ including the Planck prior.
The errors on the components drop down linearly from the best constrained ones to the worst ones.
By marginalizing over $\Omega_{\rm m}$ and $H_{\rm 0}$, the amplitudes of $\phi_{\rm i}$
become slightly smaller, which is consistent with the results of
\citet{Huterer:03}. We notice that the redshift dependence
of the eigenmodes has changed with marginalization; the eigenmodes
acquire zeros as we marginalize over $\Omega_{\rm m}$ and $H_{\rm 0}$,
indicating the impact of parameter degeneracies.

\begin{figure}
\includegraphics[width=8.0cm,angle=0]{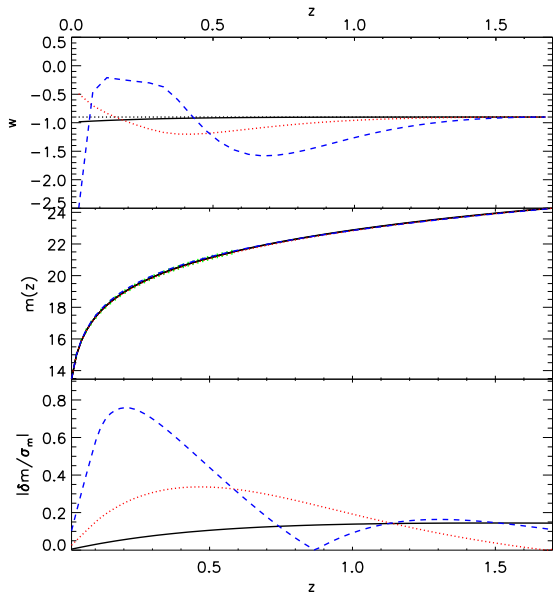}
\caption{The change of $m(z)$ when $w$ changes along the direction of
  the principal components. The upper panel shows
$w=w_{\rm fid}+6.1\lambda_{\rm i}^{-1/2}\,\ev{i}$ with $i=1,2,3$. The
  second panel from the top shows the $m(z)$ for each case. In the
  lower panels we show the absolute change $|\delta m|$ relative to $\sigma_{\rm m}=0.15$ corresponding to each case. The solid,
  dotted and dashed lines represents $i=1,2,3$, respectively. \label{fig:sne_fid}}
\end{figure}
Fig.~\ref{fig:sne_fid} shows how the apparent magnitude $m(z)$
 changes when $w$ changes along the direction of the eigenmodes.
 We take the first three components as examples.
In the upper panel, we plot $w=w_{\rm fid}+\Delta\alpha_{\rm i}\,\ev{i}$
where $\Delta\alpha_{\rm i}=6.1\lambda_{\rm i}^{-1/2}$ for the 34 free parameters.
The middle panel shows the magnitude $m(z)$. For clarity, we also show in the
 lower panel the absolute change $|\delta m|$ relative to $\sigma_{\rm m}=0.15$ for each case. The solid line is when we
vary the first eigenmode, while the dotted and dashed lines
 represent variations of the second and third eigenmodes,
 respectively.
As we expect, the error on $m$ due to these variations is less than $\sigma_{\rm m}=0.15$.

We can now proceed to compare the $\phi_{\rm i}$ from SNAP, PS4 and
DES SNe Ia surveys. PS4 will have a medium survey in the grizY
bands over 1,200 deg$^2$ and an ultra deep survey in the same bands
over 28 deg$^2$, which is two magnitudes deeper than the medium
survey. These will allow potentially the detection of $5,000$ Type Ia SNe and
$1,000$ in early type hosts\footnote{see at: {\tt pan-starrs.ifa.hawaii.edu/public/science/active.html}}. We note
that DES and PAN-Starrs will not have the capability of spectroscopic follow-up. Therefore
follow-up will have to be done with other instruments or photometric redshifts will have to be used. We do not
consider these caveats in this analysis and assume real spectra are available for redshift determination.
DES has a dedicated Supernovae program in conjunction with a spectroscopic
follow-up for a sub-sample to test purity\citep{DES}.
\begin{table}
  \begin{center}
    \begin{tabular}{|c|c|c|c|}
      \hline
      Survey &  N &  redshift range  & bins \\
      \hline
      DES & $1,900$ &(0.3,0.75) & 15  \\
      PanStarrs-4 &  $6,000$& (0.3,1) & 20  \\
      SNAP  & $2,000$  & (0.3,1.7) & 34 \\
      \hline
      \label{tb:expt_sne}
    \end{tabular}
    \caption{Experiments' Parameters For Type Ia SNe surveys. Note that we put 50 SNe Ia between $z=0$ and $0.3$. The rest is uniformly distributed
    in redshift.}
  \end{center}
\end{table}
Table~(\ref{tb:expt_sne}) lists the corresponding survey parameters and the number of bins that we use for each survey.

\begin{figure}
\includegraphics[width=8.0cm,angle=0]{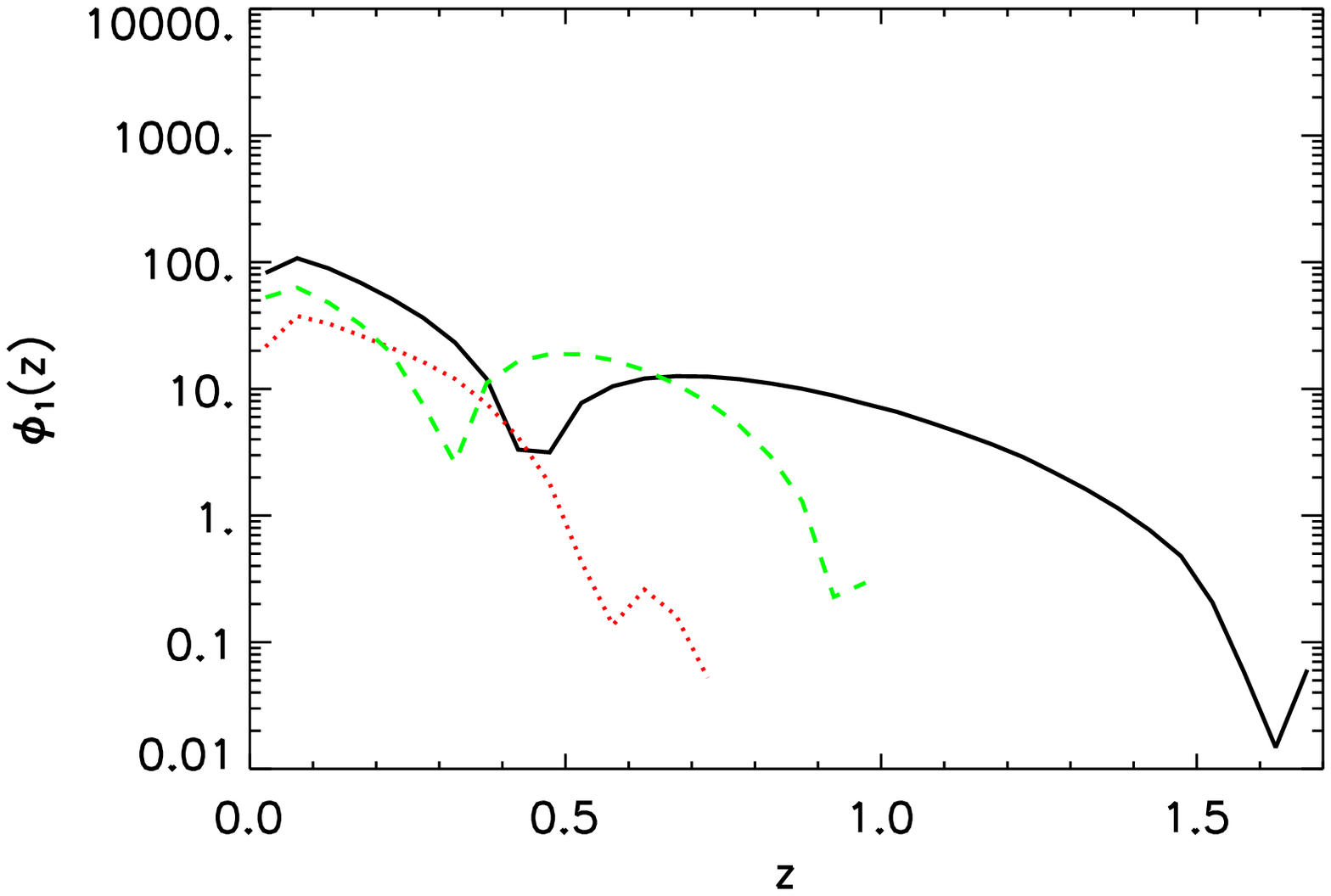}
\includegraphics[width=8.0cm,angle=0]{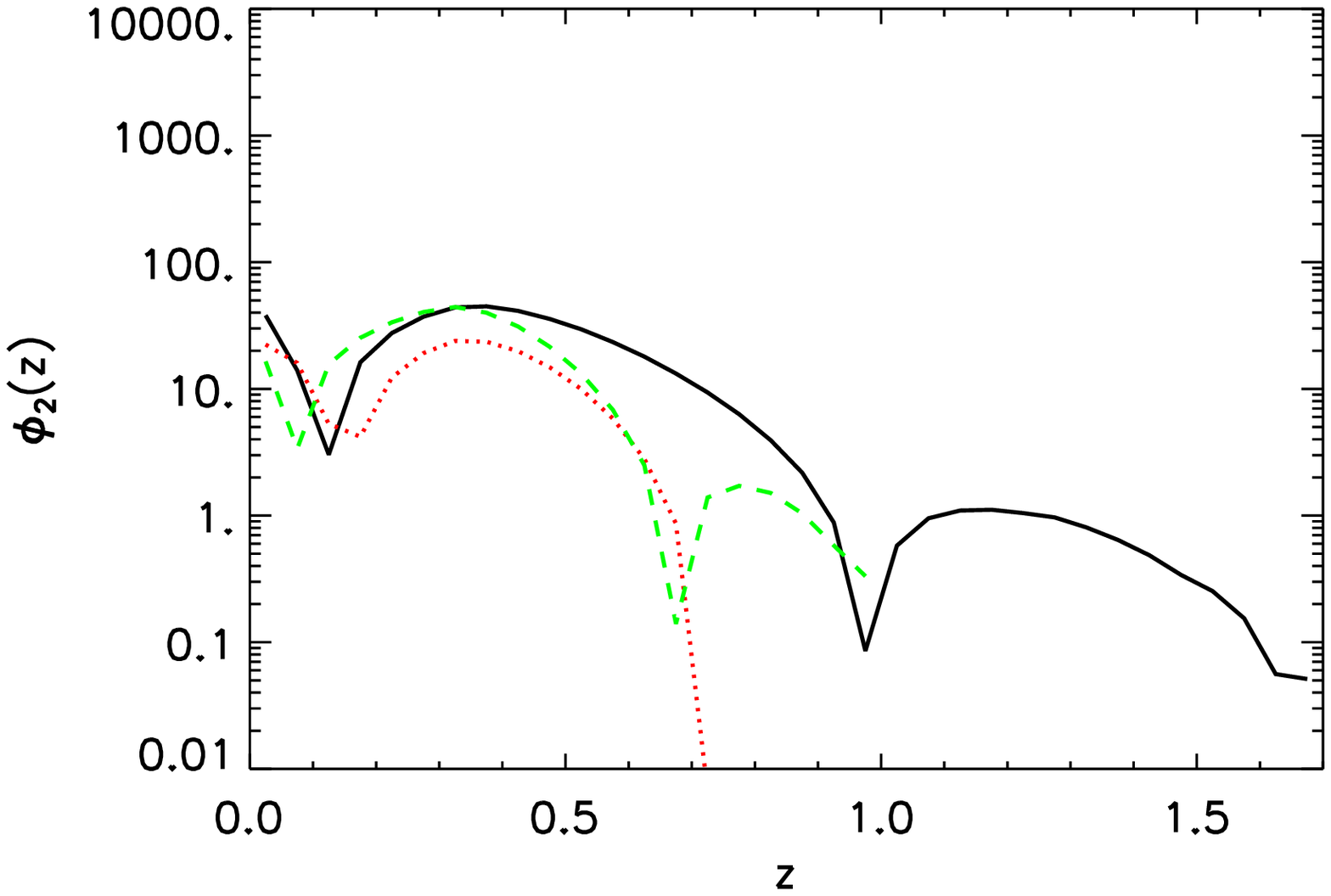}
\caption{$\phi_1$ (upper panel) and $\phi_2$ (lower panel) for different SNe Ia surveys.
The solid line, dotted line and dash line
indicate SNAP, DES and Pan-Starr4, respectively.
\label{fig:sne_com_PC}}
\end{figure}
Fig.~(\ref{fig:sne_com_PC}) shows $\phi_1$ and $\phi_2$ for each of these surveys.
The solid , dotted and dash line indicate SNAP, DES and PS4, respectively. Given the state of current Supernovae surveys, DES, SNAP and
PS4 will improve on the current situation by providing a very long
exposure time on red filters, hence being able to find Supernovae at
higher redshifts. We have not considered the PS1 Supernovae survey
as it will not be considerably deeper in the red hence not providing much
larger redshift coverage for the Supernovae detected, compared to
current surveys.
Due to the different distributions of the SNe and different minimum
and maximum redshift, the eigenmodes peak in different locations.
The relative amplitudes clearly show that surveys with more SNe Ia
give better constraints, as expected and also found by \citet{2004MNRAS.347..909K}. On the other hand, DES and SNAP
have a similar number of SNe Ia, but the constraint from DES is weaker
because of its lower maximum redshift {\em and} its higher minimum redshift.

\begin{figure}
\includegraphics[width=8.0cm,angle=0]{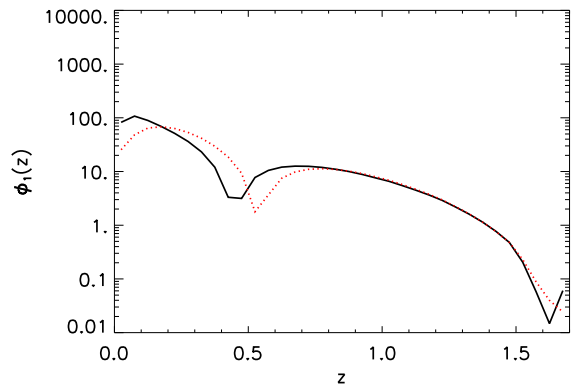}
\caption{Comparison of $\phi_1$ between different SNe Ia redshift distribution $n(z)$ for SNAP SNe Ia survey. The solid line
represents $\phi_1$ by assuming constant $n(z)$ and the dotted line shows $\phi_1$ with the
distribution from Table(1) in \citet{2004MNRAS.347..909K}.
 \label{fig:sne_distri}}
\end{figure}

In reality, the SNe Ia distribution $n(z)$ is unlikely to be constant;
instead, $n(z)$ is a result of the survey strategy. In order to find out
how the $\phi_i(z)$ depend on $n(z)$, we adopt a fiducial SNAP distribution given by Table(1) in
\citet{2004MNRAS.347..909K}. Fig.~(\ref{fig:sne_distri})
shows the $\phi_i$ with different assumptions on $n(z)$. The solid line in this plot
represents $\phi_1$ by assuming constant $n(z)$, while the dotted line shows $\phi_1$ with the
distribution from \citet{2004MNRAS.347..909K}. Beyond $z=1$, the difference between these two cases is small. The amplitude of
$\phi_1$ for the more realistic distribution is lower than the constant
$n(z)$ case due to a lower number of Type Ia SNe at low redshift.
The peak of $\phi_1$ for the more realistic distribution is located at higher redshift.

\section{Weak Lensing Tomography}
\label{sec:wl}
Weak lensing tomography is a recent technique which assumes that light
from distant galaxies travels to us in geodesics which are perturbed by
the network of dark matter present in the Universe. These perturbations
have the net result of changing the ellipticity of galaxies by roughly
a per cent. By measuring the accurate shapes of galaxies and removing
instrumental effects this cosmological effect is a nice probe of the
growth of structure in our Universe \citep{1999ApJ...522L..21H}. The techniques
to extract the shear signal by looking at the ellipticities are vast
and different for space based and ground based surveys \citep{2006MNRAS.368.1323H,2007MNRAS.376...13M}.
Different surveys will differ mainly by the quality of their image, which
will affect the quality of the shear measured, the quality of
the photometric redshifts, which will be determined by the integration
in each of the photometric bands and which will determine the
ability to extract redshift information via tomography, and the total
number of galaxies used for weak lensing which will be determined by a
combination of the depth and the mean seeing of the survey.

Cosmology is extracted by the means of the weak lensing power spectrum.
The weak lensing power spectrum is the weighted
two-dimensional projection of the three-dimensional matter power spectrum,
given by \citep{2001PhR...340..291B,2004AJ....127.3102R}
\begin{equation}
P_{\rm A,l}=\frac{9}{16}\left(\frac{H_0}{c}\right)^4\Omega_m^2\int_{0}^{z_{\rm h}}
 W_{\rm \alpha}W_{\rm \beta}P(l/d_{\rm c},z)\, \frac{\partial d_{\rm
c}}{\partial z}{\rm d}z,
\label{eq:cl_lensing}
\end{equation}
where $A$ represents the pair of tomographic slices $(\alpha,\beta)$. $d_{\rm c}$ 
is the comoving angular diameter distance given by
\begin{equation}
d_{\rm c}=\frac{d_{\rm L}}{(1+z)},
\end{equation}
with $d_{\rm L}$ from Eqn.~({\ref{eq:dl}}). 
The upper limit of the integral $z_{\rm h}$
represents the redshift of the survey depth.
The window function $W_{\rm \alpha}(z)$ is given by
\begin{equation}
W_{\rm \alpha}(z)=\frac{2}{a}\int_{z}^{z_{\rm h}}n_{\rm \alpha}
(z')\frac{d_{\rm c}(z)d_{\rm c}(z'-z)}{d_{\rm c}(z')}\frac{ \partial d_{\rm
c}}{\partial z}\mathrm{d}z', \label{eq:lensing_weight_kernal}
\end{equation}
where the weighting function $W$ represents the strength of the lensing
signal for a source at a given redshift and a distribution of
sources $n_{\rm \alpha}(z)$ in each tomographic slice $\alpha$.
Note that to distinguish the slice index form the $w$-bin index,
we use Greek letters to present the slice index in this section. $P(l/d_{\rm c},z)$ is the matter power spectrum at redshift
$z$ with $l/d_{\rm c}$ representing the wavenumber $k$ for a given harmonic $l$.

The Fisher Matrix for the weak lensing power spectrum is given by
\begin{equation}
F_{ij}=\sum_{l}\sum_{\rm A,B}\frac{\partial P_{\rm
A,l}}{\partial\theta_{\rm i}}\left[{\rm Cov}^{AB}_{\rm
l}\right]^{-1}\frac{\partial P_{\rm
B,l}}{\partial\theta_{\rm j}}, \label{eq:lensing_fisher}
\end{equation}
with $\alpha\geq \beta$. Setting
$A=(\alpha,\beta)$ and $B=(\alpha^{'},\beta^{'})$, the elements of the covariance ${\rm Cov}^{AB}_{\rm l}$ are given by \citep{2006ApJ...636...21M}
\begin{eqnarray}
{\rm Cov}^{AB}_{\rm l}&=&\left<P_{\rm A,l},P_{\rm B,l}\right>\nonumber\\
&=&\frac{1}{(2l+1)f_{\rm sky}}\left[\hat{P}_{(\alpha,\alpha^{'})}\hat{P}_{(\beta,\beta^{'})}+\hat{P}_{(\alpha,\beta^{'})}\hat{P}_{(\beta,\alpha^{'})}\right]\nonumber\\
\label{eqn:cov_lens}
\end{eqnarray}
in which
$\hat{P}$ is 
\begin{equation}
\hat{P}_{(\alpha,\beta)}(l)=P_{\rm (\alpha,\beta),l}+\delta_{(\alpha,\beta)}\frac{\sigma^2_{\rm \gamma}}{n_{\rm g}}
\end{equation}
which is a combination of the power spectrum and the noise. $f_{\rm sky}$ is  the fraction of the sky coverage of the
survey, $\sigma_{\rm \gamma}$ is the shear variance due to the
intrinsic ellipticity including other measurement errors. $n_{\rm g}$
is the average surface density of galaxies. Eqn.(\ref{eqn:cov_lens}) holds under the Gaussian assumption
on the convergence field. With $l\leq3000$, this approximation works very well \citep{2000ApJ...537....1W,2001ApJ...554...56C,2005APh....23..369H}. Therefore,
in this paper, we take $l\leq3000$. The lowest $l$ is limited by the sky coverage of the survey. Also since the cosmic variance dominates at very low $l$, low $l$ harmonics
do not contribute much to the Fisher Matrix, we take the lower limit of $l=10$.

One of the difficulties that we are facing in weak lensing tomography is to determine the  matter power spectrum
$P(k,z)$ with the $w$-binning parameterization. Dark energy modifies the growth of the matter perturbation and
the shape of the matter power spectrum. The growth factor $D(z)$ is analytically given by
\begin{equation}
D^{''}+\frac{3}{2}\frac{1}{a}[1-w(a)(1-\Omega_{\rm m}(a))]D^{'}-\frac{3}{2}\frac{1}{a^2}\Omega_{\rm m}D=0,
\end{equation}
with the prime representing the derivative with respect to the scale factor $a$. However, since future weak lensing surveys extend to
the nonlinear scale, the dependence of the shape of the nonlinear matter power spectrum on $w$ has to be considered when calculating the lensing
power spectrum.
\citet{1996MNRAS.280L..19P} and \citet{Smith:03}
developed a set of fitting formulae for calculating the nonlinear matter power spectrum
with a cosmological constant assumption. Though work has also been done to correct the fitting formulae for constant $w\neq-1$
\citep{1999ApJ...521L...1M,2006MNRAS.366..547M}, this area is still limited to a non-evolving equation of state on a small cosmological parameter space.
It requires a lot of effort calibrating the matter power spectrum considering evolving $w$ by running low resolution N-body simulations.
However, since the aim of this paper is to demonstrate the behaviour of the eigenmodes of $w$ with
different future weak lensing surveys, we simply generalize the method given by \citet{Smith:03} for $w$-bin parameterization.
We emphasize that in real data, either N-body simulation with evolving $w$ must be run or the non-linear $P(k)$ must be predicted
with a renormalisable perturbation theory model \citep{rpt} or run by a training set \citep{2007PhRvD..76h3503H}.
Here we still use the fitting formulae in \citet{Smith:03} but use the transfer function ouput from modified CAMB code (see Sec.\ref{sec:cmb}).

\begin{figure}
\includegraphics[width=8.0cm,angle=0]{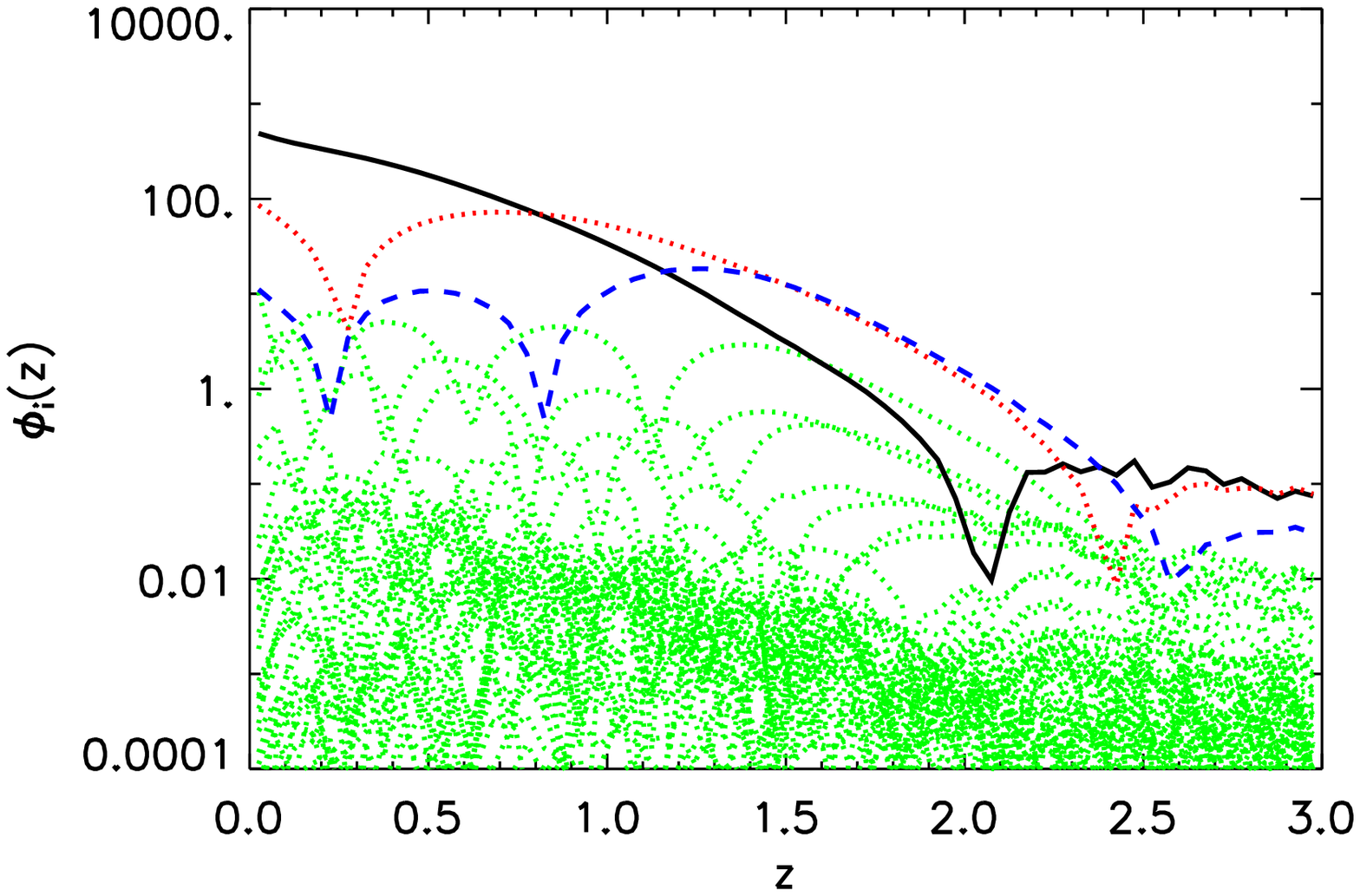}
\includegraphics[width=8.0cm,angle=0]{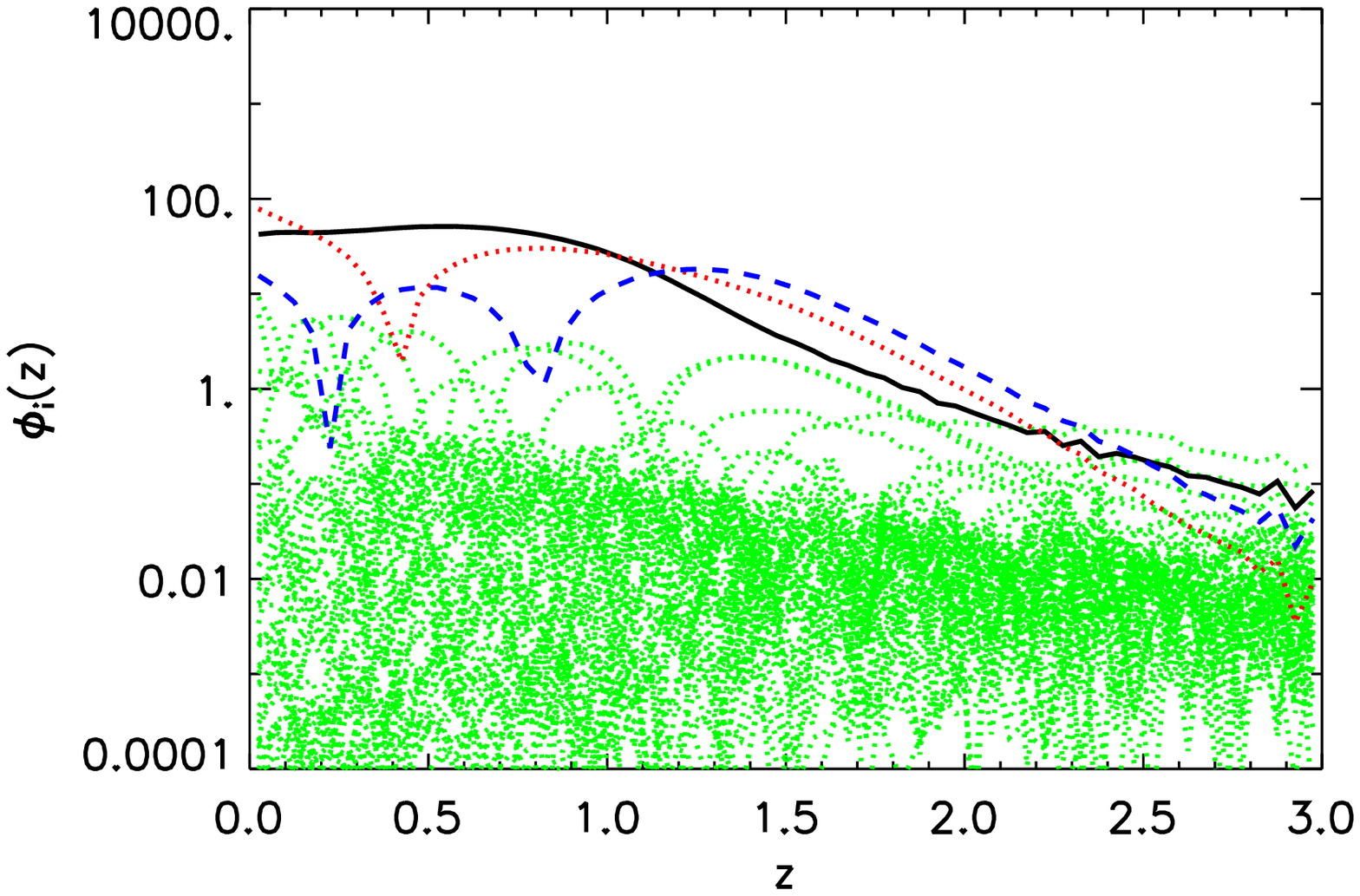}
\caption{60 $\phi_{\rm i}(z)$ for SNAP weak lensing survey.
The (black) solid line indicates the first eigenmode. The (red) dotted line
and (blue) dash line indicate the second and the third ones, respectively. The faint (green) dotted lines
show the remaining ones. The upper panel is when cosmological parameters are fixed, while the lower
panel is after cosmological parameters have been marginalized with the use of the Planck prior.
\label{fig:lensing_snap_3bins}}
\end{figure}

The SNAP weak lensing survey is one of the most representative future
surveys. Its baseline is to
image over $1,000$ square degree and approximately $100$ resolved galaxies will be found
in one square arcmin \citep{SNAP:05a} up to redshift $3$.
Six optical and three infrared filters are used to estimate the photometric redshift
of the galaxies.
In this paper, we adopt the simulated galaxy distributions in  \citet{2004AJ....127.3102R}.
We focus on the case with the three tomography slices since the constraint improves insignificantly with more slices.
We set $z_{\max}=3$ since SNAP can find galaxies up to redshift $3$.
Fig.~(\ref{fig:lensing_snap_3bins}) shows 60 $\phi_{\rm i}(z)$.
The (black) solid line represents $\phi_{\rm 1}$. The (red) dotted line
and (blue) dash line indicate $\phi_{\rm 2}$ and $\phi_{\rm 3}$, respectively. The faint (green) dotted lines
show all higher order modes. In the upper panel we fix the cosmological parameters.
The first mode dominates at low redshifts and then decays along the redshift. Since the first mode drops down four orders of
magnitude, at redshift $z>1$, the second and third modes become dominant.
The errors on the components drop down exponentially from the best constrained ones to those worst ones.
In the lower panel, we show $\phi_{\rm i}(z)$ after cosmological parameters being marginalized including Planck prior.
The amplitude of the first mode is slightly smaller in this plot. However the redshift dependences
of the modes change dramatically. $\phi_{\rm 1}$ has a peak around $z=1$. At redshift $z>2.5$, there is less information from SNAP, so $\phi_{\rm 1}$
is mainly driven by Planck Prior.
 $\phi_{\rm 2}$ and $\phi_{\rm 3}$ are one order of magnitude
lower than $\phi_{\rm 1}$ but still dominate at high redshift $z>0.8$. This is consistent with Fig.(2) in \citet{Simpson:06} where they found that
for cosmic shear, higher modes contributes significantly to weight function of $w_{\rm i}$.

\begin{figure*}
\begin{center}
\begin{minipage}[c]{1.00\textwidth}
\centering
\includegraphics[width=8.5cm,angle=0]{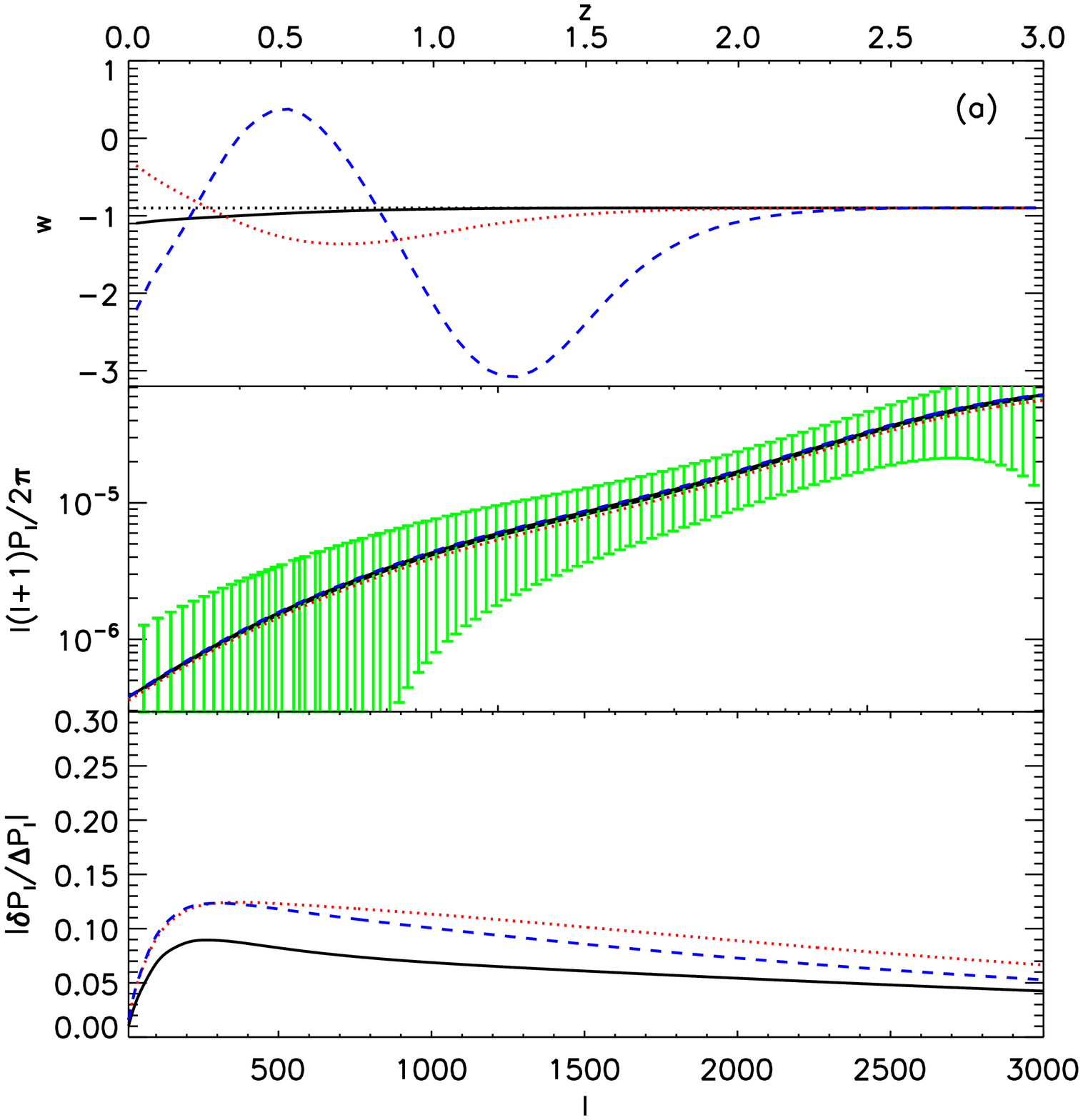}
\includegraphics[width=8.5cm,angle=0]{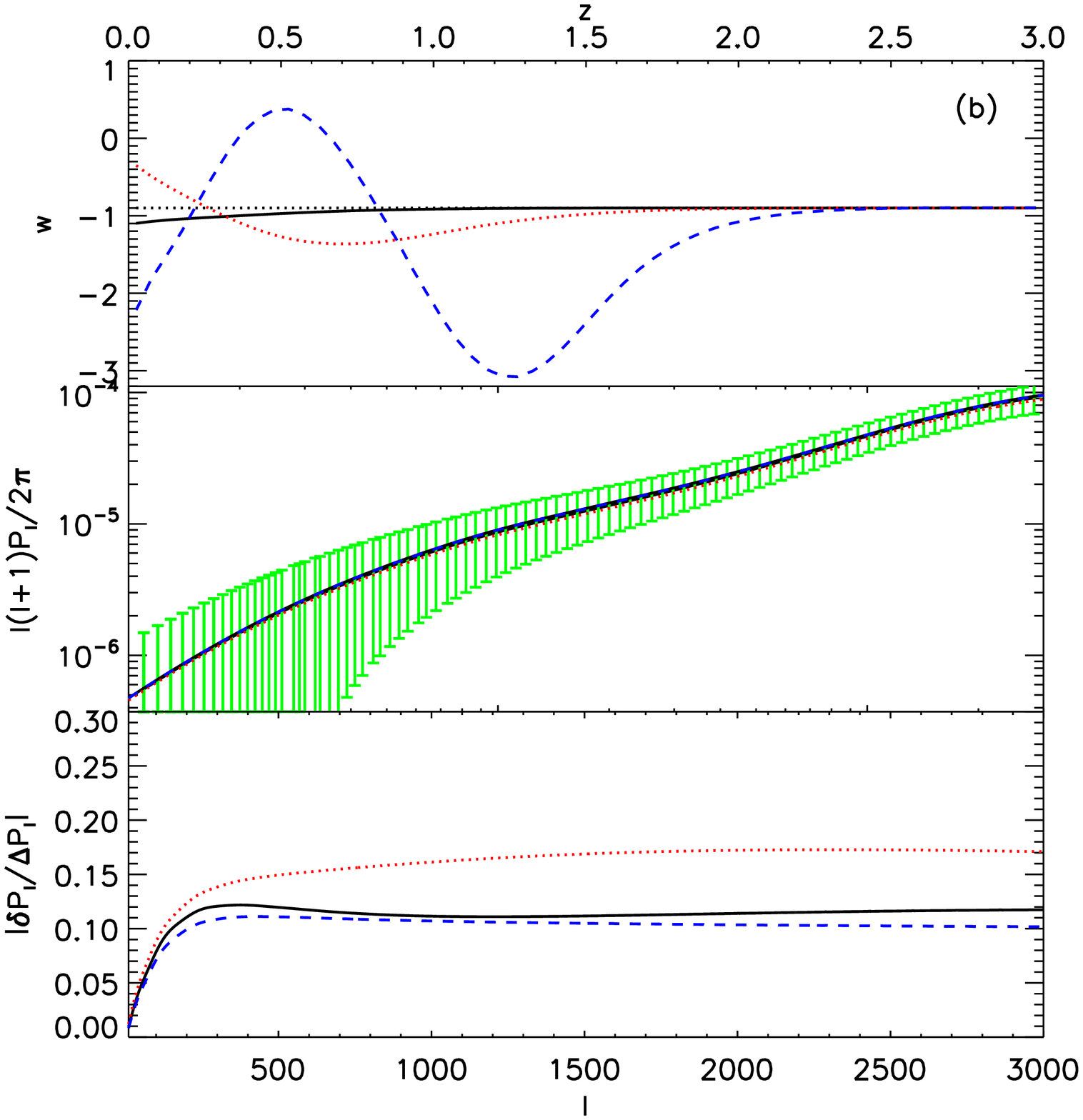}
\end{minipage}
\caption{The changes of $P_{\rm A,l}$ as we perturb $w$ around the fiducial model along the eigenvector directions.
We show the change of $P_{\rm A,l}$ with $A=(1,1)$ and $A=(1,2)$ on the left and right panel, respectively.
The (black) solid, (red) dotted and (blue) dash line represent the first three eigenmodes, respectively. In the top plot of each panel, we show $w$
given by $w=w_{\rm fid}+\Delta\alpha_{\rm i}\,e_{\rm i}$.
The faint (green) area shows the observational error on each spectrum per multipole $l$.
we present in the bottom of each panel $|\delta P_{\rm A,l}/\Delta P_{\rm A,l}|$.
\label{fig:wl_snap_3bins_fid}}
\end{center}
\end{figure*}

To test how $P_{\rm A,l}$ changes with the eigenmodes, we plot in Fig.(\ref{fig:wl_snap_3bins_fid})
$P_{\rm A,l}$ and its deviation from the fiducial value when we modify $w_{\rm i}$ along the direction of the eigenmodes.
The top plot on each panel shows $w=w_{\rm fid}+\Delta\alpha_{\rm i}\,e_{\rm i}$, where $\Delta\alpha_{\rm i}=8\sqrt{\lambda_{\rm i}^{-1}}$ for 60
free parameters.
We choose the first three best estimated modes after marginalization as the examples which are represented by (black) solid, (red) dotted and (blue)
dash lines, respectively.
In the second plot of the left panel, we show the auto correlation of the first slice, i.e. $P_{\rm A,l}$ with $A=(1,1)$.
The second plot of the right panel shows the cross correlation between the first slice and second slice, i.e. $P_{\rm A,l}$ with $A=(1,2)$.
The light shaded area is the error around the fiducial value which is presented as the black solid line.
For clarity, we show in the bottom the absolute change of $|P_{\rm A,l}|$ relative to the error bar for each mode.
The deviation is very small compared with the error bar,
which shows that the first three modes are well constrained. However,
the error in this plot only comes from a single shell. One also obtains constraints on $w$ from other auto and cross correlations.

In the following of this Section, we compare SNAP with four other weak lensing surveys: EUCLID, DES, PS1 and PS4.
SNAP and EUCLID are space based, while DES and PS (1 and 4) are ground based surveys.
As stated before ground based surveys and space based surveys have different systematic effects which enter the weak lensing analysis.
Image quality in space is far superior than on the ground. This will allow for better shape measurements in space.
This effect is hard to include in a Fisher analysis. we would be able to include this effect by producing image simulation and estimating what 
$\sigma_{\gamma}$ and $n_{\rm g}$ we would have for a given seeing and magnitude depth. Producing image simulation is beyond this work, so we
simply keep $\sigma_{\gamma}$ roughly constant and change $n_{\rm g}$ to take account for the effects of image quality.
The other main difference is the seeing. Even though PS4 will probe the sky to greater depths than EUCLID or SNAP, 
it will have less galaxies usable for weak lensing
as the seeing lowers the surface brightness of sources and many faint galaxies fall below the signal to noise required for a detection.
These differences are encoded within Table.(\ref{tb:wl}) which lists
the experiment parameters. To calculate the lensing power spectrum, one has to know the galaxy distribution. Since all these four
surveys will use photometric redshifts, one has to take into account the photometric redshift errors $\delta_{\rm z}$. A more common method to
include $\delta_{\rm z}$ is analytical convolution of the galaxy distribution $n(z)$ with the probability distribution function of the photo-z
$p(z_{\rm ph}|z)$ \citep{2006ApJ...636...21M}. For simplicity, $p(z_{\rm ph}|z)$ is assumed to be Gaussian with scatter $\sigma_{\rm z}(z)$
and bias $z_{\rm bias}(z)$.
In this paper, we use mock photometric catalogues to calculate $n(z)$ for each
tomographic slice. These photometric catalogues have been produced in the following way: galaxies from GOODS have been cloned and their
photometry has been calculated analytically. Then we have used a photometric redshift code to estimate the photo-z for these galaxies
having set aside a given number of galaxies for training purposes. A more detailed discussion regarding the creation of these
mocks can be found in \citet{2007arXiv0705.1437A} and \citet{2007arXiv0711.1059B}.
Based on these catalogues, we choose five slices for EUCLID and PS4, three slices for DES and PS1.
We also list $z_{\rm max}$ for each in Table.(\ref{tb:wl}).

\begin{figure}
\begin{center}
\includegraphics[width=8.0cm,angle=0]{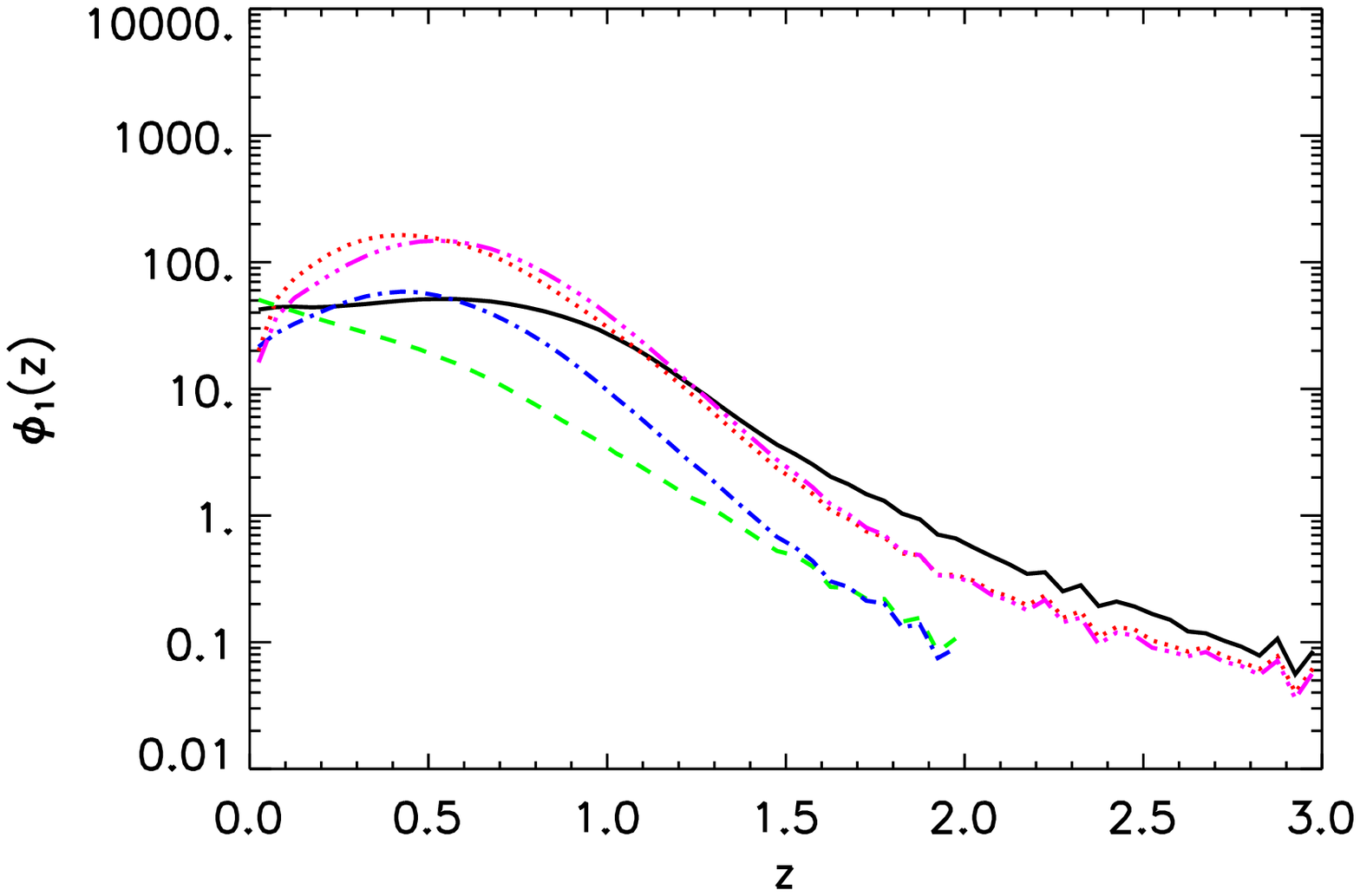}
\includegraphics[width=8.0cm,angle=0]{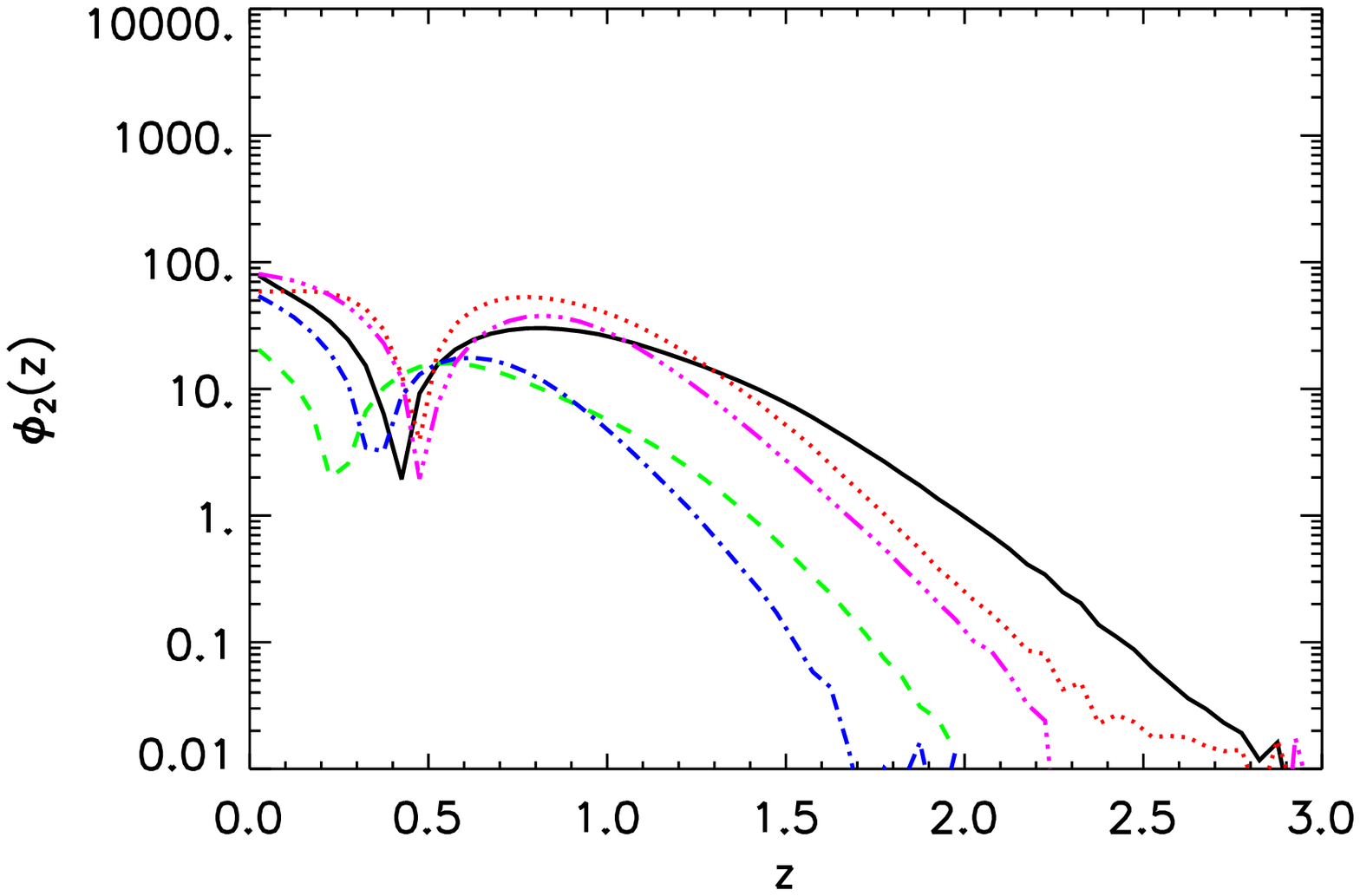}
\caption{$\phi_1$ (upper panel) and $\phi_2$ (lower panel) from different weak lensing surveys.
The (black) solid line, (red) dotted line and (green) dash line indicate SNAP, EUCLID and DES, respectively.
The (blue) dash-dotted line and (magenta) dash-dotted-dotted-dotted line represent PanStarrs 1 and 4, respectively.
\label{fig:lensing_com_PC}}
\end{center}
\end{figure}

We plot $\phi_{\rm 1}(z)$ (upper panel) and $\phi_{\rm 2}(z)$ (lower panel) in Fig.~(\ref{fig:lensing_com_PC}).
We marginalize the Fisher matrices over the cosmological parameters and $w_{\rm h}$ including a Planck prior.
As we expect, $\phi_{\rm 1}$ of DES (green dashed) and PS1 (blue dash-dotted) are about one order of magnitude
lower than $\phi_{\rm 1}$ of SNAP (black solid), EUCLID (red dotted) and PS4 (magenta dash-tripled-dotted).
DES has a relatively small sky coverage and number density of galaxies. PS1 covers half the survey, but has a relatively small number density
 of galaxies. The redshift dependences of $\phi_{\rm 1}(z)$ are different as well. $\phi_{\rm 1}$ of DES decays along the redshift; this may be due to the fact
that the values of the intrinsic Fisher matrix of DES is small so that the Planck priors have a bigger effect in the analysis.
$\phi_{\rm 1}$ of the rest peak around $z=0.5$ but stays positive in the whole redshift region.
We notice that the redshift dependence of $\phi_{\rm 2}$ do not change much for different surveys except for the amplitudes and the
redshift where $\phi_{\rm 2}=0$. We also found that with a Planck prior, $\phi_{\rm i}$ depend weakly on how we define the tomographic slices.

If we compare the eigenmodes between DES and PS1, one can also notice that PS1 has better constrained eigenmodes
than DES. Among SNAP, EUCLID and PS4, EUCLID and PS4 have better constraints on the eigenmodes than SNAP.
\citet{2007MNRAS.381.1018A} conclude that the sky coverage is the dominant parameter
on improving the figure of merit rather than $n_{\rm g}$ and $z_{\rm m}$.

\begin{table}
 \begin{center}
  \caption{Survey parameters for weak lensing observation}
\begin{tabular}{|c|c|c|c|c|c|}
  \hline
  Survey & $\sigma_{\rm \gamma}$ &  $f_{\rm sky}$ & $n_{\rm g}$ & $z_{\rm m}$ & $z_{\rm max}$  \\
  \hline
  SNAP & $0.3$ & 0.024 & $100$ & 1.23 & 3 \\
  EUCLID & $0.3$ & 0.5 & $30$ & 0.83 & 3\\
  DES & $0.3$  & 0.12 & $10$ & 0.67 & 2\\
  PS1 & $0.3$  & $0.5$ & $5$ & 0.6 & 2 \\
  PS4 & $0.3$  & $0.5$ & $20$ & 0.83 & 3 \\

  \hline
\end{tabular}
\label{tb:wl}
\end{center}
\end{table}

\section{Galaxy Cluster Counts}
\label{sec:cc}
Galaxy cluster counts probe cosmological models via the redshift
dependence of the survey volume and the sensitivity of the halo
mass function to the linear growth of structures. Forthcoming galaxy
cluster counts will be performed by selecting clusters via their
Sunyaev-Zel'dovich decrement, for example with Planck \citep{Planck}
or the South Pole Telescope \citep{Ruhl:04} or in the x-rays with, for
example, the extended ROentgen Survey with an Imaging Telescope Array
(eROSITA)\footnote{see at:
  http://www.mpe-garching.mpg.de/projects.html\#erosita}. Finally
future imaging and photometric redshift surveys will be able to
identify clusters in the optical wavebands
\citep{Koester:07,Rozo:07}. For example DES and EUCLID would be able to
identify thousands of clusters with this method.

The redshift evolution of the number of the
clusters ${\rm \Delta}N$ in the redshift interval between $z-\Delta z/2$ and
$z+\Delta z/2$ found by a Galaxy Cluster survey is given by
\begin{equation}
{\rm \Delta}N={\rm \Delta}\Omega{\rm \Delta} z \frac{{\rm d}V}{{\rm
d}z{\rm d}\Omega}\int_{M_{\rm lim}(z)}^{\infty}\frac{{\rm
d}n}{{\rm d}M}(M,z){\rm d}M,
\label{eq:deltaN_cluster}
\end{equation}
where ${\rm \Delta}\Omega$ is the sky coverage. ${\rm d}V/{\rm
d}z{\rm d}\Omega$ is the comoving volume element, $M_{\rm lim}(z)$ is the mass limit above which clusters
can be found at redshift $z$ for a given set of survey parameters. ${\rm
d}n/{\rm d}M$ is the number density of the clusters with mass $M$
and redshift $z$; in this paper we adopt the fitting function of
${\rm d}n/{\rm d}M$ presented by \citet{Jenkins:01a}
\begin{equation}
\frac{\mathrm{d}n}{\mathrm{d}M}(M,z)=-0.316\frac{\rho_m}{M\sigma_{\rm
M}}\frac{{\rm d} \sigma_{\rm M}}{{\rm
d}M}\exp{-|0.67-\log[D(z)\sigma_{\rm M}]|^{3.82}},
\end{equation}
where $\rho_m$ is the matter density at present, $D(z)$ is the
growth factor $\sigma_{\rm M}$ is the
linear overdensity at mass scale $M$ at redshift $z=0$, which is given by
\begin{equation}
\sigma^2_{\rm M}=\int_0^{\infty}W^2(kR)\Delta^2(k){\rm d}\ln k,
\end{equation}
where $R=\left(3/(4\pi)\rho_{\rm m}M\right)^{1/3}$. We use a spherical top hat window function
\begin{equation}
W(x)=3\left(\frac{\sin{x}}{x^3}-\frac{\cos{x}}{x^2}\right).
\end{equation}
The last term is the linear power spectrum $\Delta^2(k)=4\pi
k^3P(k)$. Following Sec.(\ref{sec:wl}), we use CAMB to calculate the
transfer function and the power spectrum $P(k)$.

The Fisher matrix for the cluster counts is given by
\citep{Haiman:01,2002ApJ...577..569L}
\begin{equation}
F_{ij}=\sum_{\alpha=1}^n \frac{\partial \Delta
N_{\alpha}}{\partial \theta_{\rm i}}\frac{\partial \Delta
N_{\alpha}}{\partial \theta_{\rm j}}\frac{1}{\Delta N_{\alpha}},
\end{equation}
In which $\Delta N_{\rm \alpha}$ is the number of clusters in
redshift interval $\Delta z$ at redshift $z_{\rm \alpha}$, with $z_n$
the maximum redshift of the survey.

In order to extract cosmology from a galaxy cluster count survey we
need to understand the relation between the survey parameters and the
limiting mass $M_{\rm lim}(z)$ of the survey. For example for Sunyaev-Zel'dovich
cluster surveys, $M_{\rm lim}(z)$ is a function of the flux limit
$S_{lim}$ at frequency $\nu$ \citep{Battye:03}. In this
context to estimate $M_{\rm lim}(z)$, one requires well-calibrated
mass-temperature relation for galaxy clusters and an understanding of
the scatter \citep{Lima:04}. However in order to get an estimate of
the ability of galaxy cluster counts to constrain the equation of
state for different types of cluster selection, we assume a limiting
mass, which is constant in redshift and neglect scatter. It is
straight forward to include scatter in any analysis however it is not
clear at the moment how large the intrinsic scatter is for different
survey strategies.

\begin{figure}
\includegraphics[width=8.0cm,angle=0]{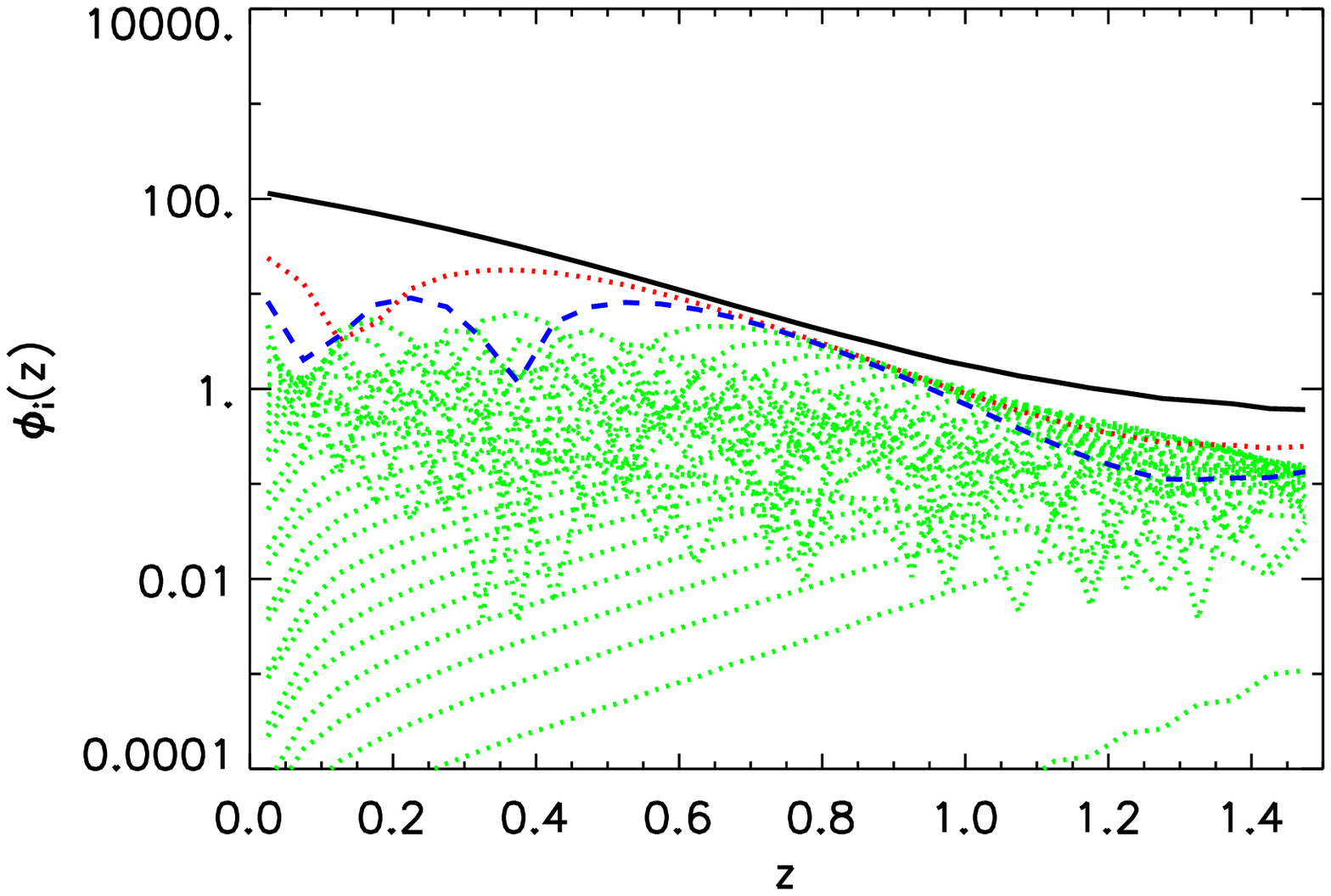}
\includegraphics[width=8.0cm,angle=0]{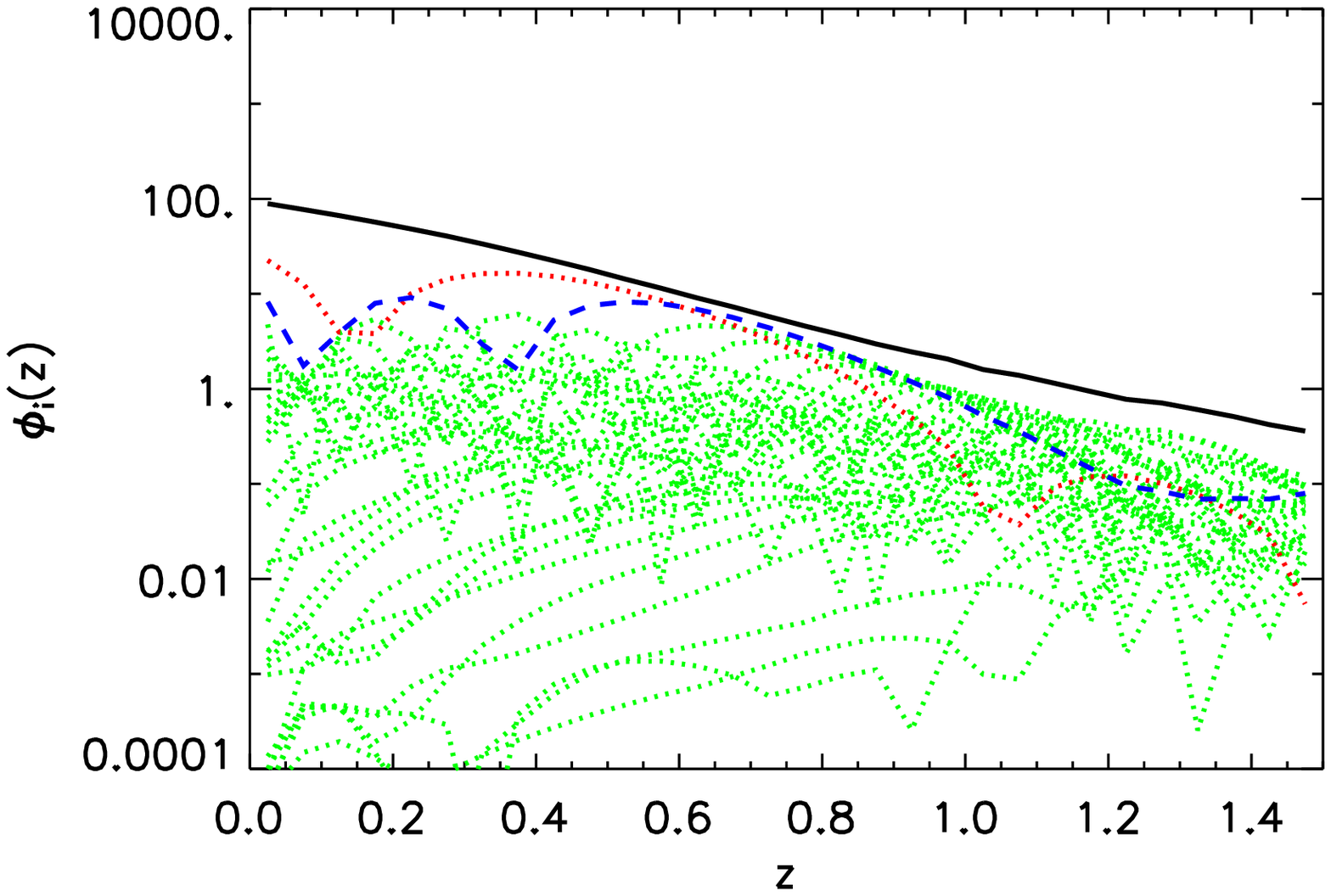}
\caption{$\phi(z)$ for a cluster survey with a mass limit of $\approx 2.4
\times10^{14}h^{-1}M_{\sun}$, which corresponds roughly to the limit
of the SPT survey. The (black) solid line indicates the first eigenmode.  The (red) dotted line
and (blue) dashed line indicate the second and the third one, respectively. The (green) faint dotted lines
show the remaining 27 modes. The upper panel is for fixed cosmological
parameters, while in the lower
panel we marginalize over cosmological parameters under the assumption
of prior information from the Planck surveyor. \label{fig:sze_spt}}
\end{figure}

The South Pole Telescope (SPT) is a bolometric array observing 4,000
square degree of the southern sky in 3-4 radio wavebands, with the
main signal coming from the 150GHz channel. It will discover thousands
of cluster roughly above a mass threshold of $2.4
\times10^{14}h^{-1}M_{\sun}$. The Dark Energy Survey is observing the
same area of sky as SPT and will provide photometric redshifts and vital weak
lensing information for the the galaxy clusters. The observed redshift
range goes roughly out to $z=1.5$.
In Fig.(\ref{fig:sze_spt}) we show $\phi_i(z)$ for fixed and
marginalized cosmological parameters. As before we highlight the first
three components. We find that only the first eigenmode is
dominating. This is similar to the SNe case, where the hierarchy of
modes is almost linear. After we marginalize over the other
cosmological parameters, the amplitudes of the eigenmodes are one order of magnitude lower. Notice that
the eigenmodes do not change significantly after the
marginalization. This is because we impose the prior from Planck,
which constrains the other cosmological parameters very tightly.

\begin{figure}
  \includegraphics[width=8.0cm,angle=0]{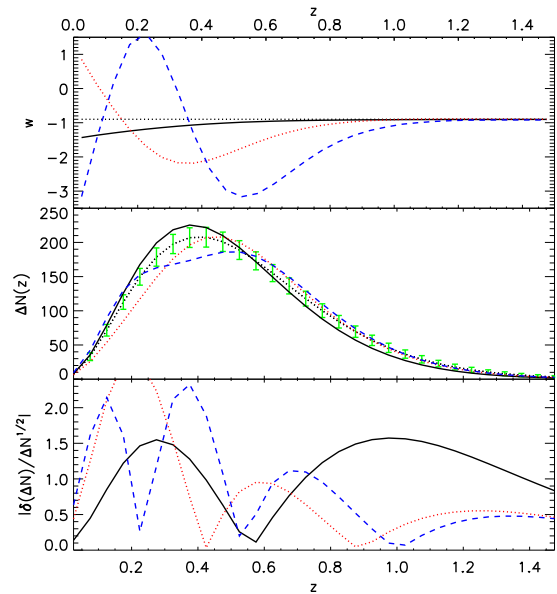}
  \caption{The change in $\Delta N(z)$ when $w$ changes along the
    principal directions of the eigenmodes. The upper panel shows
    $w=w_{\rm fid}+5.7\lambda_{\rm i}^{-1/2}\,e_{\rm i}$ with
    $i=1,2,3$. The second panel from the top shows the $\Delta N(z)$ for
    each case, where the faint dotted line is for the fiducial model
    including Poisson errors, the solid line is for the first eigenmode,
    the dotted line for the second and dashed for the third. In the lower panel
    we show $\delta(\Delta N)/\sqrt{\Delta N}$ for each mode.\label{fig:clusters_fid}}
\end{figure}

In order to test the significance of the eigenmodes, we replace the fiducial model of $w(z)$ with
$w=w_{\rm fid}+\Delta\alpha_{\rm i}\,e_{\rm i}$ with
$\Delta\alpha_{\rm i}=5.7\lambda_{\rm i}^{-1/2}$ corresponding to
the $1-\sigma$ errorbars for 30 free parameters.
In Fig.~(\ref{fig:clusters_fid}), we plot $\Delta N(z)$ for the
fiducial model and $\Delta N(z)$ when $w=w_{\rm fid}+\delta\alpha_{\rm i}\,e_{\rm i}$
for the first three modes. The solid, dotted and dashed lines
represent modes $i=1,2,3$, respectively.
In the top panel, we show $w(z)$. In the second panel from the top, we
show $\Delta N(z)$ together with Poisson errors around the fiducial
model (faint dotted). For clarity, we show the absolute change $|\delta(\Delta N)|$ relative to the error bar $\sqrt{\Delta N}$ in the lowest panel.
For some redshifts where $z<0.5$, $\delta(\Delta N)$ is larger than
expected for the third eigenmode indicating that there might be
redshift ranges where we can constrain up to three modes, although
overall the constraint on this mode is much weaker than the on the
first eigenmode. This agrees with earlier findings \citep{Battye:03} that cluster counts will be able to give a good constraint on $w$ since
$w$ can affect both the volume element ${\rm d}V/{\rm dzd\Omega}$ and the growth factor $D(z)$.

\begin{figure}
\includegraphics[width=8.0cm,angle=0]{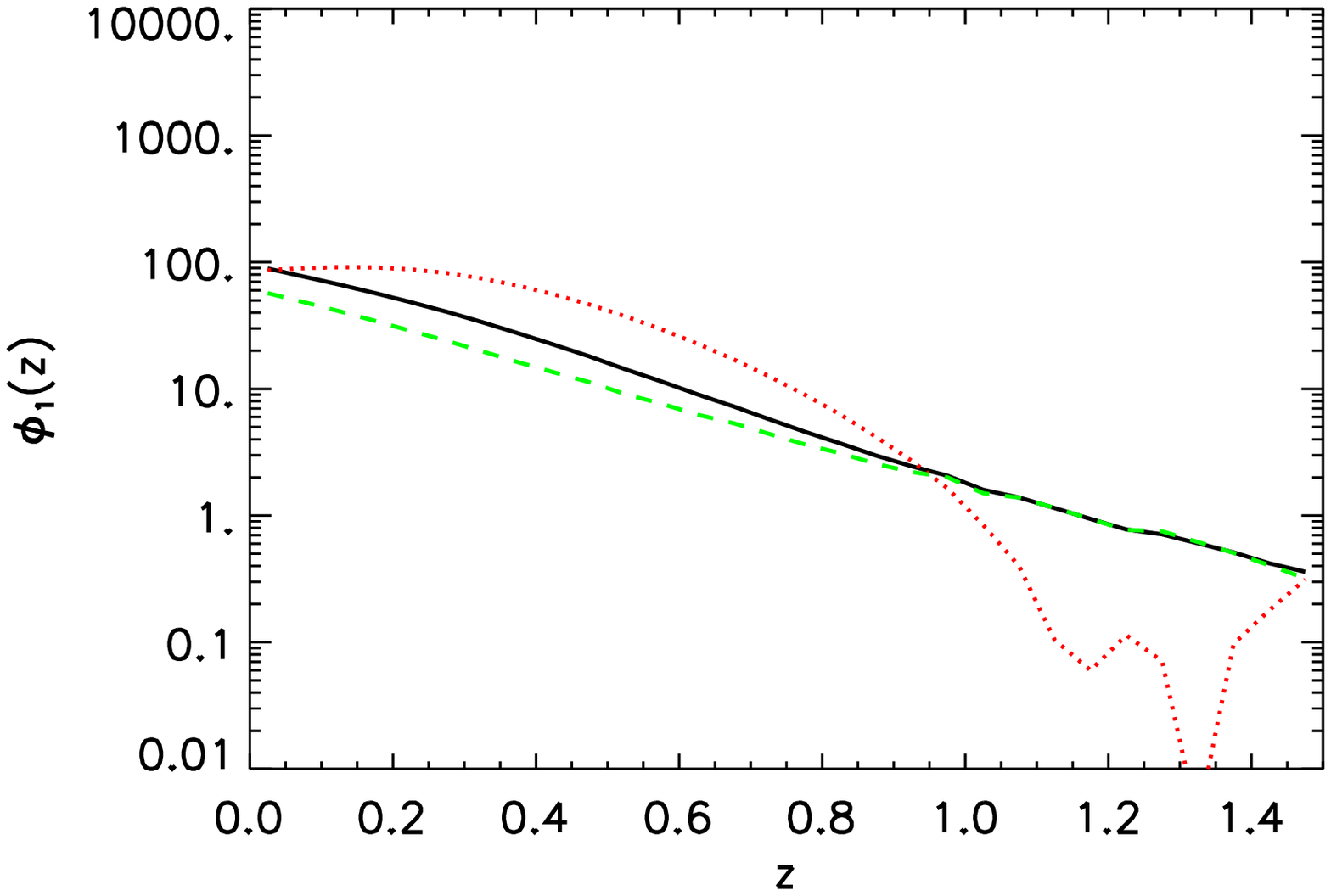}
\includegraphics[width=8.0cm,angle=0]{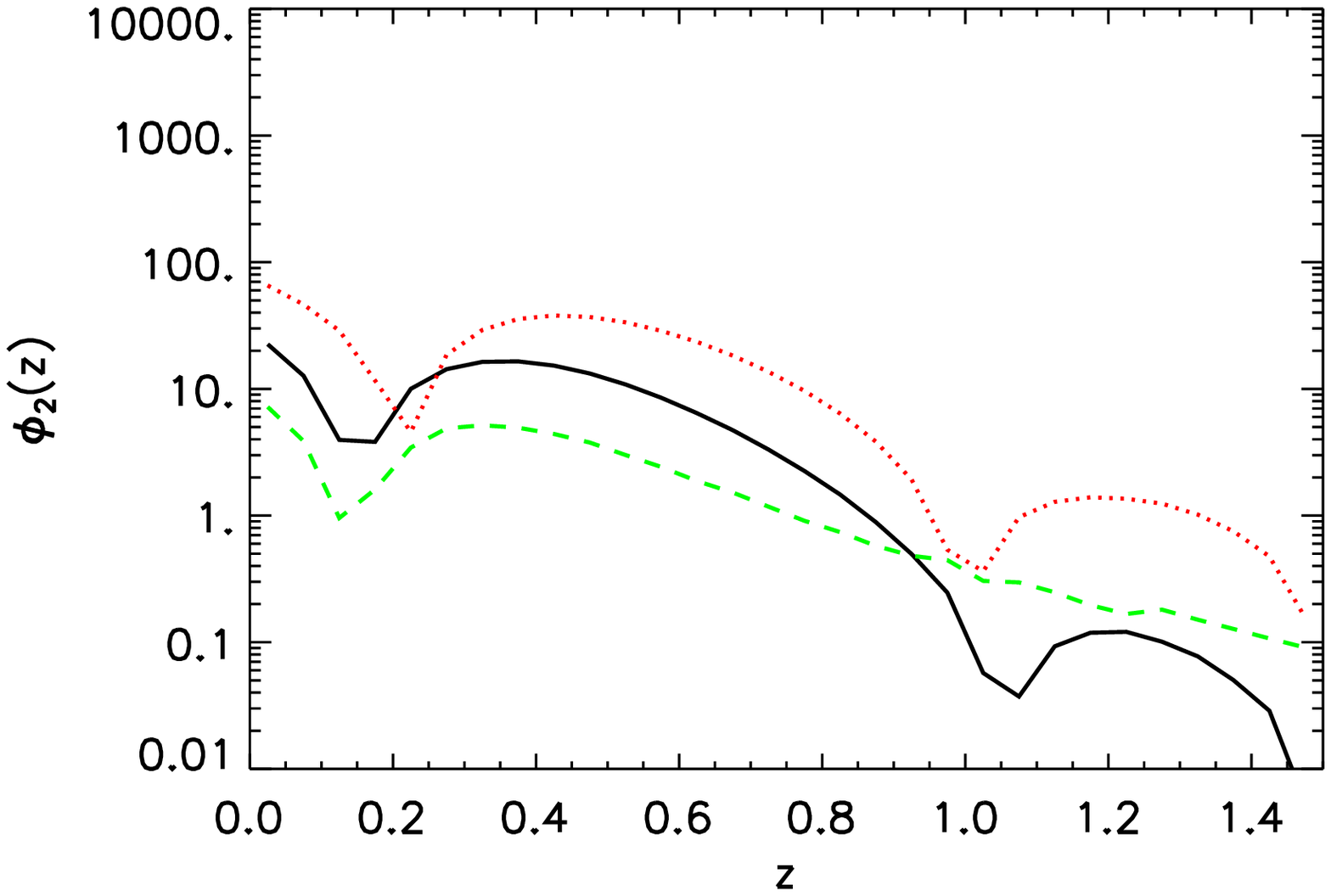}
\caption{The first (top panel) and second (lower panel) weighted
  principal component $\phi_1(z)$ and $\phi_2(z)$ for three different
  mass limits of $\Mlim(z) = 2.4\times10^{14}h^{-1}\Msun$ (solid),
  $\Mlim(z) = 10^{14}h^{-1}\Msun$ (dotted),   $\Mlim(z) = 5\times10^{14}h^{-1}\Msun$ (dashed).
\label{fig:cluster_surveys}}
\end{figure}
In order to compare different surveys we show in
Fig. (\ref{fig:cluster_surveys}) $\phi_1(z)$ and $\phi_2(z)$ for three
different cluster surveys with mass limits of $\Mlim(z) = 2.4\times10^{14}h^{-1}\Msun$ (solid),
  $\Mlim(z) = 10^{14}h^{-1}\Msun$ (dotted),   $\Mlim(z) =
5\times10^{14}h^{-1}\Msun$ (dashed), all over 4,000 deg$^2$ out to a
maximum redshift of $z=1.5$. The lower mass limit could correspond to
optical cluster selection like DES or EUCLID, albeit any realistic
treatment should include scatter in the mass limit for these
surveys. Nevertheless we clearly see that a lower mass limit leads to
much better determined modes, due to the fact the number of observed
cluster increases dramatically for lower mass limits.

\section{Baryon Acoustic Oscillations}
\label{sec:bao}
Baryon Acoustic Oscillation (BAO) arise from the fluctuation due to
the sound waves in the plasma which was composed of tightly coupled CMB photons and baryons before 
recombination. After combination, photons decoupled from baryons; thereafter the acoustic waves were
imprinted as fluctuation in the density fields of both photons and baryons. Though baryons
followed the structure evolution of the underlying dark matter field,
the spatial scale of the acoustic oscillation is well reserved as a standard ruler which can be used to
calibrate the geographical evolution of the Universe. As galaxies are tracers of baryons, galaxy
redshift surveys have been performed to be a complementary probe to constrain cosmological parameters and
the properties of dark energy \citep{2007MNRAS.381.1053P}.

The basic idea of BAO is to calibrate the length scale of the acoustic waves at different redshifts
and fit this distance-redshift relation with cosmological models. The numerical details of the methodology vary according to different authors.
In this paper, we use the methodology developed in \citet{seo:03}, where the full matter power spectrum $P({\bf k},z)$ has been used to
measure Hubble parameter $H(z)$ and the angular diameter distance $d_{\rm a}(z)$ given by
\begin{equation}
d_{\rm a}(z)=\frac{d_{\rm L}(z)}{(1+z)^2}.
\end{equation}
$P({\bf k},z)$ is estimated from galaxies in a redshift shell centered at $z$.
The fisher matrix at the shell is given by \citep{seo:03}
\begin{equation}
F_{\rm ij}=\iint \limits_{\rm \mu,k}
\frac{\partial \ln P(k,\mu)}{\partial \theta_{i}}
\frac{\partial \ln P(k,\mu)}{\partial \theta_{j}}
V_{\rm
eff}(k,\mu)\frac{2\pi k^{2}{\rm d}k{\rm d}\mu}{2(2\pi)^3},
\label{eq:bao_fisher}
\end{equation}
where $k$ is the norm of the wave vector ${\bf k}$, $\mu$ is the
cosine of the angle between ${\bf k}$ and the line of sight.
$V_{\rm eff}$ is the effective volume of the redshift shell given
by
\begin{equation}
V_{\rm
eff}=\int\left[\frac{n({\bf r})P(k,\mu)}{n({\bf r})P(k,\mu)+1}\right]^2{\rm
d}{\bf r},
\end{equation}
in which ${\bf r}$ is the vector along the line of sight and
$n({\bf r})$ is the comoving number density of the galaxies.

The observed power spectrum $P(k,\mu)$ in Eqn.~(\ref{eq:bao_fisher}) is given by \citep{seo:03}
\begin{eqnarray}
P(k_{\rm ref\bot},k_{\rm ref\|})&=&
\frac{d^2_{\rm
a}(z)_{\rm ref} H(z)}{d_{\rm a}^2
H(z)_{ref}}\,b^2\left(1+\beta\frac{k^2_{\|}}{k^2_{\bot}+
k^2_{\|}}\right)^2 \nonumber\\
&&\times D^2(z)\,P(k)+P_{\rm shot},
\label{eq:bao_p_obs}
\end{eqnarray}
in which $\bot$ and $\|$ present the traverse and line-of-sight
directions, respectively. The index ``ref'' indicates the reference
cosmology we use. In this paper, we set the reference cosmology the same as
the fiducial model that we assume. $P_{\rm shot}$ is the shot noise power.
$b$ is the bias factor and $\beta$ is $\Omega_{\rm m}^2/b$. We treat the bias
of each shell independently as an nuisance parameter and marginalize over them in the analysis. In reality,
the fiducial value of the bias depends on galaxy types that one uses to recover the power spectrum.
Here we assume $b=2$ for all galaxy surveys expect for the SKA, where we assume $b=1$.

One uncertainty in Eqn.~(\ref{eq:bao_fisher}) is the upper integral limit of $k$. This has been discussed 
by different authors \citep{Blake:03,seo:03}.  In this paper, we choose a conservative value of $k_{\rm max}=0.2hMpc^{-1}$ for shells at
$z<2$ where most surveys reach. For WFMOS deep, we choose $k_{\rm max}=0.5hMpc^{-1}$, as it is a much higher redshift survey and will 
probably have non-linearity which will prevent us to measure $P(k)$ for larger $k$.

For spectroscopic galaxy surveys, if we ignore the correlation between the shells for very small $k$ along the line of sight,
the Fisher matrix of the whole survey can be obtained simply by adding together the Fisher matrices
at different redshift shells. However, one has to be careful with the photometric redshift surveys.
The relative large photo-z error will
smear the power spectrum along the line of sight and introduce a cross correlation between different shells.
In this paper we ignore the cross correlation between different shells, but given that
the photo-z damps the power spectrum along the line of slight, i.e.
\begin{equation}
P_{\rm photoz}=P(k_{\rm ref\bot},k_{\rm ref\|})\exp(-k_{\rm ref\|}\sigma_{\rm z}c/H(z)),
\end{equation}
where $\sigma_{\rm z}$ is taken from simulations \citep{2007arXiv0705.1437A}.
We choose the width of the shells to be much larger than the photo-z error
$\sigma_{\rm z}$ of galaxies in that shell.
This strategy is also a solution for avoiding the cross correlation between shells \citep{seo:03}.

Another freedom in the analysis is how to choose the redshift shells.
The principle is to make sure that the width is large enough that
one can measure two or three wiggles in the power spectrum along the line of sight in that shell. 
At the same time, we wish to make the shell thin enough so that we have
enough information for $w_{\rm i}$ at each bin. Therefore a choice of
$\Delta z=0.2$ as the shell width means reasonable for most surveys. For WFMOS
deep, we use $\Delta z>0.2$ (see Table.(\ref{tb:bao_slice})).

\begin{table}
 \begin{center}
  \caption{Survey Parameters for BAO}
\begin{tabular}{|c|c|c|c|}
  \hline
  Survey & redshift range & $f_{\rm sky}$ &$ N_{gal} $\\
  \hline
  WFMOS(wide) &  $(0.5,1.3)$      & 0.05 & $2\times10^6  $\\
  WFMOS(deep) & $(2.3,3.3)$ & 0.0075 & $6\times10^5$ \\
  SKA  & (0.,2) & 0.5 & $1.4\times10^9$ \\
\hline
  \label{tb:expt_bao}
\end{tabular}
\end{center}
\end{table}

\begin{figure}
\includegraphics[width=8.0cm,angle=0]{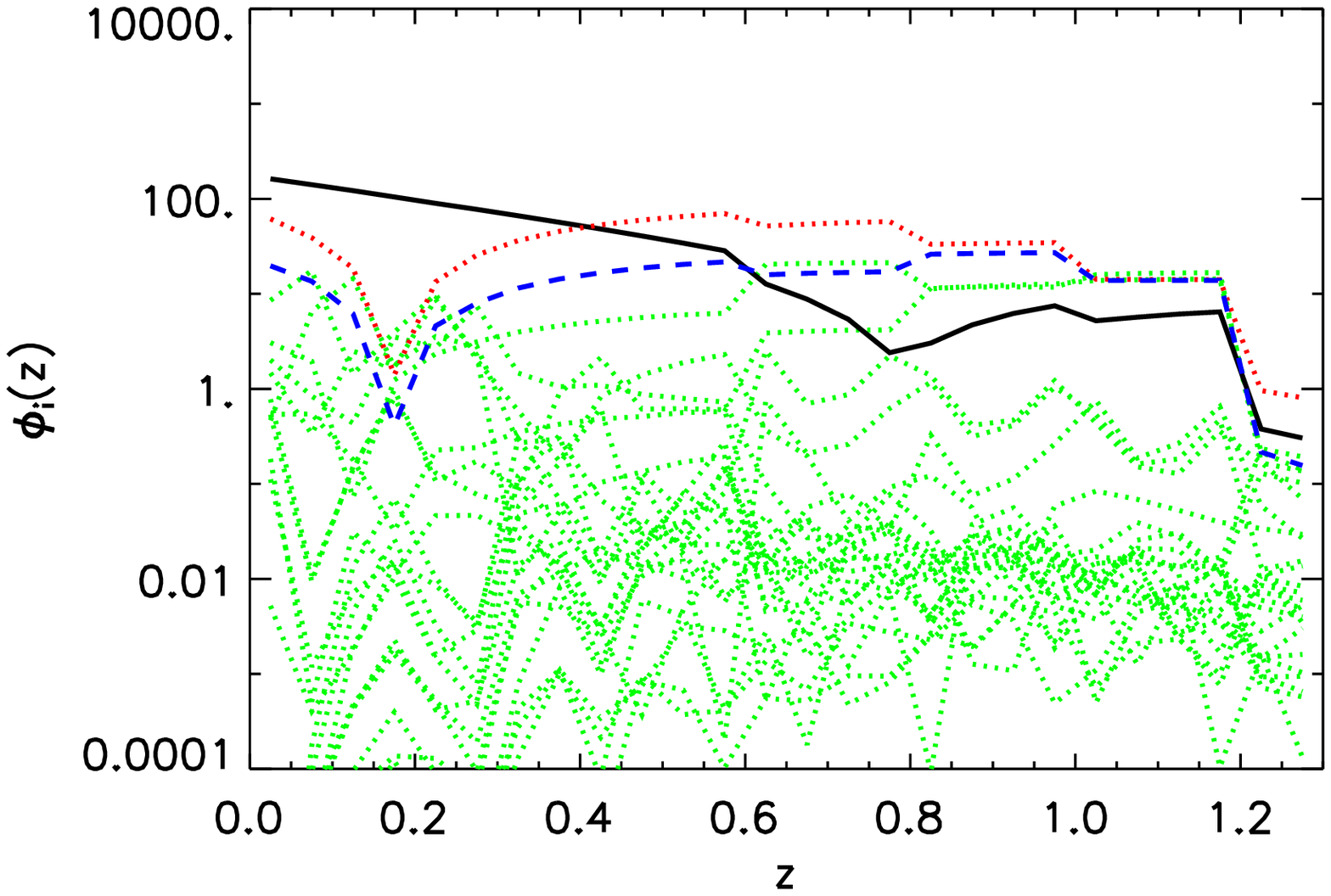}
\includegraphics[width=8.0cm,angle=0]{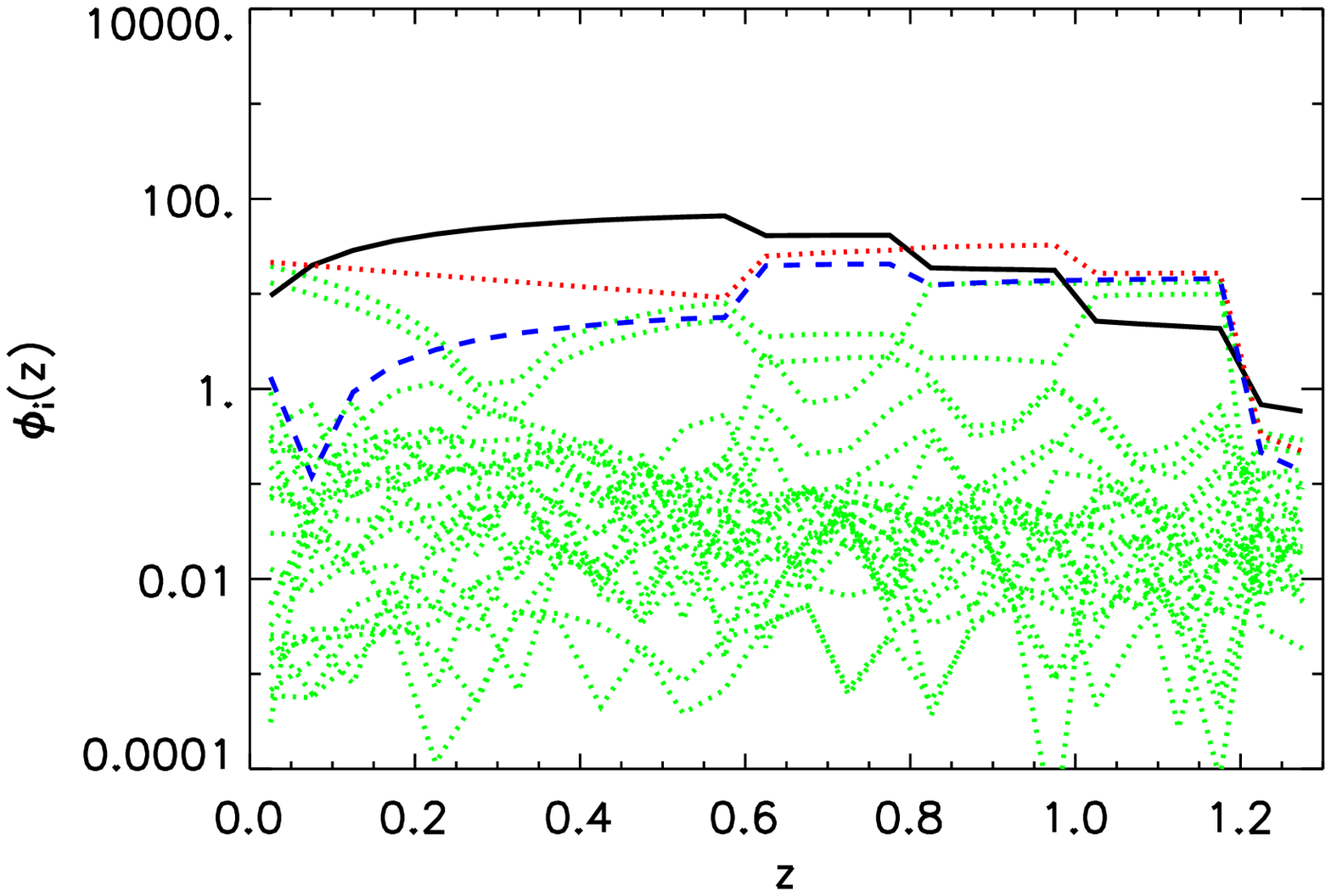}
\caption{26 $\phi_{\rm i}(z)$ for WFMOS large survey.
The (black) solid line indicates the first eigenmode.  The (red) dotted line
and (blue) dashed line indicate the second and the third ones, respectively. The (green) faint dotted lines
show the remaining modes. The upper panel is when cosmological parameters are held fix, while the lower
panel is after cosmological parameters have been marginalized with the of the Planck prior. \label{fig:bao_wfmos}}
\end{figure}

Two out of the three surveys planned by WFMOS (``wide'' and ``deep'') will provide 
a competitive constraint on
cosmology and dark energy \citep{2005astro.ph..7457G}.
In this section, we take the wide survey as an example and study how $\phi_{\rm i}$ behaves in this case.
Table.(\ref{tb:expt_bao}) shows the parameters for WFMOS survey and in
Table.(\ref{tb:bao_slice}) the locations of the slices.
We use $26$ $w$-bins for $z_{\rm max}=1.3$.
We plot $\phi_{\rm i}(z)$ in Fig.~(\ref{fig:bao_wfmos}).
The (black) solid line indicates the best estimated eigenmode.
The (red) dotted line and (blue) dash line indicate the second and the third eigenmode, respectively. 
The (green) dotted lines show the other eigenmodes. In the upper panel we 
marginalize over the biases of each of the shells. The first mode dominates at
low redshift but decays slightly along $z$ and crosses zero around $z=0.8$. At redshift $z>0.5$, the second and third modes dominate.
In the lower panel, we show $\phi_{\rm i}(z)$ after marginalization over all other parameters including a Planck prior. In this plot
the first eigenmode peaks around $z=0.5$ where the survey starts. In both panels, the value of $\phi_{\rm 1}$ is still significant,
which is different from other probes. This is also consistent with \citet{Simpson:06} where they found that the weight function of BAO
shows a high sensitivity on  $w(z)$ at high redshift due to the location of the data at high redshift. The second and third modes
are also significant at high redshift. The obvious discontinuity are found at the redshift where the shells are located.

\begin{table}
 \begin{center}
  \caption{The location of the redshift shells in WFMOS}
\begin{tabular}{|c|c|c|c|c|}
  \hline
  Survey &  1 & 2  & 3 & 4 \\
  \hline
  WFMOS(wide) & 0.6  & 0.8 & 1.0 & 1.2  \\
  WFMOS(deep) & 2.5   & 3.0 & &    \\
\hline
  \label{tb:bao_slice}
\end{tabular}
\end{center}
\end{table}

\begin{figure*}
\begin{center}
\begin{minipage}[c]{1.00\textwidth}
\centering
\includegraphics[width=8.5cm,angle=0]{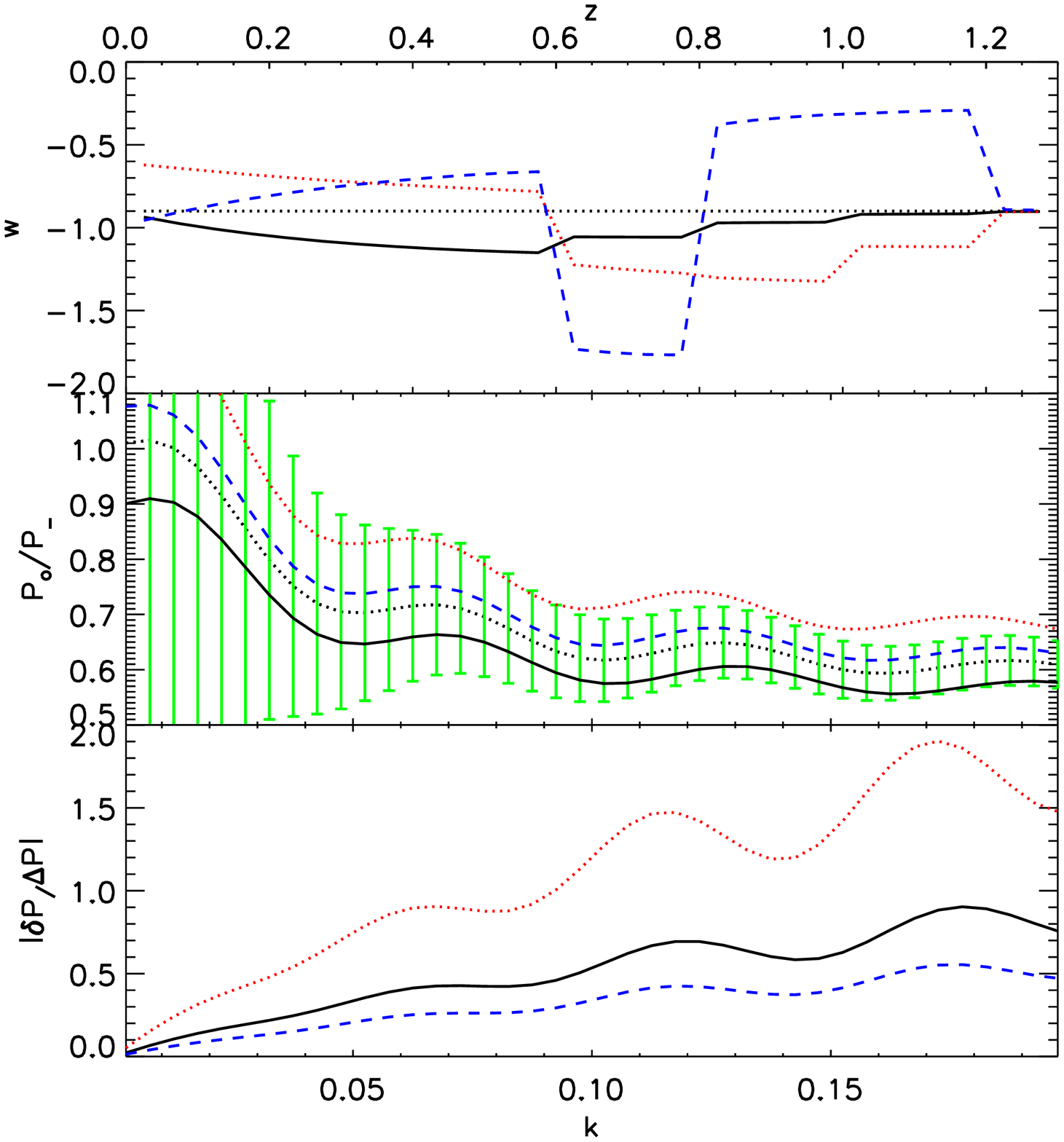}
\includegraphics[width=8.5cm,angle=0]{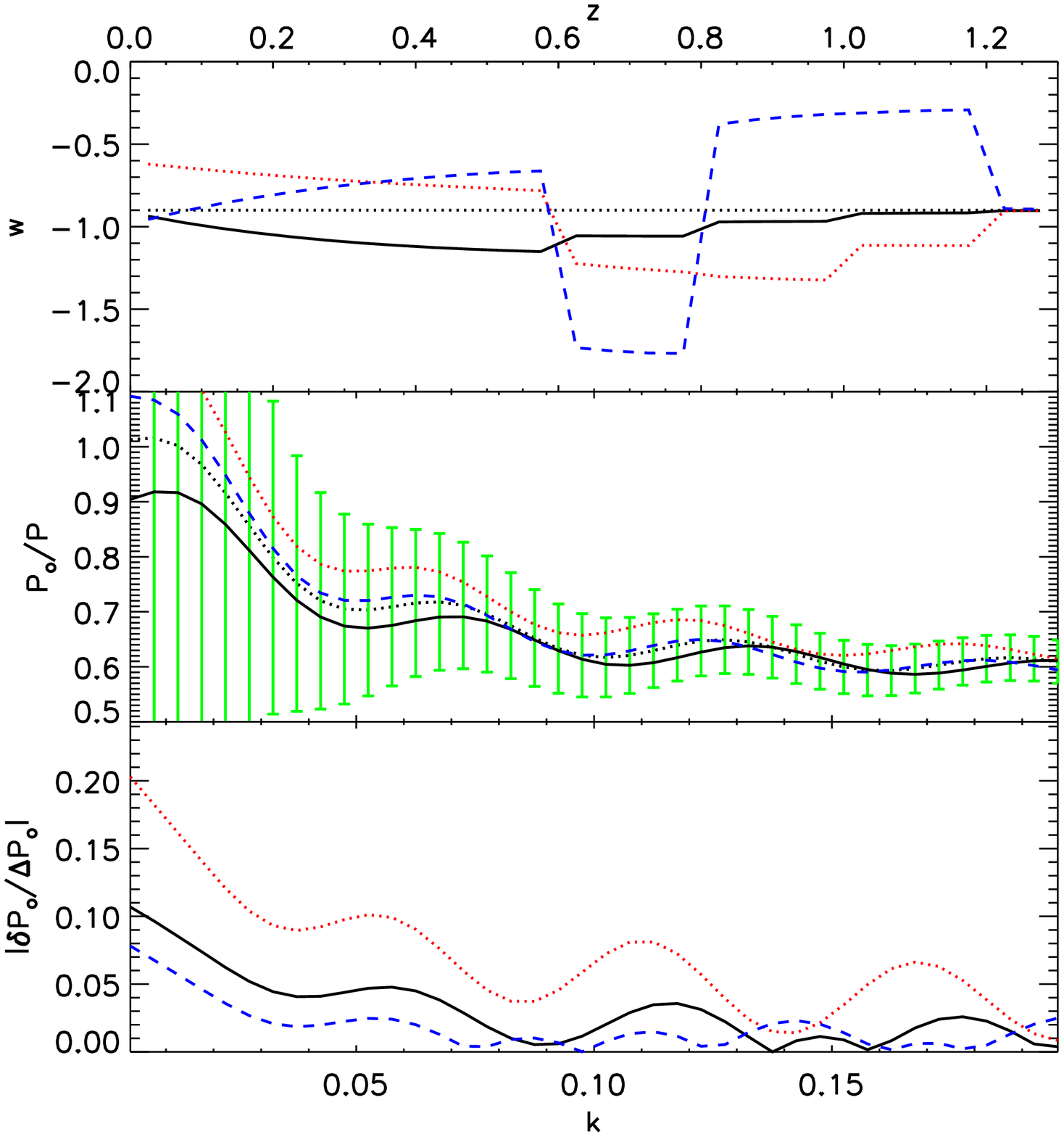}
\end{minipage}
\caption{The changes of $P(k,\mu)$ as we perturb $w$ around the
fiducial model along the eigenmodes direction after
marginalization over cosmological parameters including Planck
prior. The left panel shows the power spectrum perpendicular to
the direction of the line of sight which corresponds to $\mu=0$,
while the right panel shows the power spectrum along the line of
sight which corresponds to $\mu=1$. The (black) solid, (red) dotted,
(green) dashed lines represent the first, second and third
eigenmodes, respectively. In the top plot of each panel, we show
$w$. The second plot from the top shows $P(k,\mu)$ divided by the
power spectrum without baryons. The light area shows the
observational error on the spectrum under the same normalization. we presents in the bottom of
each panel $|\delta P/\Delta P|$. \label{fig:bao_fid}}
\end{center}
\end{figure*}

In Fig.~(\ref{fig:bao_fid}), we show how $P(k,\mu)$ changes along
with $w$. We take the first redshift shell in WFMOS wide survey as
an example. The left panel
shows the power spectrum perpendicular to the direction of the
line of sight which corresponds to $\mu=0$, while the right panel
shows the power spectrum along the line of sight which corresponds
to $\mu=1$. In all the panels, the (black) solid line indicates
the first eigenmode. The (red) dotted line and the (blue) dashed
line represents the second and third modes, respectively. The top
plot on each panel shows $w=w_{\rm fid}+\Delta\alpha_{\rm
i}\,e_{\rm i}$, where $\Delta\alpha_{\rm i}=5.4\sqrt{\lambda_{\rm
i}^{-1}}$ for $N=26$. Here we use $e_{\rm i}$ shown in the lower
panel of Fig.~(\ref{fig:bao_wfmos}). One can notice that $w$
deviates from the fiducial values dramatically at high redshift, which
is different from what we found from SNe Ia, WL and Cluster Count.
The second
plot from the top presents the $P(k,\mu)$ divided by the power
spectrum without baryons and normalized to the same value at
$k=0$. The (green) error bars represent the error $\sigma(P)$ on
the power spectrum, which is given by \citep{seo:03}
\begin{equation}
\sigma(P)=\frac{P}{V_{\rm eff}}\sqrt{\frac{2(2\pi)^2}{k^2\Delta
k\Delta \mu}}
\end{equation}
In order to be consistent, we also divide $\sigma(P)$ by the fiducial 
power spectrum without baryons. In the bottom plot we show the absolute change $|\delta|$
relative to the error bar. As we expect, $\Delta P$ is smaller than the error of the power spectrum.
One can also notice that the amplitude of the power spectrums 
changes significantly compared with the scale of the wiggle.

\begin{figure}
\includegraphics[width=8.0cm,angle=0]{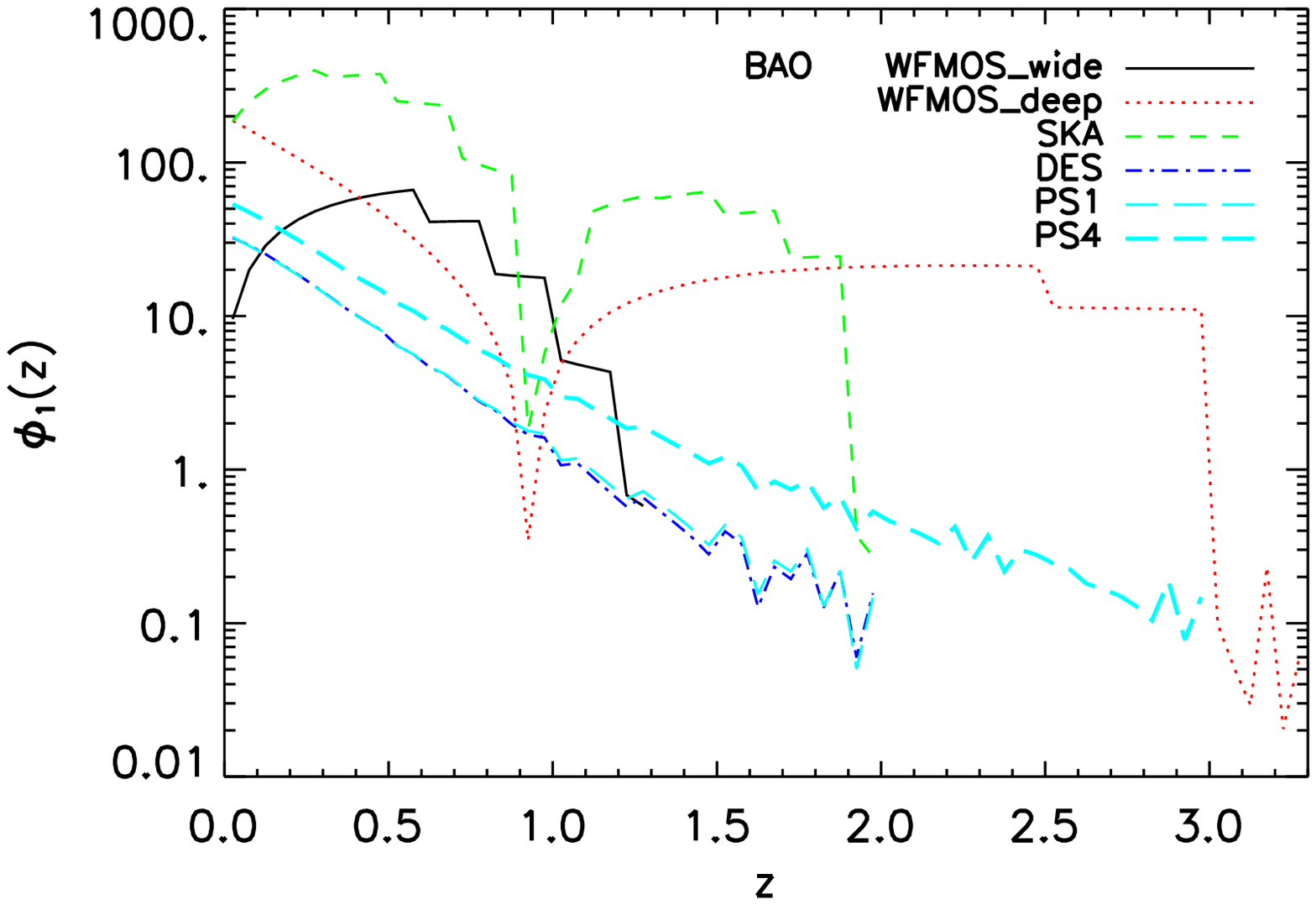}
\includegraphics[width=8.0cm,angle=0]{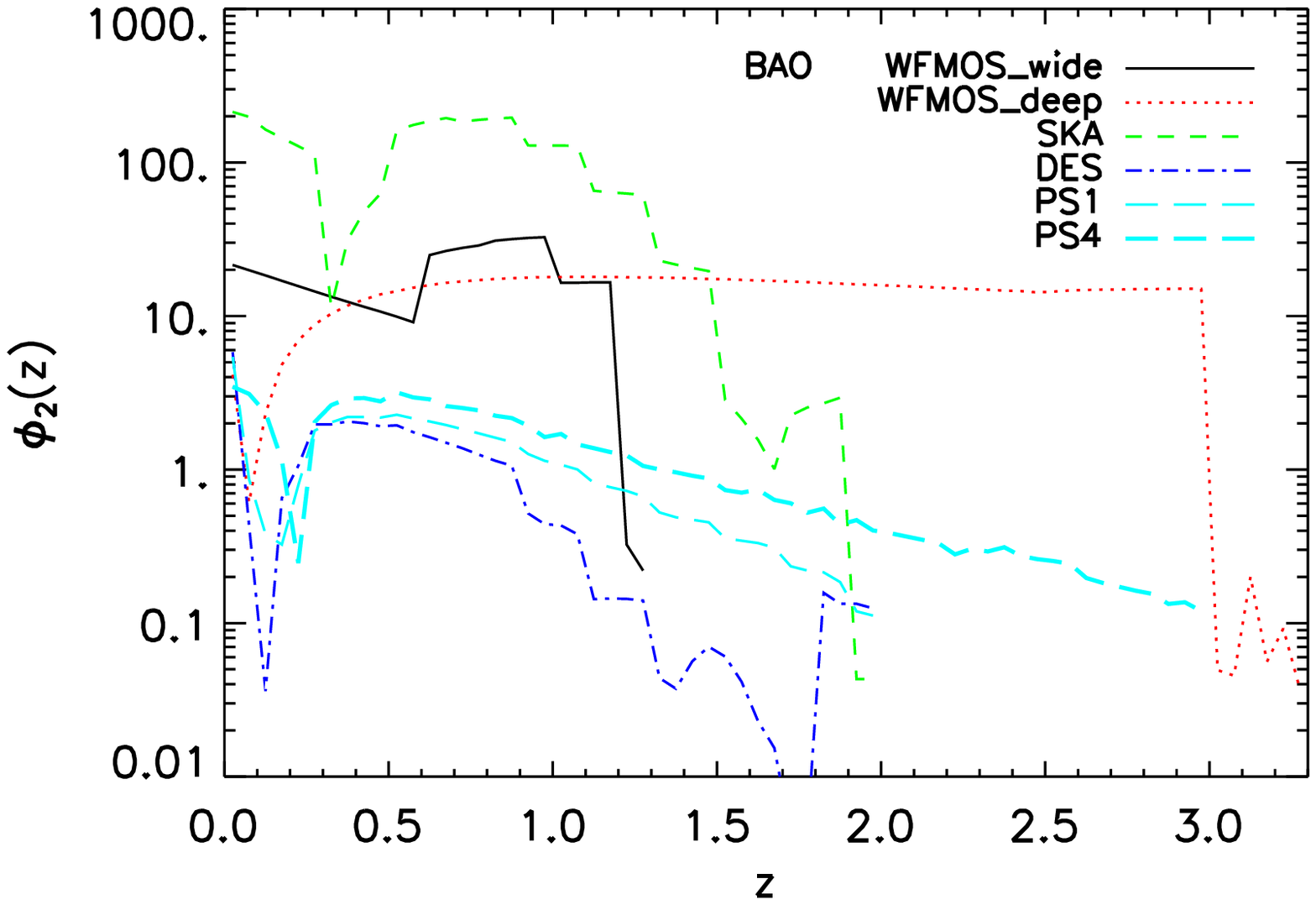}
\caption{$\phi_{\rm 1}(z)$ (upper panel) and $\phi_{\rm 2}(z)$
(lower panel) from different BAO experiments.
 The (black) solid line indicates WFMOS wide;
the (red) dotted line represents WFMOS deep. The (green) dashed line represents SKA. 
The (blue) dash dotted line shows DES, the (cyan) thin long dashed line
shows PS1 and the thick one represents PS4.
 \label{fig:bao_eigen_com}}
\end{figure}

In Fig.~(\ref{fig:bao_eigen_com}), we show $\phi_{\rm 1}(z)$ and 
$\phi_{\rm 2}(z)$ from different redshift surveys marginalized with Plank prior.
There are seven surveys shown in this plot. The (black) solid line indicates WFMOS wide;
the (red) dotted line represents WFMOS deep. The (green) dash line
represents SKA. Note that we have not analysed EUCLID BAOs in this
paper, which is a strong component of the EUCLID program.
Because of the large cosmic volume, the height of $\phi_{\rm 1}$ for the SKA is one magnitude larger than for other surveys. 
For WFMOS deep, the data comes from redshift $2.3<z<3.3$; therefore $\phi_{\rm i}$ dominates at high redshift. However, $\phi_{\rm i}$  
is significant at very low redshift as well. Both $\phi_{\rm 1}$ of SKA and WFMOS deep cross zero around $z=1$, which is 
different from $\phi_{\rm 1}(z)$ of WFMOS wide. Besides WFMOS and SKA, we also show the eigenmodes from photo-z redshift surveys. 
The (blue) dash dotted line shows DES, the (cyan) thin long dash line
shows PS1 and the thick one represents PS4. These $\phi_{\rm 1}(z)$'s
are very similar to $\phi_{\rm 1}(z)$ from Planck prior. 
This is because of the damping factor on the power spectrum 
of the line of sight due to the photo-z error. The values of the
Fisher matrices are relatively small compared with Planck prior;
therefore the Planck prior has a larger effect on $\phi_{\rm 1}(z)$.

\section{Joint Principal Components}
\label{joint}
We have performed PCA analysis on a few representative
future surveys. In the following, we will compare the eigenmodes for
different stages and discuss the
joint principal components from each stage. All the probes discussed
in this section are marginalized over the other parameters including
the Planck prior.

\begin{figure}
\includegraphics[width=8.0cm,angle=0]{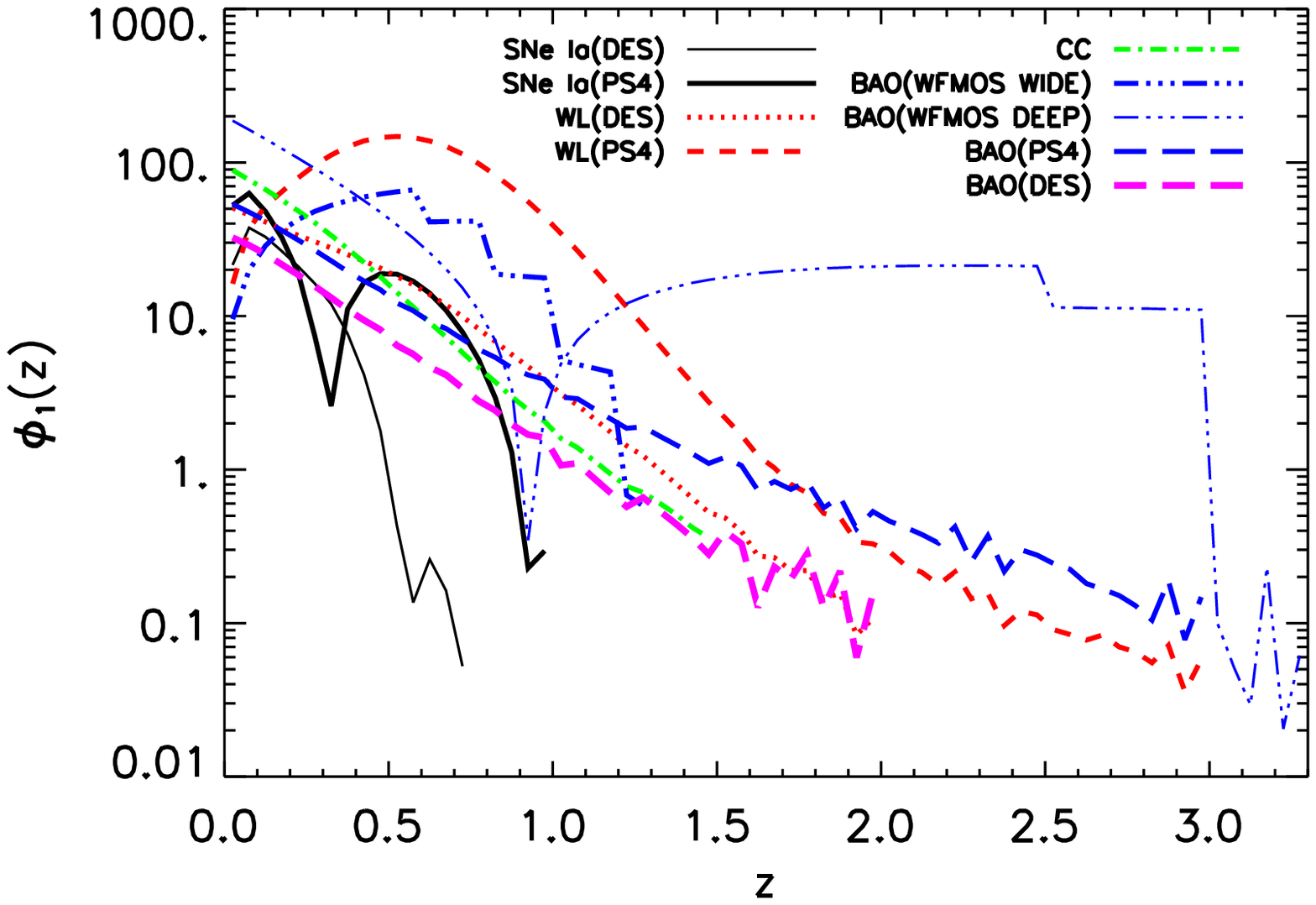}
\includegraphics[width=8.0cm,angle=0]{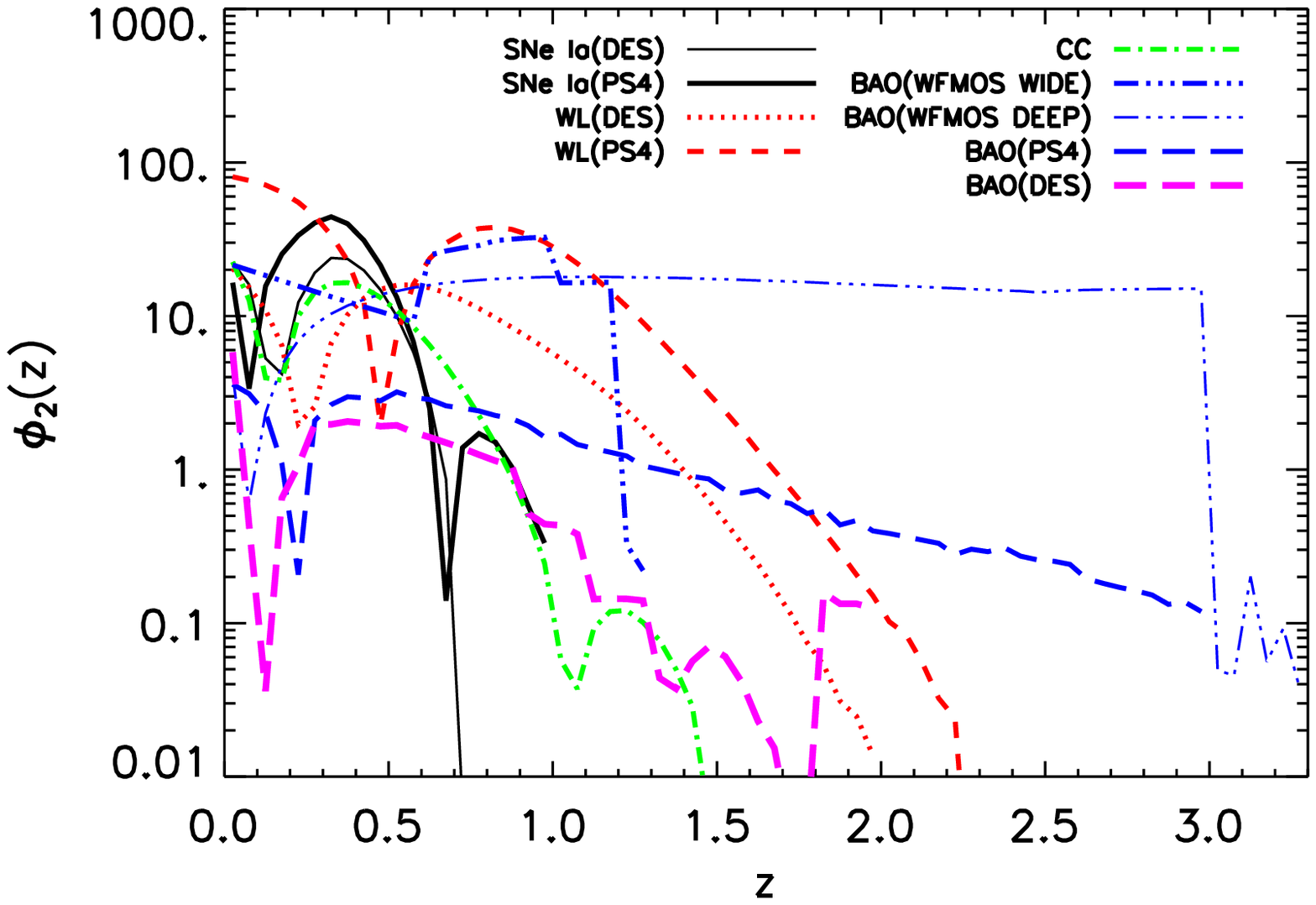}
\caption{$\phi_{\rm 1}(z)$ (upper panel) and $\phi_{\rm 2}(z)$ (lower  
panel) for future surveys
in stage III. The (black) solid line
indicates SNe Ia surveys with the thin and thick lines representing DES
and PS4, respectively. The (red) dotted and dash lines represents WL  
from DES
and PS4, respectively. The (green) dash-dotted line shows Cluster  
count. The rest lines show
BAO; the (blue) long dash line shows PS4 and the (magenta) thicker  
long dash line represents DES.
The (blue) dash dotted-dotted-dotted lines show WFMOS with the thin  
and thick lines representing
deep and wide, respectively.
\label{fig:stage_III}
}
\end{figure}

In Fig.~(\ref{fig:stage_III}), we show $\phi_{\rm 1}(z)$ (upper panel)
and $\phi_{\rm 2}(z)$ (lower panel) for the surveys that we have
analysed for stage III . The (black) solid line
indicates SNe Ia surveys with the thin and thick lines representing DES
and PS4, respectively. The (red) dotted and dash lines represents WL  
from DES
and PS4, respectively. The (green) dash-dotted line shows the cluster
count result. The remaining lines show
the result for BAO surveys; the (blue) long dashed line is for PS4 and  
the (magenta) thick long dash line represents DES.
The (blue) dash triple-dotted lines show WFMOS with the thin and thick
lines representing
deep and wide, respectively. The amplitudes of $\phi_{\rm 1}(z)$ for
this stage are roughly between ten and one hundred.
$\phi_{\rm 1}(z)$ of PS4 for WL dominates at redshift $z<1.2$; while $ 
\phi_{\rm 1}(z)$ of
WFMOS deep for BAO is dominant for higher redshifts.
The redshift dependence of $\phi_{\rm 1}(z)$
can be classified into three types. First, $\phi_{\rm 1}(z)$ behaves  
very similar to
the first mode of the Planck prior, which is represented by CC and BAO  
from DES and PS4.
This is due to the relatively strong Planck prior for these cases.
Second, $\phi_{\rm 1}(z)$ crosses zero at low redshift. $\phi_{\rm 1} 
(z)$ of DES and PS4 for SNe Ia
and WFMOS deep for BAO show this feature. However for WFMOS deep this
feature is already at intermediate redshifts around $z\approx 1$. The
third type of behaviour is encountered by PS4 for WL and WFMOS wide  
for BAO.
$\phi_{\rm 1}(z)$ stays positive and peaks at the median redshift.
For most of the probes, $\phi_{\rm 1}$ is significant at low redshift and  
then decays afterward,
while WFMOS deep has also a very significant contribution at
high redshift above $z=1.5$.
In the lower panel, we notice that the amplitude of the second mode
$\phi_{\rm 2}(z)$ is already one order of magnitude lower than the
first mode.
The dominant contribution of $\phi_{\rm 2}(z)$ is still at low
redshift, with the exception of WFMOS deep. Due to the dominance of
the first mode $\phi_{\rm 1}(z)$ the main redshift dependence of a
given survey is encoded in this mode \citep{Huterer:03}.

\begin{figure}
\includegraphics[width=8.0cm,angle=0]{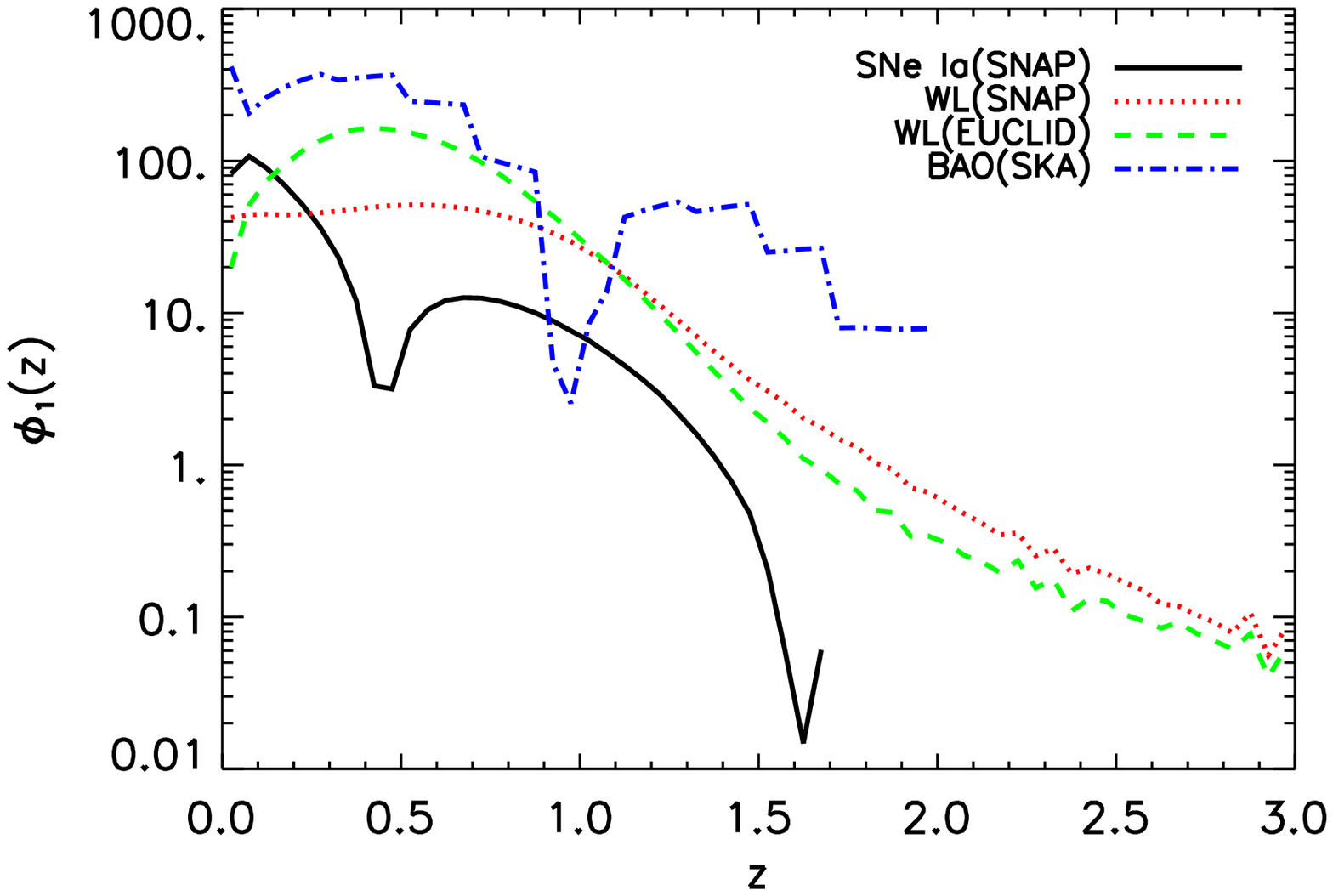}
\includegraphics[width=8.0cm,angle=0]{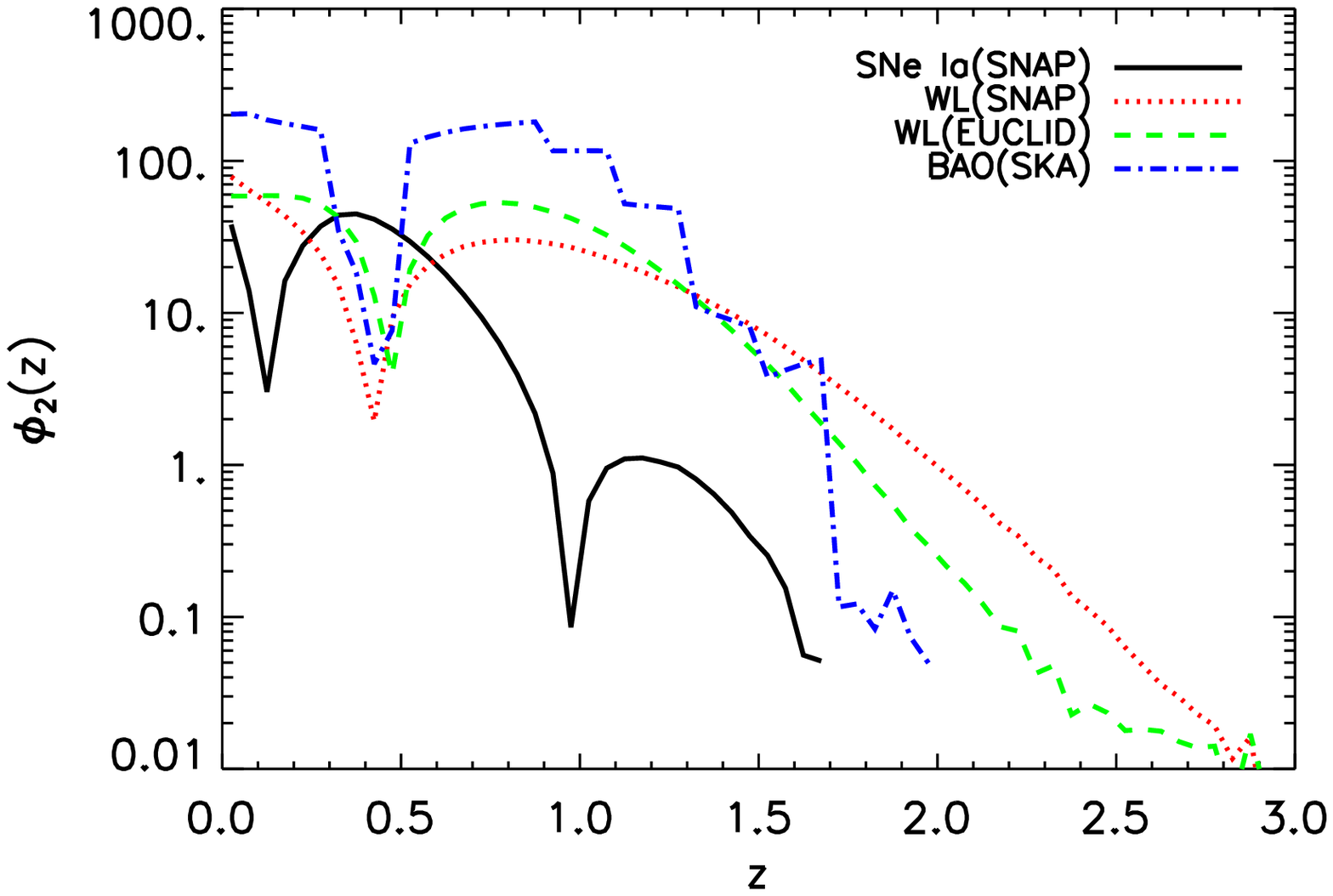}
\caption{$\phi_{\rm 1}(z)$ (upper panel) and $\phi_{\rm 2}(z)$ (lower  
panel) for future surveys
in stage IV. The (black) solid line
shows SNe Ia from SNAP. The (red) dotted line shows WL from SNAP and
the (green) dash line shows WL from EUCLID. The (blue) dash dotted
line shows BAO from SKA.
\label{fig:stage_IV}
}
\end{figure}

Fig.~(\ref{fig:stage_IV}) shows the two leading modes for the stage IV
surveys in our analysis. The (black) solid line
is for type Ia SNe from SNAP. The (red) dotted line shows WL from SNAP
and the (green) dash line shows WL from EUCLID. The (blue) dash dotted
line shows BAO from SKA. The amplitudes of $\phi_{\rm 1}(z)$ for this
stage are about one order of magnitude higher than
$\phi_{\rm 1}(z)$ in stage III; SKA for BAO dominates in its redshift
range $z<2$. $\phi_{\rm 1}$ of SNAP for SNe Ia and BAO for SKA have a
change in sign for the most dominant mode,
while $\phi_{\rm 1}(z)$ of both SNAP and EUCLID for WL have a mode
which stays positive throughout.
The amplitude of $\phi_{\rm 2}(z)$ is about one order of magnitude
lower, hence of the same significance as the primary mode for the
stage III surveys.

\begin{figure}
\includegraphics[width=8.0cm,angle=0]{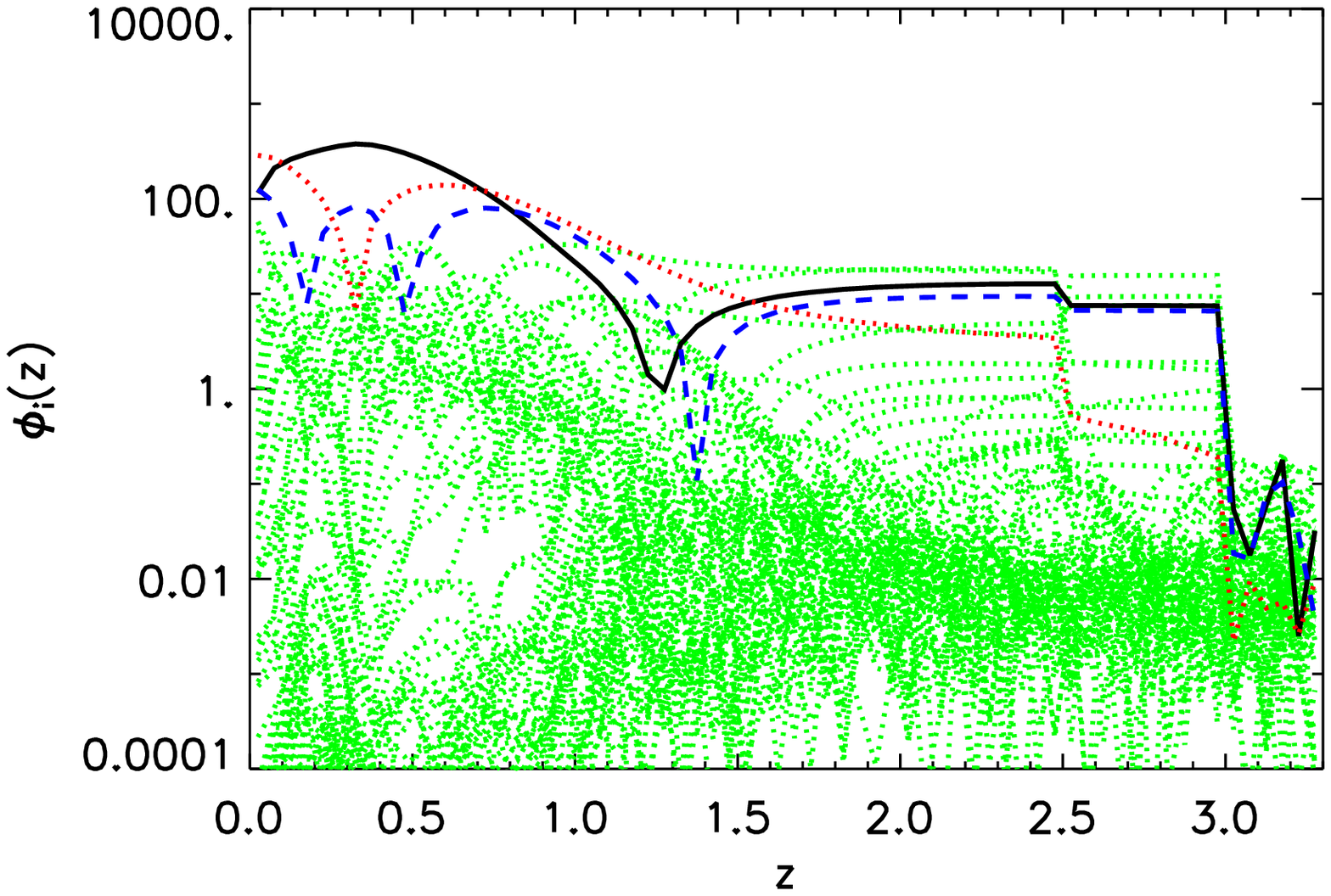}
\includegraphics[width=8.0cm,angle=0]{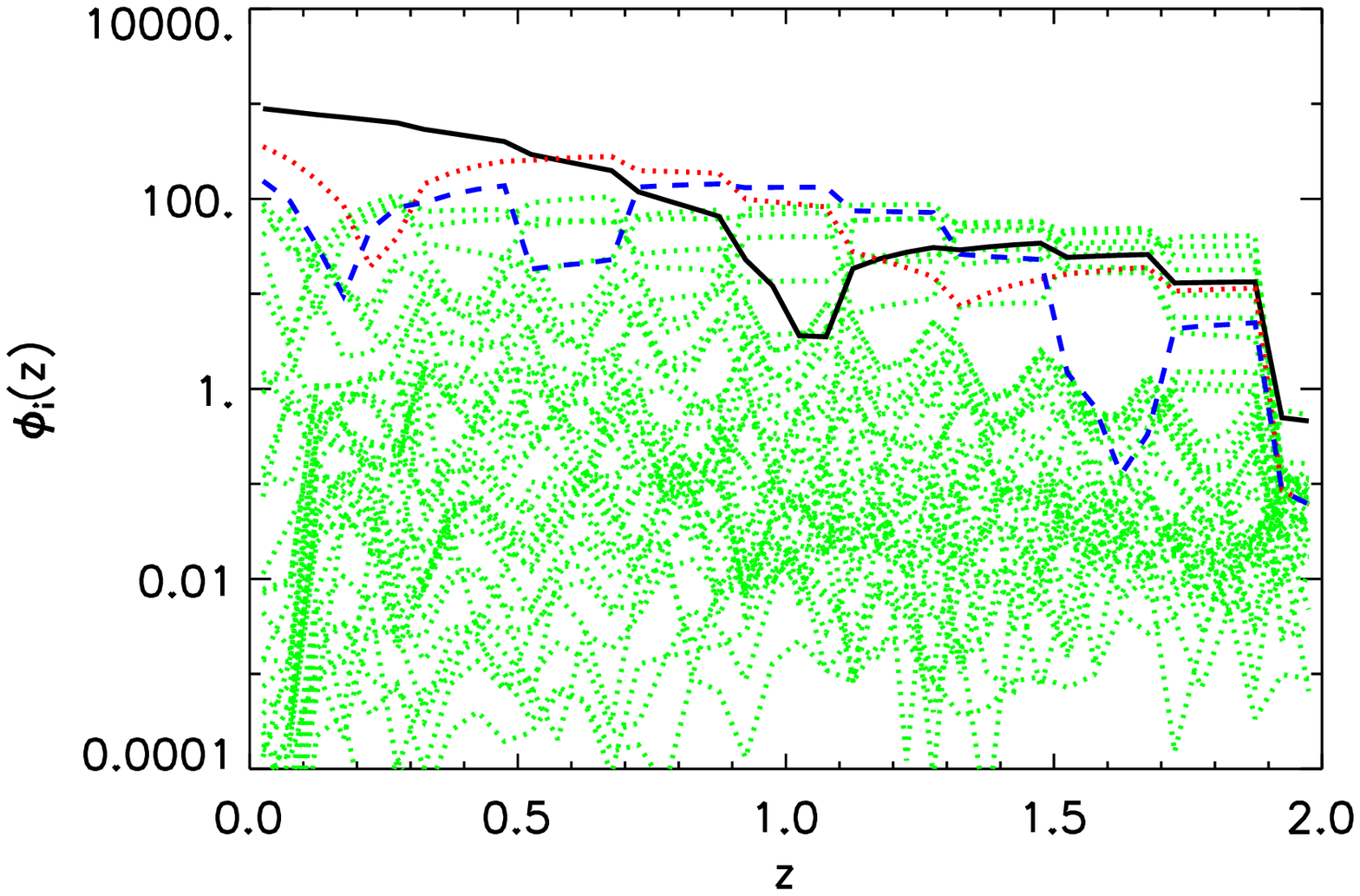}
\caption{The joint eigenmodes for stage III (the upper panel) and  
stage IV (the lower panel). The (black) solid line indicates the first  
eigenmode.
The (red) dotted line and (blue) dashed line indicate the second and  
the third ones, respectively. The (green)
faint dotted lines show the remaining modes.  For stage III,
we combine PS4(SNe Ia), PS4(WL) and WFMOS deep (BAO). For stage IV,
we combine SNAP(SNe Ia), EUCLID(WL) and SKA(BAO).
We also include Planck prior in both cases.
\label{fig:joint}
}
\end{figure}

From the four probes that we discussed, one can find that for most  
surveys the best constrained eigenmodes dominate at low redshift.
The exception are surveys which target objects at high 
redshifts as for example WFMOS deep.
$\phi_{\rm 1}$ from WFMOS deep stands out at  
high redshift in the upper panel of Fig.~(\ref{fig:stage_III}). %
In the following, we discuss how the eigenmodes behave if we combine
different probes. We perform the joint analysis based on the two
stages. For stage III, we combine PS4(SNe Ia), PS4(WL) and WFMOS deep
(BAO). For stage IV, we combine SNAP(SNe Ia), EUCLID(WL) and
SKA(BAO). We also include the Planck prior for both cases.
In Fig.~(\ref{fig:joint}) we show the joint eigenmodes with the upper
panel and the lower panel representing the Stage III and IV,
respectively.
We find that
$\phi_{\rm 1}(z)$ is significant within the whole redshift range
expect around $z=1.2$ where $\phi_{\rm 1}(z)$ changes sign. It is
interesting to note that higher order modes and namely the second mode
fill this gap in redshift. In addition boosting $\phi_{\rm i}(z)$ at
high redshift, WFMOS deep (BAO) is complementary with PS4(SNe Ia),
PS4(WL) at high redshifts.
For Stage IV we find that the equation of state will be constrained
significantly at least out to redshift $z=1.5$.

\section{$w$ Reconstruction}
\label{sec:ev}
So far, we have compared different surveys by concentrating on the  
behaviour
of $\phi_{\rm 1}(z)$ and $\phi_{\rm 2}(z)$, which allows us to explore  
the redshift
sensitivity on $w(z)$ for each probe.
To reconstruct $w(z)$, however,
one would like to include higher order eigenmodes to get less bias in the
reconstructed equation of state.
However, if we use higher order eigenmodes the error on the
reconstructed $w$ is dominated by the errors from these orders. In
order to decide how many eigenmodes to use for reconstruction we
require the bias {\em and} the variance to be low.
Hence, an alternative to the standard figure of merit of a survey
\citep{DETF:06} is to count how many eigenmodes can be well constrained
under a certain criteria.

\begin{figure}
\includegraphics[width=8.0cm,angle=0]{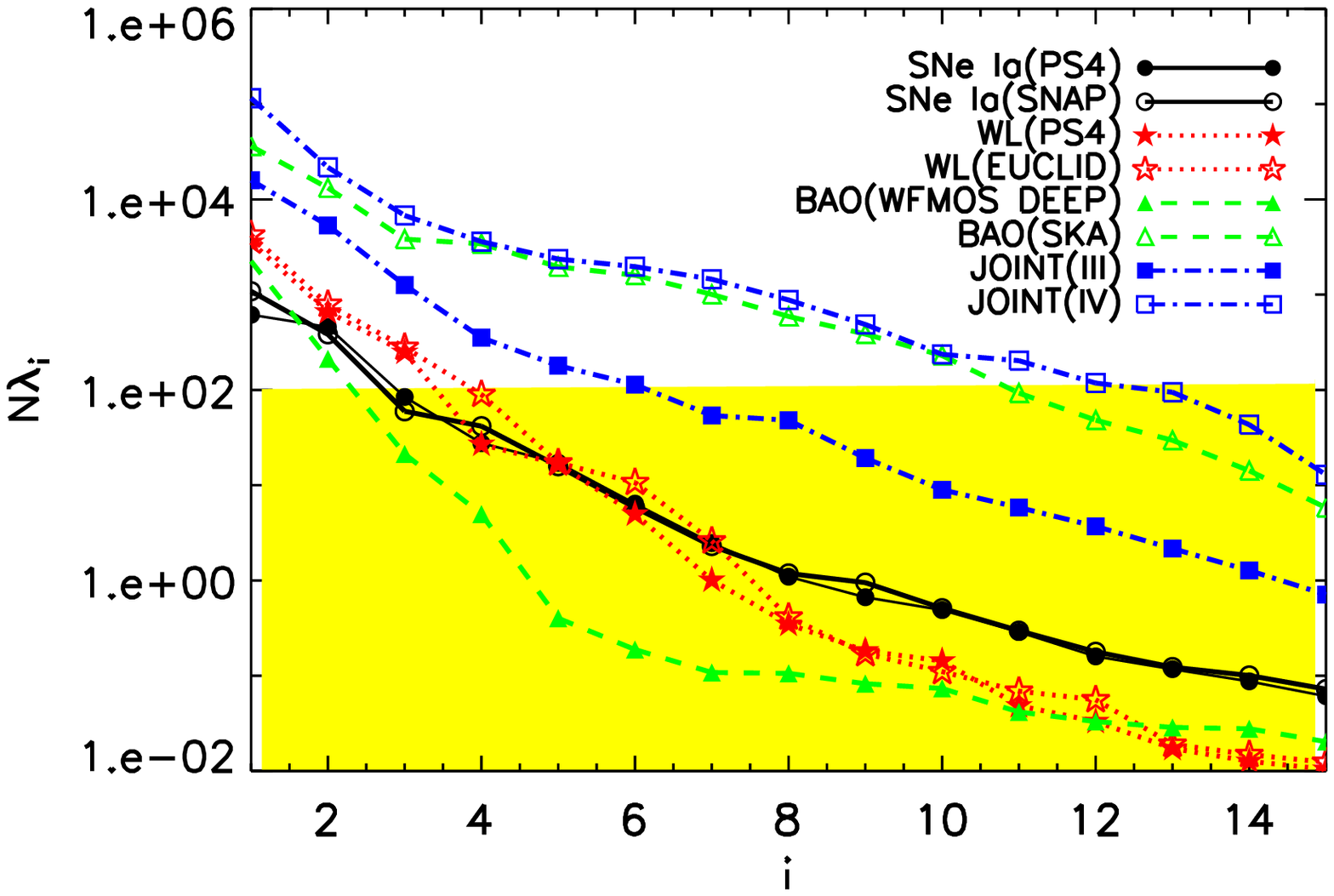}
\caption{The weighted eigenvalues for the surveys that we used in the  
joint analysis.
Every line represents a survey.
All the surveys are marginalized over other parameters including  
Planck priors. The (black) solid line shows SNe Ia surveys with the  
filled and unfilled circles indicating
PS4 and SNAP, respectively. The (red) dotted lines represents WL  
surveys with the filled and unfilled stars indicating PS4 and EUCLID,  
respectively. The (green) dash lines
represents BAO surveys with the filled and unfilled triangles  
indicating WFMOS deep and SKA, respectively. We also show the joint  
analysis with the (blue) dotted-dash
lines; the filled and unfilled squares indicating stage III and IV,  
respectively.
\label{fig:eigen_value}}
\end{figure}

In Fig.~(\ref{fig:eigen_value}), we plot the first 15 eigenvalues
weighted by the number of bins, for
the different surveys that we used in the joint analysis in the
previous section.
All the surveys are marginalized over other parameters including  
Planck priors.
The (black) solid line shows the Type Ia SNe surveys with the filled and unfilled
circles indicating PS4 and SNAP, respectively. The (red) dotted lines  
represents the WL surveys, with the filled and unfilled stars indicating  
PS4 and EUCLID,
respectively. The (green) dashed lines represents the BAO surveys with the  
filled and unfilled triangles indicating WFMOS deep and SKA,  
respectively.
We also show the joint analysis with the (blue) dotted-dash lines; the  
filled and unfilled squares indicating stage III and IV, respectively.

In order to explore the problem of the w-reconstruction
quantitatively let us assume that one only uses $M$ eigenmodes. The
reconstructed $w_{\rm rec}(z)$ is then   
given by
\begin{equation}
   w_{\rm rec}(z)=\bar{w}(z)+\sum_{\rm i=1}^{\rm M}\bar{\alpha}_{\rm  
i}e_{\rm i}(z),
   \label{eq:w_reconstruct}
\end{equation}
where $\bar{w}$ is the value around we wish to reconstruct the
equation of state. The $\bar{\alpha}_{\rm i}$ are the best fit
coefficients in the eigenmode basis $\{e_{\rm i}(z)\}$. The
reconstruction strategy given by \citet{Huterer:03} is
equivalent to setting $\bar{w}(z)=0$ and the expected
$\bar{\alpha}_{\rm i}$ is then given as the projection of $e_{\rm i}(z)$  
on the fiducial model.
\citet{Crittenden:05} define $\bar{w}=w_{\rm fid}$
under the assumption that $w_{\rm fid}$ is close to the true physical
model for any reconstruction in a real data situation and in this case
the expected $\bar{\alpha}_{\rm i}$ are zero.

We now have to employ a statistical criterion, which allows us to
gauge how many eigenmodes to use for the reconstruction of the
equation of state. There are three effects at work, which need to be
considered. First the goodness of the fit, which, of course, improves
with the number of used modes, second the degradation of the errorbars
with increasing number of modes and finally the bias between the true
underlying model and the model, ${\bar w}$, around which we
reconstruct the equation of state. As we will show below the Bayes'
factor \citep{jeffreys,Trotta:08} based on Bayesian evidence \citep{Sivia96}
provides exactly such a criterion.

We will proceed by describing a Gaussian approximation of the Bayes'
factor, based on Bayesian evidence, to decide the number of
significant modes. We will follow 
closely the discussion in \citet{Saini:04}. Under Gaussian assumptions, 
which hold for the Fisher matrix approximation of the underlying
likelihoods, the evidence for the data $D$ given the hypotheses H is
approximated by
\beq
{\cal E} = P(D|H) \approx P(D|{\mbox{\boldmath$\theta$}}_{\rm
    L},H)\exp(-C)\left(\frac{\left|{\mathbf F} + {\mathbf
    P}\right|}{\left|{\mathbf P}\right|}\right)^{-1/2}\; 
\label{eqn:evidence}
\eeq
with ${\mathbf F}$ the covariance between the w-bin parameters,
${\boldsymbol \theta}$, in the
basis of the eigenmodes, 
${\mathbf P}$ the covariance of the prior on these 
parameters and ${\boldsymbol\theta}_L$ are the parameters of the maximum
likelihood. $C$ is the term which encodes the overlap of the prior with the likelihood, i.e. this term is measuring
the bias between the prior and the likelihood and is given by
\beq
C = \frac{1}{2}\left({\boldsymbol\theta}_L-{\boldsymbol\theta}_P\right)^T{\mathbf P}\left({\mathbf F}+{\mathbf P}\right)^{-1}{\mathbf F}\left({\boldsymbol\theta}_L-{\boldsymbol\theta}_P\right)\; ,
\eeq
where ${\boldsymbol\theta}_P$ is the mean of the prior, which we
choose to be ${\bar w}(z)$ or $\alpha_i^P=0$ in the expansion given in
Eqn.~\ref{eq:w_reconstruct}. If we make the simplistic assumption that the
prior is diagonal and the same in each bin we obtain
\beq
{\mathbf P} = \frac{1}{\Delta w^2}\mathbf{1}
\eeq
Note that this prior, because of its diagonal nature, has the same
form in the original w-bin basis and the eigenmode basis. Here we have to make
an important point. We would like that the rms variance, $\delta w$, on the mean
equation state for the prior is independent of the number of
bins. Hence we have to scale the error in each bin, $\Delta w$
according to
\beq
(\Delta w)^2 = N(\delta w)^2\;. 
\eeq

Finally we will
assume that the maximum likelihood can be written as
\beq
P\left(D{|\boldsymbol\theta}_{\rm L},H\right) = {\cal
  N}e^{-\frac{1}{2}{\bar X}^2}\;.
\eeq

We will introduce the index $M$ to define the evidence for $M$
modes. Keeping in mind that we work in the basis where the entries in
the covariance matrix $\boldmath F$ are given by the eigenvalues we
can write the evidence for $M$ modes as
\begin{eqnarray}
{\cal E}_M &=& {\cal N}
\exp\left[-\frac{1}{2}{\bar
    X}_M^2\right]\times \nonumber
\\ &&\exp\left\{-\frac{1}{2}\sum_{i=1}^M\frac{{\bar\alpha}_i^2\lambda_i}{\lambda_i\Delta
    w^2+1}\right\}\times \prod\limits_{i=1}^M\frac{1}{\sqrt{\Delta w^2\lambda_i+1}}\; ,
\end{eqnarray}
where the first terms measure the 'goodness-of-fit' of the M modes,
the second term the overlap between the prior
and the likelihood and the last term is Occam's razor by comparing the posterior to the
prior volume. We can now construct the Bayes' factor \citep{jeffreys,Trotta:08}
\begin{eqnarray}
B_{M+1} &=& \left|\log_{10}\frac{{\cal E}_{M+1}}{{\cal
    E}_{M}}\right| \nonumber \\
&=&\frac{1}{2}\left|\left(\frac{{\bar X}^2_{M+1}-{\bar
    X}^2_M}{\ln 10}\right)\right. \nonumber \\
&&+\frac{1}{\ln
  10}\left[\sum_{i=1}^M\frac{\left(\bar{\beta}_i^2-{\bar\alpha}_i^2\right)\lambda_i}{\lambda_i\delta
    w^2N+1}+\frac{\bar{\beta}_{M+1}^2\lambda_{M+1}}{\lambda_{M+1}\delta
    w^2N+1}\right] \nonumber \\
&&\left.+\log_{10}\left(\lambda_{M+1}\delta
w^2N+1\right)\phantom{\frac{1}{2}}\right|\; ,
\label{eqn:bf}
\end{eqnarray}
where ${\bar X}_M$ and ${\bar X}_{M+1}$ are the the log-likelihood
values at the best fit points, if we fit for $M$ or $M+1$ modes
respectively. ${\bar \alpha}_i$ and ${\bar\beta}_i$ are the best
fit expansion parameters for these two cases.

We will first discuss the simple case with no bias. In this case
we choose ${\bar w}(z)=w_{\rm fid}$. Since we are expanding around the
fiducial model we obtain for all expansion coefficients
$\alpha_i=0$. Since there is no difference between ignoring modes or
putting $\alpha_i=0$ the best fit likelihood between $M$ and $M+1$
modes is exactly same, i.e. ${\bar X}_M={\bar X}_{M+1}$. Hence both
the best fit and the overlap term vanish for the Bayes' factor and
only Occam's razor term remains. If we further assume the prior is
wide compared to the likelihood we obtain for the Bayes' factor 
\beq
B_{M+1} \approx \left|\log_{10}\left(\frac{1}{\sqrt{\lambda_{M+1}}\sqrt{N}\delta
  w}\right)\right| = \left|\log_{10}\left[\frac{\sigma_{M+1}}{\sqrt{N}\delta
    w}\right] \right|\; .
\eeq
Hence, in this approximate scenario the evidence ratio is given by the
ratio of the likelihood uncertainty on the $M+1$st mode,
$\sigma_{M+1}$,  to the prior uncertainty on a single bin
$\sqrt{N}\delta w$. Although the equation of state at low redshifts is
determined almost to the 10\% level by current data, it is much more uncertain at
higher redshifts. We hence choose $\delta w = 1$ for the rms
uncertainty on the mean $w$ averaged over the entire redshift
range. According to Jeffrey's scale \citep{jeffreys} we 
have {\em strong} evidence if the Bayes' factor is 1-2 and substantial
evidence if it is between 0.5 and 1. We choose 1 as our evidence
level. If we employ this we find as a condition for strong
evidence
\beq
N\lambda_M \geq   100\;.
\eeq
The shaded region in Fig. \ref{fig:eigen_value} highlights the area
where this condition is violated. All points above the shaded region
are eigenmodes, which are significant. We obviously find that the
joint analysis can determine more higher order eigenmodes than the   
individual surveys, which is consistent with Fig.~(1)
in \citet{Crittenden:05}. This comes from the complementarity of the  
different dark energy probes. For most of the probes, the eigenvalues
descend exponentially.

We will now continue to analyse the full Bayes' factor expression in
Eqn.~\ref{eqn:bf} to decide for how many eigenmodes we have strong
evidence. We will discuss two cases: As before the unbiased case where
we reconstruct the equation of state around the fiducial model and a
biased case where we choose ${\bar w}(z)=-1$. The latter is what will
happen in a realistic situation, where the true underlying model is
not known a priori, although one could imagine an iterative method to
end up reconstructing around the true underlying model, but this is a
posterior statement.

\begin{figure}
   \includegraphics[width=8.0cm,angle=0]{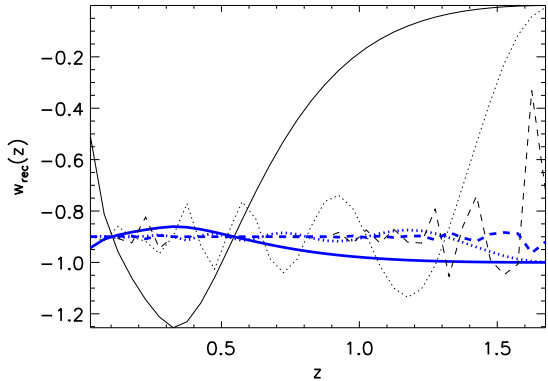}
   \caption{An example on reconstructing $w(z)$ from different $\bar{w} 
$ and $M$ for the SNAP SNe Ia survey. The solid line shows $M=2$
     and the dotted and dash lines show $M=10$ and $M=30$,
     respectively. The thin lines are for $\bar{w}=0$ and (blue)  
thick lines
     for $\bar{w}=-1.0$.
     \label{fig:w_reconstruct}}
\end{figure}

To illustrate how $\bar{w}$ will take effect on the reconstruction,
we give a very simple example for the SNAP SNe Ia survey.
We take the eigenmodes from the SNAP SNe Ia survey with fixed
cosmological parameters and use our fiducial cosmology with $w(z)=-0.9$.
In Fig.~(\ref{fig:w_reconstruct}) we plot the reconstructed $w_{\rm rec}(z)$ for
different $\bar{w}$ and $M$. The solid line is for $M=2$
and the dotted and dash lines for $M=10$ and $M=30$, respectively. The
thin (black) lines are for $\bar{w}=0$ and (blue) thick lines
represent $\bar{w}=-1.0$. One notices that no matter how we choose $ 
\bar{w}(z)$, $w_{\rm rec}(z)$ will be around
the fiducial model at low redshift, while at high redshift, $w_{\rm
   rec}(z)$ will be biased towards $\bar{w}(z)$. This is because SNe Ia
surveys have very weak constraints on $w(z)$ at high redshift.

Fig.~(\ref{fig:w_reconstruct}) also shows the oscillation on $w_{\rm  
  rec}$ at high redshift, which is an artifact of the PCA
decomposition. This is due to the limited number eigenmodes with $M<N$
that we choose to reconstruct $w(z)$ with. For $M=2$, $w_{\rm rec}$
has no constant behaviour with redshift. As we include more  
eigenmodes, $w_{\rm rec}$ starts to converge around $-0.9$
for redshifts $z<1$. The oscillation at high redshift are  
consistent with
the results in \citet{Huterer:03}, in which the analysis indicates
that it is not possible to recover $w(z)$ at redshift $z>1$ with SNAP
SNe data. It is hence
important to keep in mind the redshift evolution of the eigenmodes for
the interpretation of the reconstructed equation of state.
This simple example indicates that the choice of $\bar{w}$
will {\em make a difference} for the reconstructed $w(z)$.
If we compare the curves with $M=10$, we find that the thin line
starts oscillating around $z=0.2$, while the (blue) thick line starts  
oscillating at around $z=0.8$. The
dashed lines which show the behaviour if we include $M=30$ modes, both
the thin line and the (blue) thick line start oscillating at
$z>1$. However, the amplitude of the latter is much smaller indicating
that our initial guess value ${\bar w} = -1$ is much closer to the
true fiducial model. \citet{Huterer:03} use a mean square error (MSE)
criteria to find the optimal number of modes, which minimize 
bias and variance simultaneously. However this criteria does not
include an Occam's razor factor by comparison to the prior
information. \citet{Crittenden:05} use a similar MSE, but include the
prior variance. However their reconstruction is unbiased and they do
not use MSE to decide on the number of significant modes, they use MSE
over {\em all} modes to compare the effectiveness of different
probes. As stated above we will use the Bayes' factor to compare how
many modes are significantly constraint for different probes. We
believe that the measure provided by MSE of both \citet{Huterer:03}
and \citet{Crittenden:05} is included in our evidence expression and
incorporates both principles. Although the Jeffrey's scale is to some
extend arbitrary we can clearly identify for how many modes there is
strong evidence and for which modes there is only marginal evidence
for different surveys.

\begin{table}
   \begin{center}
     \begin{tabular}{|c|c|c|c|c|c|}
       \hline\hline
        Expt.       & SNe Ia    & WL    & CC    & BAO   & Joint     \\
        \hline\hline
        DES         &  2(3)     & 1(3)  & 1(2)  & 1(2)  &          \\
        PS1         &   -       & 2(3)  & -     & 1(2)  &          \\
        WFMOS(WIDE) &   -       & -     & -     & 3(3)  &         \\
        WFMOS(DEEP) &   -       & -     & -     & 2(4)  &         \\
        PS4         &  2(5)     & 3(3)  & -     & 1(2)  &         \\
        EUCLID      & -         & 3(4)  & -     & *     &          \\
        SNAP        &  2(5)     & 2(3)  & -     & -     &         \\
        SKA         & -         & -     &-      & 10(9) &         \\
        \hline\hline
        Joint III & -         & -     &-      &  &  6(8)        \\
        Joint IV & -         & -     &-      &  &  12(12)        \\
        
     \end{tabular}
     \caption{The number of w-eigenmodes with strong evidence
       according to Jeffrey's scale. The number in brackets are for
       the biased case, which is the realistic scenario.
	 For each survey, we marginalize over a Planck prior on the
         cosmological parameters. Note that EUCLID has a BAO
         component, which we have not analysed in the paper.}
\label{tb:optimal_M}
   \end{center}
\end{table}

In Table~\ref{tb:optimal_M}, we show the number of modes with strong
evidence for the unbiased case and in brackets for the biased
case. Note that in general the number of modes for the biased case are
larger, although there are exceptions. For each survey, we include a
Planck prior on the cosmological parameters. For most surveys, 
$M$ is around $2-4$ except for SKA. With the joint analysis, $M$ 
will be about $8$ for stage III, which indicates the complementarity
of the probes; for stage IV, $M$ is $12$, which is mainly driven by SKA.

\section{Conclusions}
\label{sec:conc}
In this paper we have studied future constraints on $w$-bins for four
dark energy probes, Type Ia Supernovae, weak lensing, cluster counts
and baryon acoustic oscillations.

Instead of assuming that the equation of state parameter $w$ can be  
modeled by a simple function with a
few parameters, we bin $w$ given by Eqn.~(\ref{eq:w_z}) and treat each
$w_{\rm i}$ as independent parameter. Throughout this article, we choose
fixed $\Delta z=0.05$ to make sure that we can compare between  
different surveys.
For each probe, we choose a few representative future surveys. We also
use a prior from the forthcoming Planck CMB experiment as prior on all
remaining cosmological parameters.
We find that in future mainly weak lensing and BAO surveys are
complementary, in a sense that weak lensing gives tight constrain at
low redshifts but BAO allows one to push to higher redshifts. However
the role of Supernovae is also complementary for low redshifts. The
high redshift sensitivity of the BAO would allow one to study also
early dark energy models \citep{2003ApJ...591L..75C} in an efficient
way. However at lower redshift weak lensing and Supernovae are the
more efficient probes, with the exception of SKA BAO measurements.
Typically forthcoming surveys constrain 2-3 modes. Since it is likely that there will be a multiple of
stage III probes it is sensible to look at the combined power of these
probes and we find that in this case five modes would be constraint. We
would advocate that the number of well constraint modes is actually a
better figure of merit than the one put forward by the DETF, since it
encompasses the ability of future surveys not just to measure the
equation of state at two redshifts, today and at the pivot
point. We address the question of the number of modes in a Bayesian model selection way
\citep{Saini:04,Liddle:04}. Although this method has recently
attracted some criticism \citep{2008arXiv0802.3185E} it provides a
well defined framework to establish the significant number of modes
for a particular survey. It takes into account the goodness of fit,
the bias and Occam's razor. However it is not only important to
establish the significant number of modes. In addition the redshift distribution of the
significant modes has to be taken into consideration to start talking
about the complementarity of surveys. The complementarity can than be
exploited to constrain eight modes for the joint stage III probes and
twelve for stage IV if we include SKA.
One question we have not addressed in this paper is how useful surveys
are to probe ``dark energy'' beyond the background evolution, for
example by probing structure formation. It is likely that in this
context future weak lensing, and to some extend galaxy cluster
surveys, will play a pivotal role.
To conclude we would argue that all four discussed probes have their
merits, but weak lensing and baryon acoustic oscillation surveys seem
to bear the largest promise for revealing the nature of the
accelerated expansion of the Universe.

\section*{Acknowledgments}
We gratefully thank Peter Capak and Huan Lin for kindly providing us
with simulation data. We sincerely thank Nick Kaiser
for providing us information related to Pan-Starrs 1. We acknowledge
Sarah Bridle, Rob Crittenden, Joshua Frieman, Gert H\"utsi, Ofer Lahav and Antony
Lewis for very
helpful discussions.

\label{lastpage}
\end{document}